%% file: vbs_ww.tex
\documentclass[a4paper,12pt]{article}
\pdfoutput=1

\usepackage{jheppub}
\usepackage{amssymb}
\usepackage{graphicx}
\usepackage[dvipsnames,table]{xcolor}
\usepackage{slashed}
\usepackage{hyperref} 
\usepackage{xspace}
\usepackage[tight]{subfigure}
\usepackage{amsmath}
\usepackage{amsfonts}
\usepackage{mathrsfs}
\usepackage{comment}
\usepackage{afterpage}
\usepackage{verbatim}
\usepackage{booktabs}
\usepackage{array}
\usepackage[title,titletoc]{appendix}
\usepackage{tabularx}

\newcolumntype{C}[1]{>{\centering\arraybackslash}p{#1}}

\newcommand{\nnb}{\nonumber}
\newcommand{\mc}{\mathcal}
\newcommand{\bw}[1]{{\rm BW}(#1)}

\newcommand{\dpatwotwo}{DPA$^{\rm (2,2)}$}
\input{macros.tex}

\title{NLO EW and QCD corrections to polarised same-sign $\PW\PW$ scattering at the LHC}
\author[a]{Ansgar Denner,}
\author[a]{Christoph Haitz,}
\author[b]{Giovanni Pelliccioli} 
\affiliation[a]{Institut f\"ur Theoretische Physik und Astrophysik,
  Universit\"at W\"urzburg, \\Emil-Hilb-Weg~22, 97074 W\"urzburg, Germany}
\affiliation[b]{Max-Planck-Institut f\"ur Physik, Bolzmannstra{\ss}e 8, 85748 Garching, Germany}
\emailAdd{ansgar.denner@physik.uni-wuerzburg.de}
\emailAdd{christoph.haitz@physik.uni-wuerzburg.de}
\emailAdd{gpellicc@mpp.mpg.de}

\abstract{
  We present the first calculation of same-sign $\PW\PW$ scattering at the LHC
  in the fully leptonic decay channel including the modelling of polarisation
  for intermediate electroweak bosons and radiative corrections up to NLO EW+QCD accuracy.
  The predictions rely on a pole expansion and on the split of
  polarisation states at matrix-element level. Doubly-polarised and unpolarised signals are
  investigated together with full off-shell results.
  A detailed phenomenlogical analysis is carried out focusing on differential
  observables that discriminate between polarisation states,
  paving the way for refined polarisation-oriented analyses of vector-boson scattering
  with Run-3 LHC data.
}%


\preprint{COMETA-2024-22, MPP-2024-178}

\begin{document}
\maketitle
\flushbottom

\section{Introduction}\label{sec:intro}
The scattering of massive electroweak (EW) bosons represents a gold-plated channel to access the
electroweak-symmetry-breaking (EWSB) mechanism realised in nature. As such, it provides a strong
probe of the EW and scalar sectors of the Standard Model (SM), as well as a crucial handle to unveil
potential new-physics effects \cite{Anders:2018oin,Covarelli:2021gyz,BuarqueFranzosi:2021wrv}.
In the SM, the scattering of longitudinally
polarised $\PW$ and $\PZ$ bosons at high energy is characterised 
by delicate cancellations among large contributions coming from pure-gauge-boson and Higgs-mediated
diagrams, which individually violate perturbative unitarity
\cite{Cornwall:1974km,Vayonakis:1976vz,Lee:1977yc,Chanowitz:1985hj}.
This motivates the strong interest in accessing vector-boson scattering (VBS) processes at the LHC.

The ATLAS and CMS collaborations have invested a huge effort to measure VBS
at the LHC, leading to the measurement of
$\PW^\pm\PW^\pm$ \cite{ATLAS:2016snd,CMS:2017fhs,ATLAS:2019cbr,CMS:2020gfh,ATLAS:2023sua},
$\PZ\PZ$ \cite{CMS:2017zmo,ATLAS:2020nlt,ATLAS:2023dkz},
$\PZ\PW^\pm$ \cite{ATLAS:2018mxa,CMS:2019uys,CMS:2020gfh,ATLAS:2024ini}, and
$\PW^+\PW^-$ \cite{CMS:2022woe,ATLAS:2024ett} scattering
in fully leptonic decay channels.
The search for new-physics effects in VBS has targeted possible deviations from the SM
quartic-gauge couplings \cite{CMS:2014mra,ATLAS:2016nmw,ATLAS:2016snd,CMS:2019qfk}.
More recently, the investigation of VBS in the semi-leptonic decay channel has started
\cite{ATLAS:2019thr, CMS:2019qfk, CMS:2021qzz}.
The first, and so far unique, measurement of polarised VBS has been achieved by CMS in the
same-sign $\PW\PW$ channel with leptonic decays \cite{CMS:2020etf}.

The calculation of higher-order corrections in the SM to VBS including
off-shell effects in the fully leptonic channel is known up to
next-to-leading-order (NLO) QCD
\cite{Jager:2006zc,Jager:2006cp,Bozzi:2007ur,Jager:2009xx,Denner:2012dz,Biedermann:2017bss,Ballestrero:2018anz,Denner:2019tmn,Denner:2020zit,Denner:2021hsa,Denner:2022pwc,Dittmaier:2023nac}
and NLO EW accuracy
\cite{Biedermann:2016yds,Biedermann:2017bss,Denner:2019tmn,Denner:2020zit,Denner:2021hsa,Denner:2022pwc,Dittmaier:2023nac}
for all production channels. For the production of $\PW^+\PW^+$ and
$\PZ\PZ$ in association with two jets, the complete NLO corrections, including those to the QCD irreducible background, have been computed \cite{Biedermann:2017bss,Denner:2021hsa,Dittmaier:2023nac}.
The matching of fixed-order corrections to parton-shower (PS) effects
has been achieved for VBS in case of leptonic decays at NLO QCD
\cite{Jager:2011ms,Jager:2013mu,Jager:2013iza,Rauch:2016upa,Ballestrero:2018anz,Jager:2018cyo,Jager:2024sij,Baglio:2024gyp,Jager:2024eet}
and more recently at NLO EW accuracy~\cite{Chiesa:2019ulk}.

The importance of VBS in unveiling possible new-physics effects is proven by
several studies in the context of the SM effective-field theory 
\cite{Brass:2018hfw,Zhang:2018shp,Gomez-Ambrosio:2018pnl,Chaudhary:2019aim,Dedes:2020xmo,Ethier:2021ydt,Bellan:2021dcy,Cappati:2022skp} and of
the Higgs effective-field theory
\cite{Delgado:2013hxa,Delgado:2015kxa,Fabbrichesi:2015hsa,Delgado:2017cls,Kozow:2019txg,Delgado:2019ucx,Quezada-Calonge:2022lop},
as well as within UV-complete models beyond the SM, like singlet extensions \cite{Ballestrero:2015jca} and
composite Higgs models \cite{BuarqueFranzosi:2017prc}.

Owing to the delicate cancellations between Higgs and pure-gauge contributions at high energies
in the longitudinal-vector-boson scattering, the LHC community has recently started to
access the polarisation state of intermediate EW bosons produced at the LHC.
Together with modern machine-learning approaches \cite{Searcy:2015apa,Lee:2018xtt,Lee:2019nhm,Grossi:2020orx,Kim:2021gtv,Li:2021cbp,Grossi:2023fqq}, the paradigm for polarisation extraction
from LHC Run-2 and Run-3 data is the so-called polarised-template method.
This has lead to the measurement of polarisation fractions and spin correlations in inclusive di-boson production \cite{Aaboud:2019gxl,CMS:2021icx,ATLAS:2022oge,ATLAS:2023zrv,ATLAS:2024qbd} and, although to a lesser extent owing to limited statistics, also in VBS \cite{CMS:2020etf}.
Sensitivity studies for longitudinally polarised VBS with upcoming LHC runs are promising \cite{Azzi:2019yne,Roloff:2021kdu}.
To enable the measurement of polarised cross-sections, the SM theoretical predictions have to be tailored to specific
helicity states of intermediate EW bosons, modelling properly their production and decay mechanisms.
This task has been achieved in recent years with a focus on inclusive
di-boson production, including radiative corrections up to (N)NLO QCD \cite{Denner:2020bcz,Denner:2020eck,Poncelet:2021jmj,Denner:2022riz}
and NLO EW \cite{Denner:2021csi,Le:2022lrp,Le:2022ppa,Dao:2023pkl,Denner:2023ehn,Dao:2023kwc} accuracy, and the (N)LO matching to PS \cite{Hoppe:2023uux,Pelliccioli:2023zpd,Javurkova:2024bwa}.
The predictions for polarised bosons in VBS are so far limited to LO
accuracy
\cite{Ballestrero:2017bxn,BuarqueFranzosi:2019boy,Ballestrero:2019qoy,Ballestrero:2020qgv,Hoppe:2023uux}
in the publicly available event generators \phantommc
\cite{Ballestrero:2007xq}, \madgraphnlos \cite{Alwall:2014hca} and
\Sherpa \cite{Bothmann:2019yzt}. Reaching NLO accuracy for polarised
VBS signals is urgently needed in view of the increased statistics of
the upcoming LHC data, which will soon enable refined VBS analyses and
more accurate extraction of polarisation fractions. 

Motivated by this urgency, in this work we have achieved for the first time NLO QCD + EW accuracy for (doubly) polarised
$\PW^+\PW^+$ production in association with two jets, specifically in the decay channel with two same-sign, opposite-flavour charged leptons. The calculation, based on an extension of methods introduced for inclusive di-boson production \cite{Denner:2020bcz,Denner:2021csi,Le:2022lrp,Denner:2023ehn},
represents a milestone for the investigation of di-boson systems produced at the LHC, paving the way for a deeper understanding of the spin structure and the consequent implications for the EWSB mechanism.

The article is structured as follows.
In \refse{sec:calcdetails} we describe the strategy employed to compute polarised signals,
tailored but not limited to VBS processes.
In \refse{sec:numericalres} we show the numerical results obtained for
doubly polarised $\PW^+\PW^+$ scattering at the LHC, including fiducial cross-sections,
polarisation fractions, and differential distributions. The presented results are relevant for the comparison with LHC Run-3 data. In \refse{sec:conclu} we draw the conclusions of our work.

\section{Details of the calculation}\label{sec:calcdetails}
In this section we describe the details of a general pole-expansion approach to obtain
a theoretically sound definition of polarised-vector-boson signals at NLO accuracy in the EW and QCD couplings.
While all techniques are fully general and applicable to any multi-boson collider process,
we take as an example a VBS process at the LHC,
\ie the production of two same-sign $\PW$~bosons in association with two jets, 
\beq\label{eq:ourproc}
\Pp\Pp\to\Pe^+\nu_{\Pe}\mu^+\nu_{\mu} + \Pj\Pj.
\eeq
As it is clear from \refeq{eq:ourproc}, the fully leptonic decay
channel with different final-state lepton flavours is considered in the
phenomenological analysis presented in this work, but the formalism applies also to hadronic decays
\cite{Denner:2022riz}.

In this work, we consider all LO contributions of
$\order{\alpha^6}$, $\order{\alphas\alpha^5}$, and
$\order{\alphas^2\alpha^4}$ to the process in \refeq{eq:ourproc}. At NLO we
only calculate the NLO EW and QCD corrections to the EW LO, \ie the
$\order{\alpha^7}$ and $\order{\alphas\alpha^6}$. The
$\order{\alpha^7}$ contributions are pure EW corrections to the LO
 purely EW cross-section of $\order{\alpha^6}$. The
$\order{\alphas\alpha^6}$ contributions, on the other hand, consist of QCD
corrections to $\order{\alpha^6}$ and EW corrections to the LO
interference $\order{\alphas\alpha^5}$, which cannot be separated
unambiguously. We nevertheless call these corrections QCD corrections
for simplicity,
as this is the dominant contribution \cite{Biedermann:2017bss}.

Following the general procedure used in inclusive di-boson processes
\cite{Denner:2020bcz,Poncelet:2021jmj,Denner:2021csi,Le:2022ppa,Denner:2023ehn},
our polarised-vector-boson calculation
in VBS boils down to two main stages: (1) the selection of
doubly-resonant topologies contributing to the SM amplitude at tree
level and at one-loop level, and (2) the selection of individual polarisation states in such doubly-resonant contributions.
We start the discussion from stage (1), \ie the selection of resonant diagrams and the use of a
pole approximation (PA) to render the resonant amplitude gauge invariant. In this regard, owing to the NLO accuracy of
the calculation, we separately consider the treatment of Born-like and real-radiation contributions in \refses{sec:BornDPA} and \ref{sec:realcorr}, respectively.
The application of the pole approximation to the Catani--Seymour (CS)
subtraction terms is discussed in \refses{sec:prod_counterterms} and \ref{sec:decay_counterterms}.
The selection of polarisation states, \ie stage (2), is described in \refse{sec:polsel}.

While the discussion is tailored to the VBS process in \refeq{eq:ourproc}, the calculation strategy is general and applies to any multi-boson process, provided a
suitable choice for an on-shell approximation is made, depending on the number of resonances and decay products. This can become especially involved when
more than one resonant structure contributes to the same final state,
as recently shown in tri-boson production \cite{Denner:2024ufg} and
semi-leptonic VBS \cite{Denner:2024xul}. A general algorithm for an
on-shell projection has been formulated in \citere{Denner:2024xul}.

\subsection{Born-like contributions in the pole approximation}\label{sec:BornDPA}
\begin{figure*}
   \centering
   \begin{tabular}{cc}
     \includegraphics[scale=0.4]{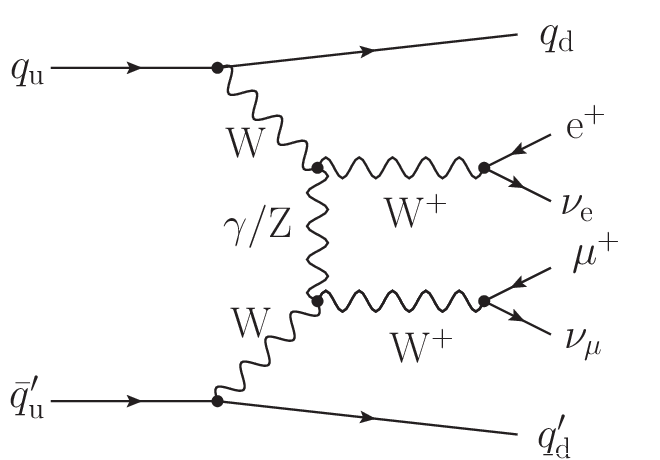} &
     \includegraphics[scale=0.4]{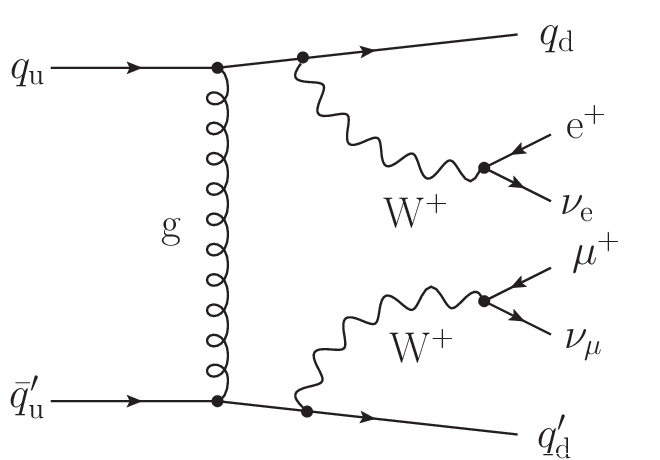} \\[0.2cm]
     \includegraphics[scale=0.4]{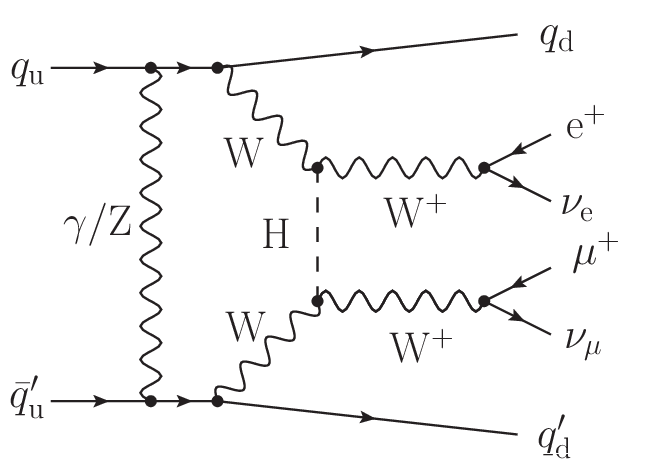} &
     \includegraphics[scale=0.4]{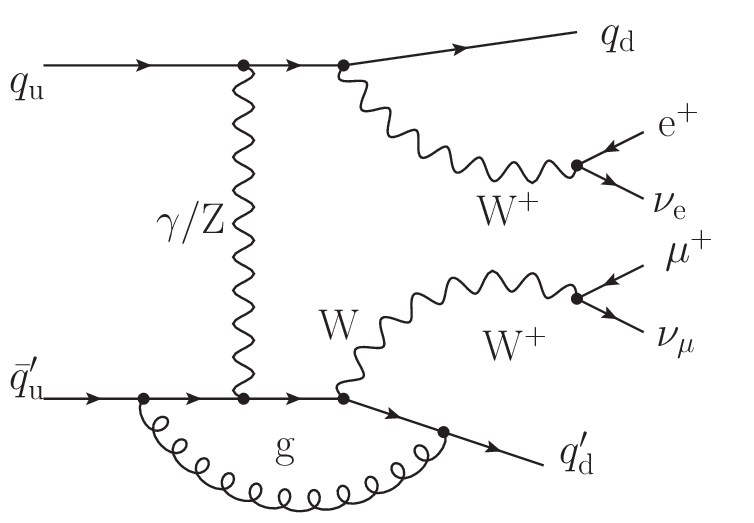} 
   \end{tabular}
   \caption{Sample doubly-resonant contributions to $\PW^+\PW^+$ scattering at LO EW (top left), NLO EW (bottom left) and NLO QCD (bottom right) accuracy, and to the QCD background at LO (top right). Both tree-level and one-loop diagrams are characterised by a Born-like kinematics.}\label{fig:diagsNbody}
\end{figure*}

The pole approximation (PA) \cite{Stuart:1991cc,Stuart:1991xk,Aeppli:1993rs,Denner:2000bj}, tailored to fully leptonic $\PW^+\PW^+$ scattering at the LHC, relies on the selection of diagram topologies that are characterised by two intermediate $s$-channel $\PW^+$ bosons which are produced in association with two jets and then undergo leptonic decays.
Owing to the resonance topology with two $\PW^+$~bosons, we refer interchangeably to the PA or to the double-pole approximation (DPA).
At NLO, both contributions with Born and real kinematics are
present. Those with Born kinematics come from Born matrix
elements, one-loop corrections of EW or QCD type, and integrated
subtraction counterterms needed to treat the infrared (IR) singularities. In the DPA such contributions are characterised by both $\PW^+$~bosons undergoing two-body decays.
In \reffi{fig:diagsNbody} we show sample doubly-resonant diagrams contributing in the PA to VBS at Born-level and at one loop. A tree-level contribution to the QCD background is also depicted.
The treatment of Born-like contributions in the PA follows straightforwardly from previous
literature results obtained for VBS \cite{Ballestrero:2017bxn,Ballestrero:2019qoy,Ballestrero:2020qgv}
and inclusive di-boson production \cite{Denner:2020bcz,Denner:2020eck,Denner:2021csi,Denner:2022riz,Denner:2023ehn}.
Sticking to the notation of \citere{Denner:2021csi}, we dub the PA applied to Born-like contributions \dpatwotwo.
In short, the \dpatwotwo\, procedure amounts to applying an on-shell
mapping to the Born kinematics
in the numerator of doubly-resonant matrix elements, preserving the off-shellness of intermediate EW bosons
in the denominator. In formulas, this reads
\begin{align}
\cM_{\rm res}
={}&
\cM^{\mu\nu}_{\rP}(q_1,q_2)
\frac{-\ri g_{\mu\alpha}}{q^2_{1}-\MW^2+\ri \GW \MW}\,
\frac{-\ri g_{\nu\beta}}{q^2_{2}-\MW^2+\ri \GW \MW}
{\cM_{\rD_1}^{\alpha}(q_1)}
\cM^{\beta}_{\rD_2}(q_2)\nnb\\
={}&
-\frac{\cM^{\mu\nu}_{\rP}({q}_1,{q}_2)
{\cM_{\rD_1,\mu}({q}_1)}
\cM_{\rD_2,\nu}({q}_2)}{
[q^2_{1}-\MW^2+\ri \GW \MW]
\, [q^2_{2}-\MW^2+\ri \GW \MW]}\nnb\\
\longrightarrow&
-\frac{\cM^{\mu\nu}_{\rP}(\tilde{q}_1,\tilde{q}_2)
{\cM_{\rD_1,\mu}(\tilde{q}_1)}
\cM_{\rD_2,\nu}(\tilde{q}_2)}{
[q^2_{1}-\MW^2+\ri \GW \MW]
\, [q^2_{2}-\MW^2+\ri \GW \MW]}\,=\cM_{\rm DPA}\,,
\end{align}
where $\cM_{\rP}$ represents the production amplitude for two $\PW^+$ bosons in association with two jets,
while $\cM_{\rD}$ is the two-body decay amplitude for
$\PW^{+}\rightarrow \ell^+\nu_\ell$. We have explicitly written
the momenta of the intermediate bosons before ($q_1,q_2$) and after the on-shell mapping ($\tilde{q}_1,\tilde{q}_2$).
While the on-shell mapping is not unique, the size of possible differences among various choices is expected to be
of the order of magnitude of the intrinsic DPA uncertainty, \ie $\mc
O(\GW/\MW)$ \cite{Denner:2000bj}, for observables that are inclusive
in the decay products of the unstable particles.
In the present work, we choose the same mapping as in
\citere{Denner:2021csi} (see Section~2.1 therein), which has been
generalised in \citere{Denner:2024xul}.

\subsection{Real corrections in the pole approximation}\label{sec:realcorr}
The most involved part of PA calculations concerns
the real-radiation contributions and, in particular, real-photon corrections to production and decay of EW bosons.
These corrections, in the specific case at hand, are part of the NLO
EW corrections to both the $\mc O(\alpha^6)$ and $\mc
O(\alphas\alpha^5)$ contributions.

We stress that in our PA calculation we only consider factorisable EW corrections, both in virtual and
in real-photon corrections. The impact of universal, non-factorisable corrections of soft-photon origin
\cite{Beenakker:1997bp,Denner:1997ia,Beenakker:1997ir,Dittmaier:2015bfe}
is known to be small if both real and virtual corrections are treated
in the PA. In fact, the sum of virtual and real non-factorisable
corrections cancels in observables that are inclusive with respect to
the decay products of the resonance \cite{Fadin:1993dz,Melnikov:1993np,Fadin:1993kt}.
We postpone their treatment to future investigations.

\begin{figure*}
   \centering
   \begin{tabular}{ll}
     \includegraphics[scale=0.45]{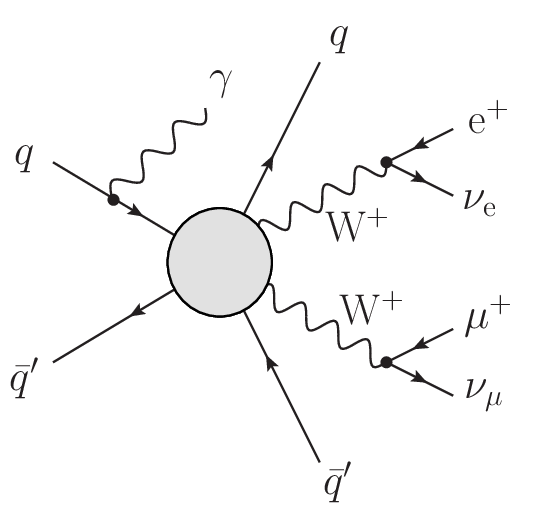} &
     \includegraphics[scale=0.45]{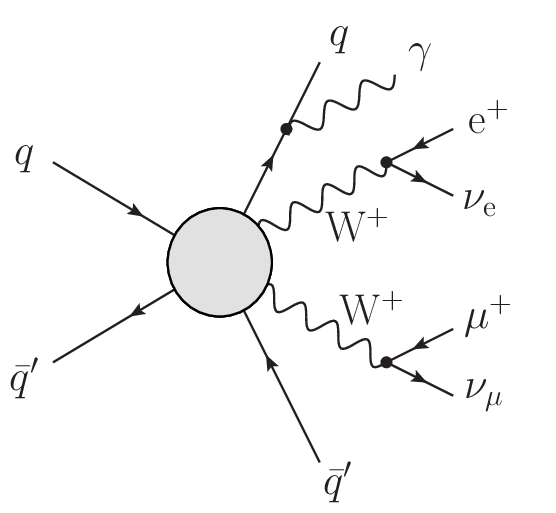} \\
     \includegraphics[scale=0.45]{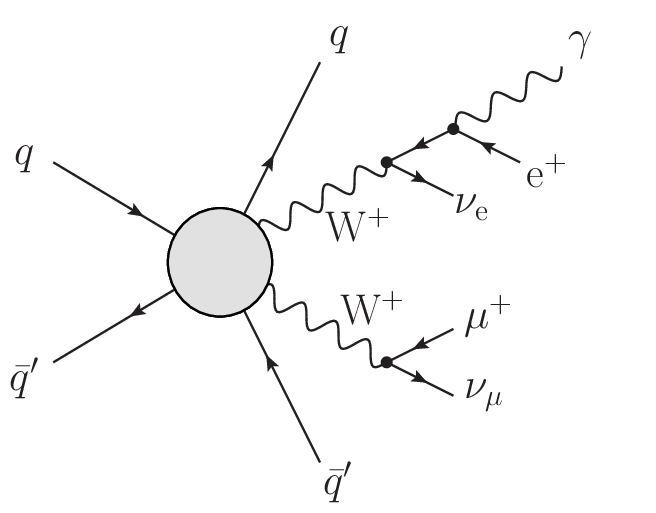} &
     \includegraphics[scale=0.45]{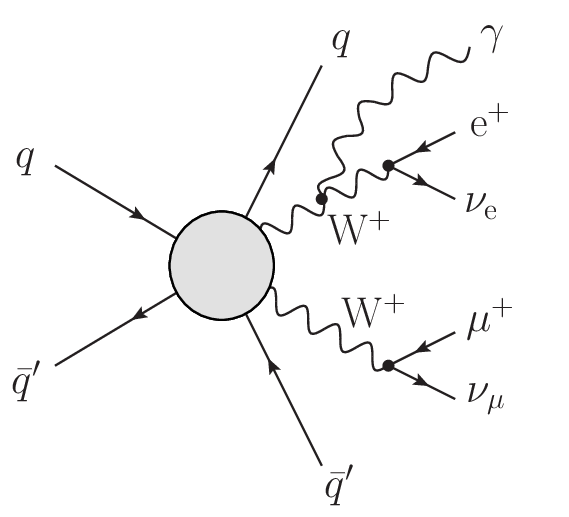} \\
   \end{tabular}
   \caption{Sample photon-radiation diagrams contributing to $\PW^+\PW^+$ scattering at NLO EW accuracy.}\label{fig:diags}
\end{figure*}

For simplicity, without loss of generality, we consider the real-photon contributions at
$\mc O(\alpha^7)$. Sample diagrams are shown in \reffi{fig:diags}.
The diagrams in the upper row of \reffi{fig:diags} involve photons
radiated off initial-state (IS) or final-state (FS) quarks, and can
therefore uniquely be associated to photon radiation off the
\emph{production process} of two $\PW$~bosons. The bottom--left
diagram is obviously associated to emission from the \emph{decay process} of a $\PW^+$~boson.
The bottom--right diagram cannot be associated to either the
{production} or the {decay sub-process} of VBS but embeds contributions to both.
On the other hand, the genuine QCD corrections to the LO EW process
are uniquely associated
to the {production} sub-process, owing to the leptonic decay of the bosons.
Their treatment in the PA is identical to the one of EW corrections to the {production} sub-process.

\begin{figure*}
  \centering
  \includegraphics[scale=0.48]{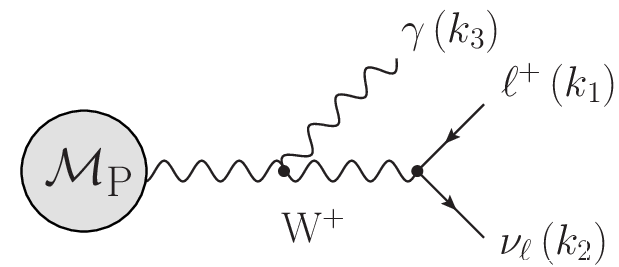} 
  \caption{Photon-radiation off a $\PW$-boson propagator.}\label{fig:photonW}
\end{figure*}

In order to allow for a separate treatment of real corrections
to {production} and {decay}
and of the corresponding IR-subtraction counterterms in the PA, the first step to take
is the split of contributions with a photon radiated off
a $\PW$-boson propagator via partial fractioning. This step, while essential for charged
resonances, \ie $\PW$~bosons at NLO EW and top~quarks at both NLO QCD and NLO EW,
is obviously not needed for neutral resonances like $\PZ$ and Higgs bosons. 
In \reffi{fig:photonW} we depict
the generic production (sub-amplitude $\cM_{\rP}$) of a $\PW$~boson
with momentum $q=k_1+k_2+k_3$ that radiates a photon with momentum $k_3$ and then
decays into a pair of leptons (with momenta $k_1$ and $k_2$).
The corresponding amplitude (which we label $\cM_{\rm prop}$) can be written as
\begin{align}\label{eq:pf1}
  \cM_{\rm prop} ={}& \quad\mc{N}_{\rm prop}(k_1,k_2,k_3)\,\,\frac1{s_{123}-\MW^2+\ri\MW\GW}\,\cdot\,\frac1{s_{12}-\MW^2+\ri\MW\GW}\nnb\\
  &= -\frac{\mc{N}_{\rm prop}(k_1,k_2,k_3)}{s_{13}+s_{23}}\left(\frac1{s_{123}-\MW^2+\ri\MW\GW}-\frac1{s_{12}-\MW^2+\ri\MW\GW}\right)\,,
\end{align}
where the invariants are defined as $s_{ab}=2k_a\!\cdot\!k_b$ and $s_{abc}=s_{ab}+s_{ac}+s_{bc}$.
If the PA is applied to the two-body decay of the $\PW$~boson (labelled PA$^{(2)}$), the resulting amplitude reads
\begin{align}\label{eq:pf2}
  \tilde{\cM}^{(2)}_{\rm prop} ={}&\frac1{s_{12}-\MW^2+\ri\MW\GW}\nnb\\
  &\, \times\left[\frac{\mc{N}_{\rm prop}({k}_1,{k}_2,{k}_3)}{s_{13}+s_{23}}\left(1-\frac{s_{12}-\MW^2+\ri\MW\GW}{s_{123}-\MW^2+\ri\MW\GW}\right)\right]_{s_{12}=\MW^2,\GW=0}\nnb\\
  ={}&\frac1{s_{12}-\MW^2+\ri\MW\GW}\left[\frac{\mc{N}_{\rm prop}(\tilde{k}^{(12)}_1,\tilde{k}^{(12)}_2,\tilde{k}^{(12)}_3)}{\tilde{s}^{(12)}_{13}+\tilde{s}^{(12)}_{23}}\right]\,,
\end{align}
where the tilde indicates on-shell-projected momenta within the approximation PA$^{(2)}$.
If the PA is applied to the three-body decay of the $\PW$~boson (labelled PA$^{(3)}$), the amplitude reads
\begin{align}\label{eq:pf3}
  \tilde{\cM}^{(3)}_{\rm prop} ={}& \frac1{s_{123}-\MW^2+\ri\MW\GW}\nnb\\
  &\, \times\left[-\frac{\mc{N}_{\rm prop}({k}_1,{k}_2,{k}_3)}{s_{13}+s_{23}}\left(1-\frac{s_{123}-\MW^2+\ri\MW\GW}{s_{12}-\MW^2+\ri\MW\GW}\right)\right]_{s_{123}=\MW^2,\GW=0}\nnb\\
  ={}& \frac1{s_{123}-\MW^2+\ri\MW\GW}\left[-\frac{\mc{N}_{\rm prop}(\tilde{k}^{(123)}_1,\tilde{k}^{(123)}_2,\tilde{k}^{(123)}_3)}{\tilde{s}^{(123)}_{13}+\tilde{s}^{(123)}_{23}}\right]\,,
\end{align}
where the tilded momenta correspond to the on-shell-projected momenta within the approximation PA$^{(3)}$.
Note that the partial fractioning performed in \refeq{eq:pf1} gives exactly the sum of the two contributions
that must be treated with PA$^{(2)}$ and PA$^{(3)}$, respectively, and which correspond to the results of
\refeqs{eq:pf2} and \eqref{eq:pf3}.
We recall that the combination of one PA$^{(2)}$ mapping (two-body
decay of one $\PW$~boson) with one PA$^{(3)}$ mapping (three-body
decay of the other $\PW$~boson) gives rise to the DPA$^{(2,3)}$ and DPA$^{(3,2)}$ procedures in the notation introduced in \citere{Denner:2021csi}, while two PA$^{(2)}$ mappings correspond the DPA$^{(2,2)}$ mapping in the same reference.

\subsection{Subtraction dipoles for production sub-processes}\label{sec:prod_counterterms}
\begin{figure}[t]
   \centering
   \subfigure[FF$_{m,m'}$\label{dipm1}]{
     \includegraphics[scale=0.52]{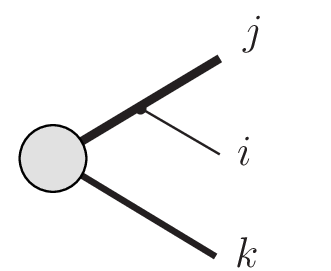}} 
   \subfigure[FF$_{m,0}$\label{dipm2}]{
     \includegraphics[scale=0.52]{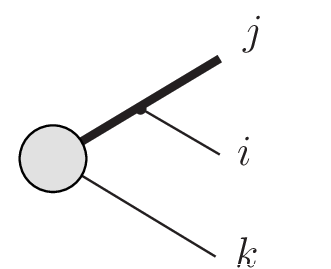}} 
   \subfigure[FF$_{0,m}$\label{dipm3}]{
\includegraphics[scale=0.52]{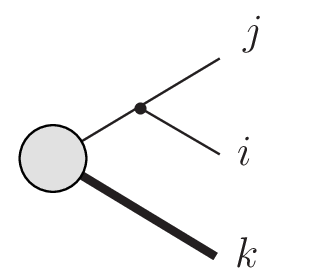}}\\ 
   \subfigure[FI$_{m,0}$  \label{dipm4}]{
     \includegraphics[scale=0.52]{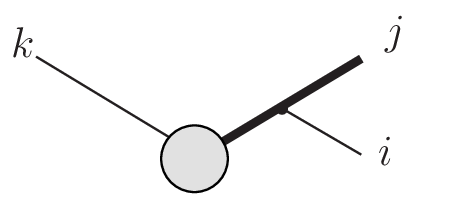}} 
   \subfigure[IF$_{0,m}$  \label{dipm5}]{
     \includegraphics[scale=0.52]{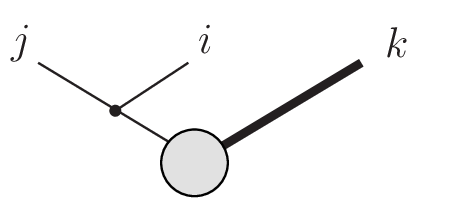}} 
   \caption{QED subtraction dipoles involving one or two massive
     external particles (highlighted as bold lines). In $\PW\PW$
     scattering at the LHC the massive particles are on-shell
     $\PW$~bosons, but in general they can be any massive charged 
     particles. Analogous QCD dipoles appear in the presence of
     massive coloured particles, like top quarks. The indices indicate
     the masses of emitters (first indices) and spectators (second
     indices) and the labels F and I
     stand for final and initial state.}\label{fig:massdipole}
\end{figure}

Since in VBS the QCD real corrections to the LO EW signal can only affect the quark lines, 
the subtraction counterterms needed to regulate QCD IR singularities
are of the same kind as those
appearing in the full off-shell calculation (no emission from
particles affected by the on-shell projection). This holds for gluon radiation and gluon-induced real processes.
A similar reasoning applies to the photon--quark-initiated
processes. In our calculation, the initial-state collinear singularity
associated to the $\gamma\rightarrow q\bar{q}$ splitting is regulated
by a unique dipole where the other incoming parton plays the role of the spectator, as also done in inclusive $\PW^+\PW^-$~production \cite{Denner:2023ehn}.
When the real photon appears in the final state and is emitted off the
production sub-process, the subtraction counterterms for the
on-shell process
\beq\label{eq:wwjja}
qq \longrightarrow \PW^+\PW^+qq + \gamma
\eeq
are needed to properly subtract soft and collinear singularities.
In the dipole formalism \cite{Catani:2002hc}, the list of subtraction counterterms associated to the process in \refeq{eq:wwjja} include massless dipoles with a quark as an emitter and another quark as a spectator, but also dipoles with $\PW$~bosons acting as emitters and/or spectators.
Such dipoles are depicted in \reffi{fig:massdipole}.
At LHC energies, the photon radiation off a $\PW$~boson leads only to
soft divergences which are independent of the spin of the emitting
particle, while the collinear singularities are absent owing to the
finite $\PW$-boson mass.
Therefore, we can safely use dipoles devised for massive fermions \cite{Catani:2002hc}
upon replacing the massive fermion with a $\PW$~boson.
Note that massive dipoles of the type of \reffis{dipm1}, \ref{dipm4},
and \ref{dipm5} are also present in inclusive $\PW^+\PW^-$ production
in the DPA \cite{Denner:2023ehn,Dao:2023kwc} and are used here
following the same implementation as in \citere{Denner:2023ehn}.
Dipole types shown in \reffis{dipm2} and \ref{dipm3} appear for the
first time in DPA calculations of boson-pair production in association
with jets and have been newly introduced for this calculation in the
context of polarised predictions.

For the proper subtraction of the IR singularities, the local counterterms must embed the same singular behaviour as the real-emission matrix element. Additionally, an exact correspondence between the integrated and local counterterms must be ensured to avoid unphysical dependences on the subtraction scheme.
These unavoidable requirements entail important consequences for the application of the
on-shell projections in the PA and the subtraction mappings.
The soft-photon emission off a $\PW$~boson leads to a
single IR pole in dimensional regularisation only if the bosons are
on~shell, complicating the interplay between the
subtraction counterterms for the production sub-process and the DPA$^{(2,2)}$ procedure, as we are
going to show in the following.

In order to further detail this delicate aspect, let us consider the partonic process of \refeq{eq:wwjja} with the momentum assignment
\beq
q\,q
\rightarrow
\underbrace{\mu^+(k_1)\,\nu_\mu(k_2)}_{\PW^+(k_{12})}\,
\underbrace{\Pe^+(k_3)\,\nu_{\Pe}(k_4)}_{\PW^+(k_{34})}\,
q(k_5)\,q(k_6)\,
\gamma(k_7)\,,
\eeq
and focus on the dipole with a massive $\PW$~boson as an emitter (with momentum $k_{12}$, decaying into two leptons with momenta $k_1,k_2$), a massless final-state quark as spectator (with momentum $k_5$), and an emitted photon with momentum $k_7$. This corresponds to the dipole type FF$_{m,0}$ in \reffi{fig:massdipole}.
Although tailored to a specific dipole, the following discussion can
be straight-forwardly extended to other processes and dipoles with a charged resonance as emitter and/or spectator.
Since we consider the DPA contribution from the sub-process for the
production of an off-shell $\PW$~boson, the singularity associated to
this dipole becomes only manifest after setting the $\PW$~boson on its mass shell.
Momenta resulting from a DPA projection are indicated with a tilde,
while those originating from CS mappings get a bar.
For simplicity we abbreviate the denominators of the off-shell propagators of the two $\PW$~bosons as
\beq
\bw{k_{ij}}=\left(k_{ij}^2-\MW^2\right)^2+\bigl(\GW\MW\bigr)^2\,.
\eeq
We write the real-photon contribution $\mc R$ to the differential cross-section
with the off-shell phase-space measure factorised in a production
phase-space measure, off-shell virtualities of the \PW~bosons, and two
phase-space measures for the two-body decays as (suppressing flux
factor and symmetry factors)
\begin{align}\label{eq:realdpa_2}
  \mc R \propto{}& \frac{\left|\cM^{(7)}\left(Q; \tilde{k}_1,..\,, \tilde{k}_4, k_5,k_6,k_7\right)\right|^2}{\bw{k_{12}}\,\bw{k_{34}}}\,\rd\Phi_7\left(Q;{k}_1,..\,, {k}_5,k_6,k_7\right)\,\notag\\
  ={}&
  \frac{\left|\cM^{(7)}\left(Q; \tilde{k}_1,..\,, \tilde{k}_4, k_5,k_6,k_7\right)\right|^2}{\bw{k_{12}}\,\bw{k_{34}}}\,\rd\Phi_5\left(Q;{k}_{12},{k}_{34}, k_5,k_6,k_7\right)\notag\\
  & \times\frac{\rd k_{12}^2}{2\pi}\frac{\rd k_{34}^2}{2\pi} \,\rd\Phi_2\left(k_{12};{k}_1,k_2\right)
  \rd\Phi_2\left(k_{34};{k}_3,k_4\right)\,.
\end{align}
The numerator of the real matrix element is then factorised as
\begin{align}
  \left|\cM^{(7)}\left(Q; \tilde{k}_1,..\,, \tilde{k}_4,
      k_5,k_6,k_7\right)\right|^2 &{}= 
\sum_{\lambda_{12},\lambda_{34}}\left|\cM^{(5)}_{\rP,\mu\nu}\left(Q; \tilde{k}_{12}, \tilde{k}_{34}, k_5,k_6,k_7\right) {\tilde{\varepsilon}}^{\mu,*}_{12}\tilde{\varepsilon}^{\nu,*}_{34}\right|^2\nnb\\
  &\times \,\left|{\tilde{\varepsilon}}^\mu_{12}\cM_{\rD,\mu}^{(2)}\left( \tilde{k}_{12}; \tilde{k}_1,\tilde{k}_2\right) \right|^2
  \,\left|{\tilde{\varepsilon}}^\mu_{34}\cM_{\rD,\mu}^{(2)}\left(
      \tilde{k}_{34}; \tilde{k}_3,\tilde{k}_4\right) \right|^2 ,
\end{align}
where the sum runs over the polarisations $\lambda_{12}$,
$\lambda_{34}$ of the \PW~bosons with momenta $\tilde k_{12}$,
$\tilde k_{34}$, respectively, 
[see also \refse{sec:polsel}, in particular \refeq{eq:polsum}]
leading to the following contribution to the cross-section:
\begin{align}
                \mc R \propto{}& \sum_{\lambda_{12},\lambda_{34}}\dfrac{\left|\cM^{(5)}_{\rP,\mu\nu}\left(Q; \tilde{k}_{12}, \tilde{k}_{34}, k_5,k_6,k_7\right) {\tilde{\varepsilon}}^{\mu,*}_{12}\tilde{\varepsilon}^{\nu,*}_{34}\right|^2}{\bw{k_{12}}\,\bw{k_{34}}}\notag\\
                & \times \left|{\tilde{\varepsilon}}^\mu_{12}\cM_{\rD,\mu}^{(2)}\left( \tilde{k}_{12}; \tilde{k}_1,\tilde{k}_2\right) \right|^2\,\left|{\tilde{\varepsilon}}^\mu_{34}\cM_{\rD,\mu}^{(2)}\left( \tilde{k}_{34}; \tilde{k}_3,\tilde{k}_4\right) \right|^2\notag\\
                & \times\,\rd\Phi_5\left(Q;{k}_{12},{k}_{34},
                  k_5,k_6,k_7\right)\frac{\rd k_{12}^2}{2\pi}\frac{\rd
                  k_{34}^2}{2\pi} \,
  \rd\Phi_2\left(k_{12};{k}_1,k_2\right)
  \rd\Phi_2\left(k_{34};{k}_3,k_4\right)\, .
\end{align}
The two-body phase-space measure for the $\PW$-boson decay reads in $d=4-2\epsilon$ dimensions
\beq
\rd\Phi_2\left(k_{12};{k}_1,k_2\right) = \frac{\left(k_{12}^2\right)^{-\epsilon}}{2\,(4\pi)^{2-2\epsilon}}\,\rd\Omega_2^{(2-2\epsilon)}\,,
\eeq
and after applying the \dpatwotwo
\beq
\rd\Phi_2\left(\tilde{k}_{12};\tilde{k}_1,\tilde{k}_2\right) = \frac{\left(\tilde{k}_{12}^2\right)^{-\epsilon}}{2\,(4\pi)^{2-2\epsilon}}\,\rd\Omega_2^{(2-2\epsilon)}\,,
\eeq
as the \dpatwotwo\, conserves the angles ($\Omega_2^{(2-2\epsilon)}$) of final state leptons in the $\PW$-boson rest frame \cite{Denner:2021csi}. This means that the
off-shell and on-shell-projected decay phase-space measures only
differ by a term $(k_{12}^2/\tilde{k}_{12}^2)^{\epsilon}$. Since the
decay parts (matrix element and phase-space measures) of the
DPA-projected real contribution are treated at LO, they are always
evaluated in $d=4$, where the difference disappears.

Turning to the discussion of the local dipole, we choose to first
apply (generalised) CS mappings to the real off-shell kinematics, and 
thereafter apply the DPA on-shell projection to the reduced Born kinematics.
This choice is especially important as the reverse order (DPA
projection first, CS mapping second) leads to a mismatch between local
and integrated dipoles within the accuracy of the DPA, as discussed in
detail in Appendix \ref{sec:appendix_dpa_cs}.  

The real-subtracted contribution $\mc R - \mc D$, where $\mc R$ is
the real contribution to the cross-section and $\mc D$ the one of the
considered dipole, before application of the DPA reads:
\begin{align}\label{eq:rsubtr_2_cs}
  \mc R-\mc D \propto{}& \sum_{\lambda_{12},\lambda_{34}}
\biggl[
\left|\cM^{(5)}_{\rP,\mu\nu}\left(Q;  {k}_{12},  {k}_{34},
    k_5,k_6,k_7\right) { {\varepsilon}}^{\mu,*}_{12} {\varepsilon}^{\nu,*}_{34}\right|^2 \notag\\
                &\times \dfrac{1}{\bw{k_{12}}}\left|{ {\varepsilon}}^\mu_{12}\cM_{\rD,\mu}^{(2)}\left(  {k}_{12};  {k}_1, {k}_2\right) \right|^2 \notag\\
                & \times\,\rd\Phi_5\left(Q;{k}_{12},{k}_{34}, k_5,k_6,k_7\right)\frac{\rd k_{12}^2}{2\pi} \rd\Phi_2\left(k_{12};{k}_1,k_2\right) \notag\\
                &-{\mc D}_{[12]7,5}(\bar{ {k}}_{12},\bar{k}_5; {y},  {z},  {\phi})\, \left|\cM^{(4)}_{\rP,\mu\nu}\left(Q; \bar{ {k}}_{12},  {k}_{34}, \bar{k}_5,k_6\right) { \bar{\varepsilon}}^{\mu,*}_{12} {\varepsilon}^{\nu,*}_{34}\right|^2 \notag\\
                &\times \dfrac{1}{\bw{\bar{k}_{12}}}\left|{ \bar{\varepsilon}}^\mu_{12}\cM_{\rD,\mu}^{(2)}\left(  \bar{k}_{12};  \bar{k}_1, \bar{k}_2\right) \right|^2 \notag\\
                & \times\,\rd\Phi_{\rm rad}\left(\bar{k}_{12}+\bar{k}_5;z,y,\phi\right)\rd\Phi_4\left(Q;\bar{k}_{12},{k}_{34}, \bar{k}_5,k_6\right)
                \frac{\rd \bar{k}_{12}^2}{2\pi}
                \rd\Phi_2\left(\bar{k}_{12};\bar{k}_1,\bar{k}_2\right)
\biggl]
\notag\\
&\times\dfrac{1}{\bw{k_{34}}}\,\left|{ {\varepsilon}}^\mu_{34}\cM_{\rD,\mu}^{(2)}\left(  {k}_{34};  {k}_3, {k}_4\right) \right|^2
\frac{\rd k_{34}^2}{2\pi}
               \, \rd\Phi_2\left(k_{34};{k}_3,k_4\right)
\,,
\end{align}
where the generalised CS mapping has been applied to off-shell
momenta, projecting the momenta of the 
decay leptons $\{k_1,k_2\}$ in such a way that the lepton angles
in the rest frame of the decaying boson are preserved
\cite{Le:2022ppa,Denner:2024xul}. 
The dipole functions ${\mc D}_{[12]7,5}(\bar{ {k}}_{12},\bar{k}_5;
{y},  {z},  {\phi})$ refer to the emitter $[12]$ the spectator $5$ and
the emitted particle $7$ and depend on the three corresponding
momenta, where
the variables $ {y},  {z},  {\phi}$
parametrise the momentum of the radiated particle.
In order to have a proper correspondence between subtraction and
integrated dipoles, we need to add back the integrated contribution that reads (before the DPA projection),
\begin{align}\label{eq:IntDip}
  \mc  I ={}& 
  \sum_{\lambda_{12},\lambda_{34}}
{\mc I}_{[12]7,5}(\bar{ {k}}_{12},\bar{k}_5)\dfrac{1}{\bw{\bar{k}_{12}}\,\bw{k_{34}}}{\,\left|\cM^{(4)}_{\rP,\mu\nu}\left(Q; \bar{ {k}}_{12},  {k}_{34}, \bar{k}_5,k_6\right) { \bar{\varepsilon}}^{\mu,*}_{12} {\varepsilon}^{\nu,*}_{34}\right|^2}\notag\\
                & \times\,\left|{ \bar{\varepsilon}}^\mu_{12}\cM_{\rD,\mu}^{(2)}\left(  \bar{k}_{12};  \bar{k}_1, \bar{k}_2\right) \right|^2\,\left|{ {\varepsilon}}^\mu_{34}\cM_{\rD,\mu}^{(2)}\left(  {k}_{34};  {k}_3, {k}_4\right) \right|^2\notag\\
                & \times\, \rd\Phi_4\left(Q;\bar{k}_{12},{k}_{34}, \bar{k}_5,k_6\right)\frac{\rd \bar{k}_{12}^2}{2\pi}\frac{\rd k_{34}^2}{2\pi} \rd\Phi_2\left(\bar{k}_{12};\bar{k}_1,\bar{k}_2\right) \rd\Phi_2\left(k_{34};{k}_3,k_4\right),
\end{align}
where
\begin{equation}\label{eq:Idef1}
        \begin{split}
                {\mc I}_{[12]7,5}(\bar{ {k}}_{12},\bar{k}_5) =& \int_{d=4-2\epsilon} {\mc D}_{[12]7,5}(\bar{ {k}}_{12},\bar{k}_5; {y},  {z},  {\phi})\, \rd\Phi_{\rm rad}\left(\bar{k}_{12}+\bar{k}_5;z,y,\phi\right)\,
        \end{split}
\end{equation}
contains explicit $\epsilon$ poles only in the on-shell limit of the $\PW$-boson virtuality.

After applying the DPA to both \refeqs{eq:rsubtr_2_cs} and \eqref{eq:IntDip},
and upon re-writing the phase-space measures without the factorisation into production $\times$ decay,
the real-subtracted and integrated-dipole contributions read
\begin{align}\label{eq:rsubtr_2_dpa}
                \mc R-\mc D \propto{}& 
  \sum_{\lambda_{12},\lambda_{34}}\biggl[
\left|\cM^{(5)}_{\rP,\mu\nu}\left(Q;  \tilde{k}_{12},  \tilde{k}_{34}, k_5,k_6,k_7\right) { \tilde{\varepsilon}}^{\mu,*}_{12} \tilde{\varepsilon}^{\nu,*}_{34}\right|^2 \notag\\
                &\times \frac{1}{\bw{k_{12}}}\left|{ \tilde{\varepsilon}}^\mu_{12}\cM_{\rD,\mu}^{(2)}\left(  \tilde{k}_{12};  \tilde{k}_1, \tilde{k}_2\right) \right|^2\,\rd\Phi_7\left(Q;{k}_{1},\ldots,k_7\right)  \notag\\
                & -{\mc D}_{[12]7,5}(\tilde{\bar{k}}_{12},\bar{k}_5; \tilde{y},  \tilde{z},  \tilde{\phi})\,\left|\cM^{(4)}_{\rP,\mu\nu}\left(Q; \tilde{\bar{k}}_{12},  \tilde{k}_{34}, \bar{k}_5,k_6\right) \tilde{ \bar{\varepsilon}}^{\mu,*}_{12} \tilde{\varepsilon}^{\nu,*}_{34}\right|^2 \notag\\
                &\times \frac{1}{\bw{\bar{k}_{12}}}\left|\tilde{ \bar{\varepsilon}}^\mu_{12}\cM_{\rD,\mu}^{(2)}\left(  \tilde{\bar{k}}_{12};  \tilde{\bar{k}}_1, \tilde{\bar{k}}_2\right) \right|^2 \notag\\
                &\times \rd\Phi_{\rm
                  rad}\left(\bar{k}_{12}+\bar{k}_5;z,y,\phi\right)\rd\Phi_6\left(Q;\bar{k}_{1},\bar{k}_{2},{k}_{3}, k_4, \bar{k}_5,k_6\right)\biggr]\, \notag\\
 & \times \frac{1}{\bw{k_{34}}}\left|{ \tilde{\varepsilon}}^\mu_{34}\cM_{\rD,\mu}^{(2)}\left(  \tilde{k}_{34};  \tilde{k}_3, \tilde{k}_4\right) \right|^2 \,
, 
\end{align}
\begin{align}\label{eq:newInteg_dip}
  \mc I \propto{}& 
  \sum_{\lambda_{12},\lambda_{34}}\biggl[
{{\mc I}_{[12]7,5}(\tilde{ \bar{k}}_{12},\bar{k}_5)\, \left|\cM^{(4)}_{\rP,\mu\nu}\left(Q; \tilde{ \bar{k}}_{12},  \tilde{k}_{34}, \bar{k}_5,k_6\right) \tilde{\bar{\varepsilon}}^{\mu,*}_{12} \tilde{\varepsilon}^{\nu,*}_{34}\right|^2} \notag\\
                &\times \frac{1}{\bw{\bar{k}_{12}}\,\bw{k_{34}}} \left|\tilde{ \bar{\varepsilon}}^\mu_{12}\cM_{\rD,\mu}^{(2)}\left(  \tilde{\bar{k}}_{12};  \tilde{\bar{k}}_1, \tilde{\bar{k}}_2\right) \right|^2\,\left|{ \tilde{\varepsilon}}^\mu_{34}\cM_{\rD,\mu}^{(2)}\left(  \tilde{k}_{34};  \tilde{k}_3, \tilde{k}_4\right) \right|^2 \notag\\
                &\times \rd\Phi_6\left(Q;\bar{k}_{1},\bar{k}_{2},{k}_{3}, k_4, \bar{k}_5,k_6\right)\biggr],
\end{align}
respectively, where
\beq
{\mc I}_{[12]7,5}(\tilde{ \bar{k}}_{12},\bar{k}_5) = \int_{d=4-2\epsilon}\!{\mc D}_{[12]7,5}(\tilde{\bar{k}}_{12},\bar{k}_5;\tilde{y},\tilde{z},\tilde{\phi})\,\rd\Phi_{\rm rad}\left(\tilde{\bar{k}}_{12}+\bar{k}_5;\tilde{z},\tilde{y},\tilde{\phi}\right)\,
\eeq
has the same functional structure as \refeq{eq:Idef1} but depends on
the on-shell-projected momenta. 
The change from off-shell to on-shell radiation variables $ \tilde{y},  \tilde{z},
\tilde{\phi}$ in the $d$-dimensional integration only amounts to a relabelling of the variables and does not give a Jacobian.

Disregarding the phase-space measure and matrix elements for decays
(which are not modified by CS mappings and by the DPA up to
terms that equal 1 for $d=4$),
in the limit of a soft/collinear photon we have
$\{{\bar{k}}_{12}, {k}_{34}, \bar{k}_5,k_6\}\rightarrow \{{k}_{12}, {k}_{34},{k}_5,k_6\}$,
and therefore,
\begin{align}
        &\dfrac{\left|\cM^{(5)}_{\rP,\mu\nu}\left(Q;  \tilde{k}_{12},  \tilde{k}_{34}, k_5,k_6,k_7\right) { \tilde{\varepsilon}}^{\mu,*}_{12} \tilde{\varepsilon}^{\nu,*}_{34}\right|^2}{\bw{k_{12}}\,\bw{k_{34}}}\nnb\\
        &-\dfrac{{\mc D}_{[12]7,5}(\tilde{\bar{k}}_{12},\bar{k}_5; \tilde{y},  \tilde{z},  \tilde{\phi})\, \left|\cM^{(4)}_{\rP,\mu\nu}\left(Q; \tilde{\bar{k}}_{12},  \tilde{k}_{34}, \bar{k}_5,k_6\right) \tilde{ \bar{\varepsilon}}^{\mu,*}_{12} \tilde{\varepsilon}^{\nu,*}_{34}\right|^2}{\bw{\bar{k}_{12}}\,\bw{k_{34}}}\nnb\\
        \rightarrow{}&\notag\\
        &\dfrac{\left|\cM^{(5)}_{\rP,\mu\nu}\left(Q;  \tilde{k}_{12},  \tilde{k}_{34}, k_5,k_6,k_7\right) { \tilde{\varepsilon}}^{\mu,*}_{12} \tilde{\varepsilon}^{\nu,*}_{34}\right|^2}{\bw{k_{12}}\,\bw{k_{34}}}\nnb\\
        &-\dfrac{{\mc D}_{[12]7,5}(\tilde{{k}}_{12},{k}_5; \tilde{y},  \tilde{z},  \tilde{\phi})\,
    \left|\cM^{(4)}_{\rP,\mu\nu}\left(Q; \tilde{{k}}_{12},  \tilde{k}_{34}, {k}_5,k_6\right) \tilde{ {\varepsilon}}^{\mu,*}_{12} \tilde{\varepsilon}^{\nu,*}_{34}\right|^2}{\bw{k_{12}}\,\bw{k_{34}}},
\end{align}
giving the correct local subtraction of the production-level phase-space singularity in the real-photon contribution.

Next, we analyse the correspondence between the subtraction
counterterms and integrated counterterms.
The difference between the two equations,
\beqn
&\rm \refeq{eq:rsubtr_2_dpa}:&\int_{(d=4)}\!{\mc D}_{[12]7,5}(\tilde{\bar{k}}_{12},\bar{k}_5; \tilde{y},  \tilde{z},  \tilde{\phi})\, \rd\Phi_{\rm rad}\left(\bar{k}_{12}+\bar{k}_5;z,y,\phi\right),\nnb\\
&\rm \refeq{eq:newInteg_dip}:&\int_{(d=4-2\epsilon)}\!{\mc D}_{[12]7,5}(\tilde{\bar{k}}_{12},\bar{k}_5;\tilde{y},\tilde{z},\tilde{\phi})\,\rd\Phi_{\rm rad}\left(\tilde{\bar{k}}_{12}+\bar{k}_5;\tilde{z},\tilde{y},\tilde{\phi}\right)\,\\
&\qquad &\qquad = \,{\mc I}_{[12]7,5}(\tilde{\bar{k}}_{12},\bar{k}_5)\,,\nnb
\eeqn
only resides in the phase-space measures, which are related by the
application of the DPA on-shell projection.
In fact, the radiation variables $\tilde{z},\tilde{y},\tilde{\phi}$ are obtained applying the
DPA to the original off-shell real kinematics and the momentum  $\tilde{\bar{k}}_{12}+\bar{k}_5$ is obtained applying the
DPA to the off-shell mapped momentum ${\bar{k}}_{12}+\bar{k}_5$. 
The discrepancy in the phase-space measure results in a Jacobian factor of the on-shell projection. When the integration over the off-shell masses $k_{12}^2$ and $k_{34}^2$ is carried out this only gives a contribution that is beyond the DPA intrinsic accuracy. 
Practically, in most of the considered differential distributions this effect is negligible, and small effects are only visible
in parts of distributions that are not dominated by resonant $\PW$-boson pairs.

Although applying the CS mappings before the DPA on-shell projection
enables a more straightforward subtraction of IR singularities,
it results in having to deal with some phase-space configurations
where the real off-shell momenta cannot be on-shell projected
(owing to the $2\MW$ threshold of the DPA procedure),
while the CS-mapped Born-level momenta can be projected on~shell.
Such configurations can only appear when the photon is hard,
as the DPA on-shell projection and the CS mappings commute
in the soft-photon limit. Thus, the treatment of these
events does not hamper the local cancellation of IR singularities. 
While for standard phase-space configurations the dipole kernels are evaluated
with the on-shell-projected momenta, we choose to evaluate the dipole
kernels with off-shell kinematics in these peculiar cases.
Although this choice introduces a discrepancy between local and
integrated dipoles, this is beyond the accuracy of the PA and 
its numerical impact is negligible.

In the above discussion, the CS subtraction mapping
has only been applied to the momenta of the resonant $\PW$~bosons.
In order to evaluate the decay amplitudes, CS-mapped momenta for
the decay products are required, and we need a prescription
to map the decay-product momenta accordingly. The most natural choice
is to boost the decay-particle momenta into
the rest frame of the corresponding $\PW$-boson momentum using the original
real kinematics and boost them back using instead the CS-mapped momentum of
the boson. 
For the example dipole (massive final-state emitter, massless final-state spectator)
discussed so far, this gives the mapped decay momenta
\begin{equation}
        \bar{k}_{i} =\Lambda(\bar{k}_{12})^{-1}\cdot\Lambda({k}_{12})\cdot {k}_{i}\quad{\textrm{for }} i=1,2\,,
\end{equation}
where $\Lambda(p)$ is the boost into the rest frame of the momentum $p$. 

As a last comment of this section, we stress that in the above discussion,
only the dipoles with a massive final-state emitter and a massless
final-state spectator have been considered.
The argument for using the CS subtraction mapping before applying the DPA on-shell projection
holds also for the other dipole types.
In addition, while our reasoning is tailored to W$^{+}$W$^{+}$ scattering in the PA, the
whole analysis of this section applies to the general case of any charged resonances.
In Appendix~\ref{sec:production_counterterms} we list the results for all local and
integrated dipoles involving generic electrically charged massive particles as emitters and/or spectators.
They represent the abelian version of the analogous results for QCD-charged massive partons
\cite{Catani:2002hc}.

\subsection{Subtraction dipoles for decay sub-processes}\label{sec:decay_counterterms}
In this section, we focus on the subtraction of QED IR divergences
originating from corrections to the decays of the vector bosons, \ie
arising from the lower left diagram in \reffi{fig:diags} and from the
lower right one after partial fractioning the photon radiation off the
$\PW$-boson propagator (see \refse{sec:realcorr}). Note that both soft
and collinear singularities are associated to the radiation off the
decay products (leading to a double pole in dimensional
regularisation), while the photon radiated off the charged resonance only leads to additional soft singularities (single pole).

The general treatment of IR divergences in top-quark decays is known in the literature \cite{Basso:2015gca} and such an approach can be applied to $\PW$~bosons as well \cite{Le:2022lrp,Denner:2023ehn}, since the soft-photon singular structure is independent of the spin of the emitting resonance. Note that quasi-collinear configurations can be safely neglected, owing to the typical energy scale of the hard process at the LHC, which is not much larger than the resonance mass \cite{Schonherr:2017qcj}.
As in \citere{Denner:2023ehn}, instead of using the dipole structure of \citere{Basso:2015gca}, we employ a subtraction counterterm reproducing the exact structure of the radiative decay $\PW^+\rightarrow \ell^+\nu_\ell\gamma$, relying on a final--final dipole mapping with a massless spectator (the neutrino) and a massless emitter (the charged lepton).
Let $k_1, q$, and $k_3$ be the momenta of the charged lepton, the decaying $\PW$~boson, and the radiated photon, respectively. 
The momentum  $k_2$ of the neutrino plays the role of the spectator. 
The subtraction mapping takes the usual form of final--final massless dipoles \cite{Catani:1996vz},
\begin{align}\label{eq:mappdecdip}
\bar{k}^\mu_{1}=k^\mu_{1}+k^\mu_3-\frac{s_{13}}{q^2-s_{13}}k^\mu_2\,,\qquad\bar{k}^\mu_{2}=k^\mu_{2}\frac{q^2}{q^2-s_{13}}\,.
\end{align}
Defining the usual radiation variables,
\beq
y=\frac{s_{13}}{q^2}\,,\qquad z = \frac{s_{12}}{s_{12}+s_{23}} = \frac{s_{12}}{q^2-s_{13}}\,,
\eeq
we write the local counterterm as 
\begin{equation}\label{eq:dsubdef2}
        \begin{split}
                {\mc D}^{\rm (dec)}(q^2,y,z) = \,\frac{8 \pi \alpha \,  }{q^2\,y\,} 
                \, \frac{\left(1-y\right)\left(1-z\right)\left[1+(1-y)(z^2-y(2-2z+z^2))\right]}{[1-z(1-y)]^2}\,.
        \end{split}
\end{equation}
This kernel reproduces the soft and collinear singularities
originating from photon emission from both the final-state charged
lepton and the on-shell \PW~boson.
After integration in $4-2\epsilon$ dimensions over the radiative phase space, the integrated counterterm reads
\begin{equation}
\label{eq:intsubdef2}
        \begin{split}
                \int\!\rd\Phi_{\rm rad} \,{\mc D}^{\rm (dec)}(q^2,y,z)\,  = \frac{\alpha}{2\pi}\,\,\frac{(4\pi)^\epsilon}{\Gamma (1-\epsilon )}\,\left(\frac{\mu ^2}{q^2}\right)^{\!\epsilon}
                \left[\frac1{\epsilon^2}+\frac5{2\epsilon}+\left(\frac{95}{12} - \frac{\pi^2}{2}\right)+\mc{O}(\epsilon)\right]\,,
        \end{split}
\end{equation}
reproducing the pole structure consistent with \citere{Basso:2015gca}.
Since there is a single charged particle in the final state, no
additional final--final dipoles contribute, and all divergences are
cancelled by a single counterterm. This would not be the case for the
hadronic decay of a $\PW$~boson, where two dipoles of the
type described in \refeq{eq:dsubdef2} (one per each decay quark)
and two final--final massless dipoles are needed to properly subtract all
soft and collinear singularities.
Additionally, when investigating processes with charged resonances
decaying into more than three particles (like the top quark), the
spectator momentum (sum of momenta of more than one massless particle) is massive,
giving rise to a more complicated subtraction mapping \cite{Basso:2015gca}.

In \refeqs{eq:mappdecdip}--\eqref{eq:intsubdef2} we have retained
$q^2$ as the squared mass of the $\PW$~boson. In more detail, we
choose to apply the subtraction mapping to the off-shell real
kinematics, and then to apply to the reduced kinematics the
DPA$^{(2,2)}$ mapping in order to set the $\PW$~boson on its mass
shell. The real matrix element is instead treated with the
DPA$^{(3,2)}$ or DPA$^{(2,3)}$ mapping. Since these mappings do not
change the ratio of invariants of the final-state particles, the
integration measure and the radiation variables in
\refeqs{eq:dsubdef2} and \eqref{eq:intsubdef2} are untouched by the DPA mappings, and the pole structure in \refeq{eq:intsubdef2} remains the same whether $q^2=\MW^2$ or not.
Since in the DPA the numerical integration of the real matrix element
and of the associated local counterterm are carried out using the
off-shell phase space, there is a mismatch between the local
counterterm and its integrated counterpart, consisting of the
Jacobian determinant of the DPA on-shell projection. However, upon
phase-space integration, this mismatch boils down to a contribution
beyond the accuracy of the DPA and can therefore safely be neglected.
We stress that applying the DPA mapping first and the subtraction
mapping second would also lead only to a mismatch that is beyond the
DPA accuracy for the decay-dipole contribution. This was indeed the approach used for inclusive $\PZ\PZ$
production \cite{Denner:2021csi}. We have proven for the decay-dipole
contributions that the two mapping sequences lead to results (integrated and differential) that are fully compatible within numerical uncertainties.
This represents a substantial difference compared to production-level dipoles, where instead the application of the mappings in different sequences may lead to numerically sizeable discrepancies.

\subsection{Polarised-signal definition}\label{sec:polsel}
At this point of the discussion, we can introduce the concept of
polarised signal, once all matrix elements entering the NLO
calculation have been projected on shell using the pole approximation and therefore are gauge invariant. 
In VBS, this results in the production of two intermediate on-shell $\PW$~bosons in $s$~channel, which undergo leptonic decays.
Focusing on one $\PW$~boson with momentum $q$, in the `t Hooft--Feynman gauge an amplitude in the PA reads
\beq\label{eq:prodxdec}
\cM_{\rm PA} =\cM_{\rm P}^{\mu}(\tilde{q})\,\frac{-\ri g_{\mu\nu}}{q^2-\MW^2+\ri\GW\MW} \cM_{\rm D}^{\nu}(\tilde{q})\,,
\eeq
where $\cM_{\rm P}$ and $\cM_{\rm D}$ are the production and decay
sub-amplitudes with polarisation vectors stripped off. These
amplitudes depend on the on-shell-projected momentum $\tilde{q}$ for the boson, while the off-shell momentum $q$ is retained in the denominator of the propagator. The tensorial part of the propagator in \refeq{eq:prodxdec} can be written in terms of polarisation vectors associated to the on-shell boson,
\beq\label{eq:polsum}
-g_{\mu\nu} = \sum_{\lambda'=\rL,\pm, \rm A}\varepsilon^{(\lambda')}_{\mu}(\tilde{q})\,\varepsilon^{*\,(\lambda')}_{\nu}(\tilde{q})\,,
\eeq
where the sum runs over three physical states, longitudinal ($\rL$),
left ($-$) and right handed ($+$), and an auxiliary state ($\rm
A$). The latter is unphysical and cancels against Goldstone-boson
contributions for any gauge choice for on-shell matrix elements. 
From the unpolarised amplitude of \refeq{eq:prodxdec}, it is then natural to retrieve a polarised amplitude by replacing the complete sum on the right side of \refeq{eq:polsum} with one single term for a definite physical polarisation state $\lambda$ \cite{Ballestrero:2017bxn,Denner:2020bcz},
\beq
\sum_{\lambda'}\varepsilon^{(\lambda')}_{\mu}(\tilde{q})\,\varepsilon^{*\,(\lambda')}_{\nu}(\tilde{q})\,\longrightarrow\,\varepsilon^{(\lambda)}_{\mu}(\tilde{q})\,\varepsilon^{*\,(\lambda)}_{\nu}(\tilde{q})\,,
\eeq
leading to the equality,
\beq\label{eq:sumpolamp}
\cM_{\rm PA} = \sum_{\lambda'} \cM_{\rm PA}^{(\lambda')}\,,
\eeq
where the sum only runs over the three physical polarisation states.
Cross-sections depend on squared matrix elements. Squaring the sum of polarised
amplitudes we obtain
\beq\label{eq:sqamp}
\left|\cM_{\rm PA}\right|^2 = \sum_{\lambda'} \left|\cM_{\rm PA}^{(\lambda')}\right|^2 + \sum_{\lambda' \neq \lambda''} \cM_{\rm PA}^{(\lambda')}{\cM_{\rm PA}^{(\lambda'')}}^*\,,
\eeq
where the first term represents the incoherent sum of polarised squared matrix elements, while
the second one includes all contributions resulting form the interference between two different
polarisation states. The interference term only vanishes if a complete integration over the decay-product phase-space measure is performed. This is usually prevented by the selection cuts on decay products, leading to non-vanishing interferences.

\begin{figure}
  \centering
  \includegraphics[scale=0.50]{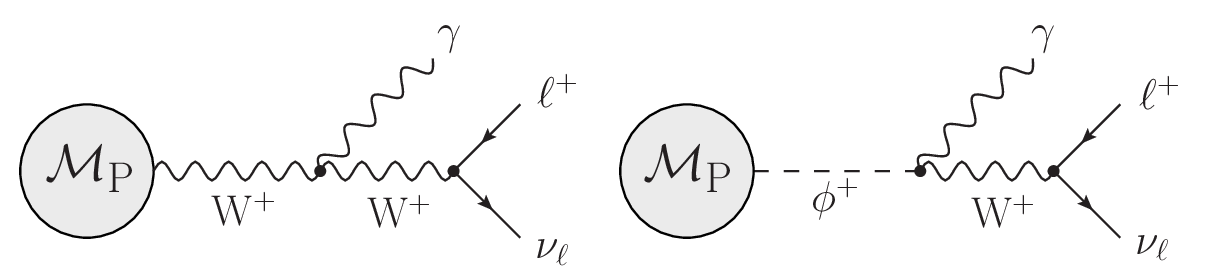}
  \caption{Real-photon correction to the $\PW^+$-boson decay (left)
    and its Goldstone-boson counterpart (right), with $\cM_{\rP}$ denoting a generic production amplitude.}\label{fig:goldstone}
\end{figure}

In the sum of \refeq{eq:sumpolamp} we have excluded the contribution of the unphysical polarisation state, as it cancels against Goldstone-boson contributions. This cancellation turns out to be trivial when the EW bosons undergo tree-level decays into massless fermions, since the Goldstone-boson contributions are proportional to the fermion masses. It is slightly more involved when considering NLO EW real and virtual corrections to the EW-boson decay, owing to non-vanishing Goldstone-boson contributions, like the ones depicted in \reffi{fig:goldstone}.

While this cancellation is expected to happen at any perturbative
order in the EW coupling, we verified this numerically using
\recola version 1.4.4 by comparing different contributions in the employed
amplitudes. 

It is essential to recall that, while the complete sum in
\refeq{eq:polsum} is Lorentz invariant, the individual terms depend on
the specific frame in which the boson momentum $\tilde{q}$ is evaluated. In other words, the polarised amplitudes are Lorentz-frame dependent. It is then crucial to evaluate all matrix elements entering the NLO calculations in the same Lorentz frame, in order to have a consistent definition of the polarised signal \cite{Denner:2020eck,Denner:2021csi}.
For VBS processes, different choices are possible
\cite{Ballestrero:2020qgv}, but the di-boson centre-of-mass
frame is best motivated from a theoretical point of view. In this
frame, the unitarity cancellations between Higgs and pure-gauge-boson
contributions at tree level are maximal for on-shell
longitudinal-boson scattering and
there is only one
reference axis for the definition of polarisation vectors of the two bosons (which are in a back-to-back kinematic configuration).

\begin{figure}
  \centering
  \includegraphics[scale=0.50]{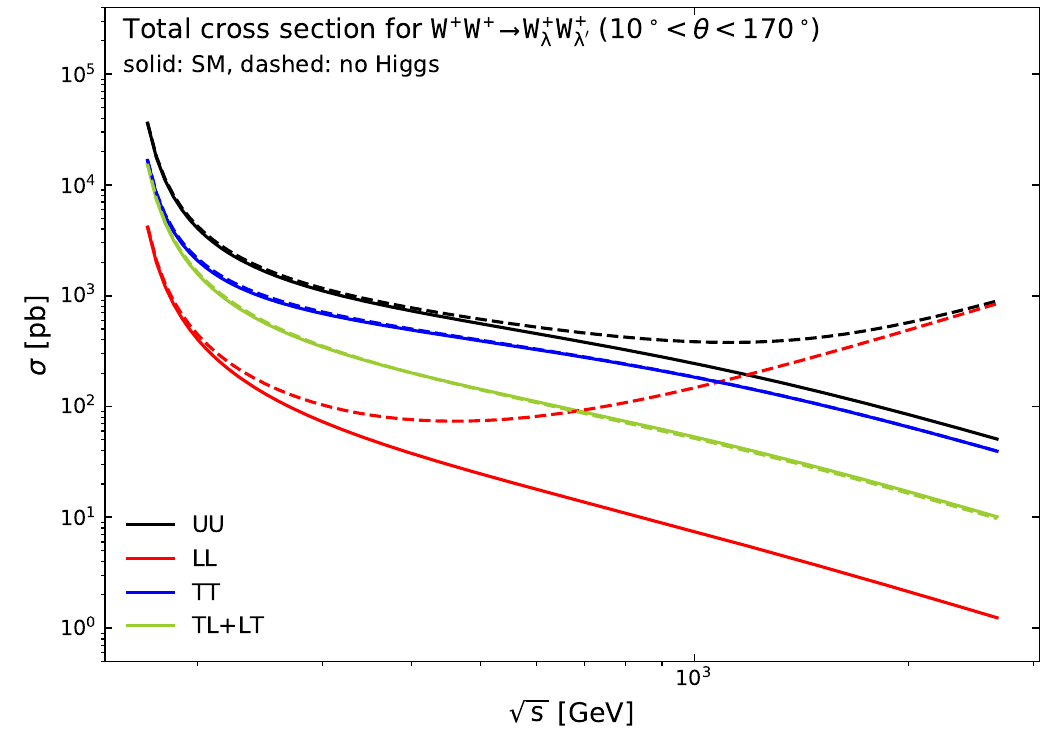}
  \caption{Total cross-sections for on-shell $\PW^+\PW^+\rightarrow\PW^+_{\lambda}\PW^+_{\lambda'}$ scattering at tree level, as functions of the di-boson
centre-of-mass energy. The two initial bosons are unpolarised, while the two final ones can be in a longitudinal (L) or transverse (T) polarisation state. Numerical results have been obtained with \recola1 \cite{Actis:2012qn,Actis:2016mpe} for the complete SM (solid curves) and in the absence of Higgs-mediated diagrams (dashed curves).}\label{fig:OSxs}
\end{figure}

It is well known
\cite{Veltman:1989vw,Dawson:1989up,Passarino:1990hk,Denner:1997kq}
that in the scattering of longitudinal bosons the unitarity-violating
behaviour in the high-energy limit of pure-gauge-boson diagrams is
regularised by the inclusion of Higgs-mediated diagrams, recovering
perturbative unitarity in the tree-level amplitude. In
\reffi{fig:OSxs} we consider the energy growth of the total cross-section for $\PW^+\PW^+\rightarrow\PW^+\PW^+$ with possibly polarised final-state bosons, both in the complete SM and excluding Higgs-mediated contributions.
The cross-sections are shown as functions of the di-boson
centre-of-mass energy
($\sqrt{s}$) and have been integrated over the range $10^\circ<\theta<
170^\circ$ of the scattering angle to avoid the physical singularity
associated with $t$-channel photon exchange. While for transverse and
mixed polarisations the energy growth is almost absent when excluding
Higgs diagrams, for two  longitudinally polarised final-state bosons tree-level unitarity is fulfilled through delicate cancellations of large contributions from pure gauge and Higgs diagrams.
It is also known
\cite{Veltman:1989vw,Dawson:1989up,Passarino:1990hk,Denner:1997kq}
that in the SM the cross-section decreases with energy as $1/s$ for
all polarisation combinations, apart from those with an odd number of external longitudinal bosons, which are suppressed by one further power of $s$.
The NLO EW corrections to the on-shell scattering are large and
negative \cite{Denner:1997kq} and become quickly unreliable for
scattering energies in the 10\TeV range, where resummation of large
logarithms is clearly mandatory. Beyond LO, the calculation becomes
tricky if a real-mass scheme is used, requiring the consistent
inclusion of finite-width effects \cite{Denner:1997kq}.
These problems are overcome by embedding \PW scattering in a physical
process as is done in the present paper.

\subsection{Overlap with tri-boson production}\label{sec:overlap}
\begin{figure*}   \centering
     \includegraphics[scale=0.50]{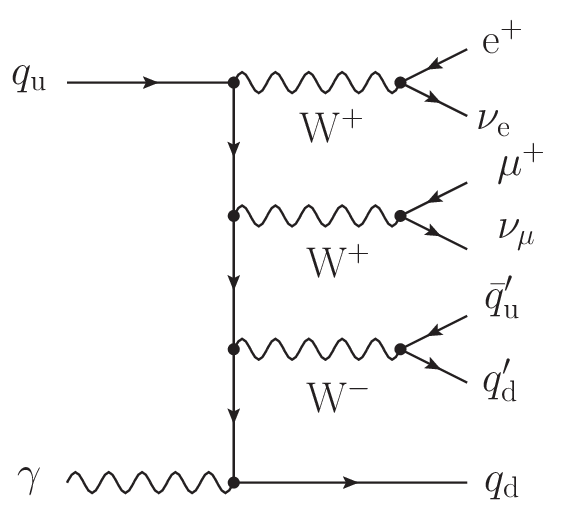}
     \includegraphics[scale=0.50]{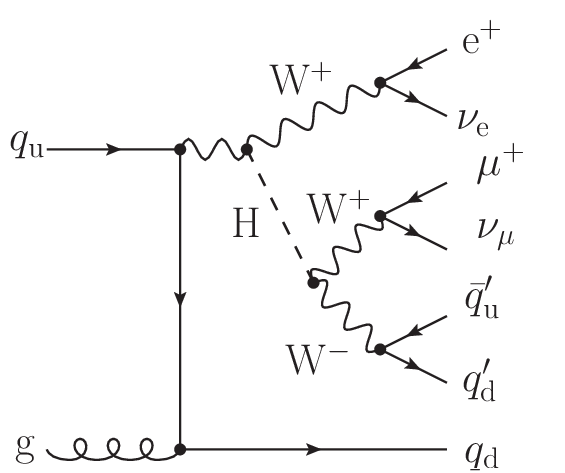}
     \includegraphics[scale=0.50]{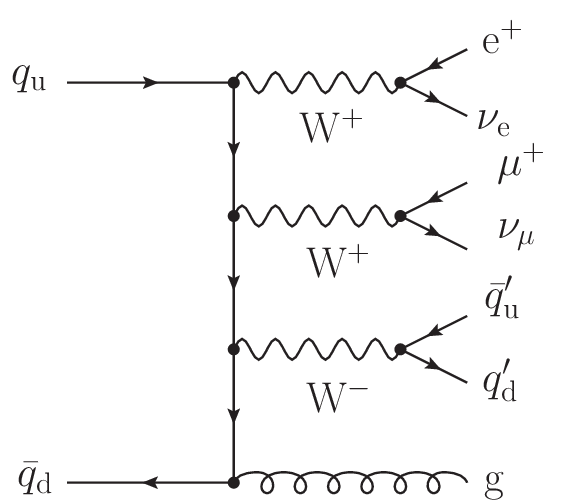}
   \caption{Sample real-radiation diagrams of order $\mc O (\alpha^7)$ (left) and $\mc O (\alphas\alpha^6)$ (middle, right), embedding triple-$\PW$-boson production and contributing to the same final state as W$^+$W$^+$ scattering at the LHC.}\label{fig:diags_triple_W}
\end{figure*}

While the application of typical VBS selections enhances VBS topologies in our case,
a non-trivial overlap with tri-boson production and Higgs-strahlung shows up in real corrections.
In particular, the application of a DPA strategy targeting two leptonically decaying $\PW^+$ bosons
requires to set the EW-boson widths to zero everywhere apart from the two target propagators, in order
to preserve gauge invariance. This leads to a divergence if an additional EW-boson $s$-channel
propagator gets close to the EW-boson pole mass, namely in the case of $\Pp\Pp\rightarrow\PW^+\PW^+\PW^-$ contributions with
$\PW^-\rightarrow\Pj\Pj$. At LO [$\mc O(\alpha^6)$], the VBS invariant-mass cut $M_{\Pj\Pj}>M_{\rm cut}\gg \MW$
completely suppresses contributions of this type and avoids the singularity.
At NLO the possible presence of three jets after clustering
leads to an ambiguity in the identification of the two VBS tagging jets.
A common choice is to use the two hardest jets, sorted in transverse momentum \cite{Jager:2009xx,Denner:2012dz}. This is the choice made in
our work as well. If one of the two tagging jets does not originate from a $\PW^-$ decay, the tri-boson topologies can
lead to a $\PW^-$ propagator with an invariant equal to $\MW$, rendering the DPA amplitude
divergent.
This situation appears in photon- and gluon-induced real-emission
processes as well as gluon-emission processes, as shown in
\reffi{fig:diags_triple_W}.
To properly take into account the third resonance for these processes the standard DPA cannot be used.
While in general it would be possible to subtract the tri-boson
contribution through a triple-pole approximation
\cite{Denner:2024xul}, we choose to regularise possibly unprotected
$\PW^-$-boson $s$-channel propagators restoring for those a finite
width, thereby introducing a gauge dependence. This gauge dependence
is not enhanced by any large factor in our
case. The comparison with the full off-shell calculation shows that
the size of the gauge-dependent contributions is numerically
small, \ie within the uncertainty of the DPA.
Since integrated subtraction counterterms are based on a Born
kinematics, they do not suffer from a potentially unprotected
$\PW$-boson propagator, and are therefore computed in the standard DPA.
Although this introduces a slight mismatch between local and
integrated counterterms for tri-boson configurations, its numerical
impact is completely negligible owing to the application of VBS-like
invariant-mass cuts (see remarks on checks in \refse{sec:setup}). 

We also note that tri-boson diagrams include contributions from
Higgs-associated production with the Higgs boson undergoing a
semi-leptonic decay $\PH\rightarrow \PW^+\PW^-\rightarrow
\ell^+\nu_\ell\Pj\Pj$ (middle diagram in \reffi{fig:diags_triple_W}).
While such contributions are completely cut away by VBS cuts at LO
and at NLO EW, real-QCD-radiation events can pass these cuts if
one of the tagging jets does not result from the Higgs boson, leading to enhancements in the cross-section owing to the $s$-channel Higgs boson.
Although these configurations are suppressed (and regulated by the
finite Higgs-boson width) as they appear only
in partonic processes with small contributions, their interplay with the
DPA procedure targeting two on-shell $\PW^+$ bosons is delicate.  The
impact of this is numerically negligible for our purposes, however,
a dedicated PA approach is in general needed \cite{Denner:2024xul}.

\section{Numerical results}\label{sec:numericalres}

\subsection{Setup and tools}
\label{sec:setup}
We consider an LHC centre-of-mass energy of $\sqrt{s} = 13.6\TeV$. The on-shell masses and widths of weak bosons take the values \cite{Workman:2022ynf},
\begin{align}
  \MZOS    &{}=  91.1876\GeV,\quad &
  \GZOS    &{}=  2.4952 \GeV,\nonumber \\
  \MWOS    &{}=  80.377 \GeV,\quad &
  \GWOS    &{}=  2.085  \GeV\,,
\end{align}
which are converted to  pole values according to \citere{Bardin:1988xt}:
\beq
 M_V = \frac{\MVOS}{\sqrt{1+\left(\GVOS/\MVOS\right)^2}}\,,\qquad \Gamma_V = \frac{\GVOS}{\sqrt{1+\left(\GVOS/\MVOS\right)^2}}\,,\qquad V=\PW,\,\PZ\,.
\eeq%
The masses and widths of the top quark and of the Higgs boson are set to \cite{Workman:2022ynf}
\begin{align}
\Mt &{}= 172.69\GeV,\quad &
\Gt &{}= 1.42\GeV, \nnb\\ 
\MH &{}= 125.25\GeV,\quad & 
\GH &{}= 0.0041\GeV\,. 
\end{align}
Note that in the DPA calculations $\Gt$ is set to zero, while $\GH$ is kept finite. We employ the $G_\mu$ scheme \cite{Denner:2000bj,Dittmaier:2001ay} to identify the EW coupling.
In particular, the off-shell calculation is performed in the complex-mass scheme \cite{Denner:2005fg,Denner:2006ic,Denner:2019vbn},
\begin{align}\label{eq:alphadef}
  &\alpha = \frac{\sqrt{2}}{\pi}\,G_\mu\,\left|\mu_{\PW}^2\, \left(  1 - \frac{\mu_{\PW}^2}{\mu_{\PZ}^2} \right)\right|\,,\qquad \mu_V^2=M_V^2-\ri M_V\Gamma_V \quad(V=\PW,\,\PZ)\,,
\end{align}
where the Fermi constant reads $\GF = 1.16638\times 10^{-5}\GeV^{-2}$ and the parameters $\MZ,\MW$ and $\GW, \GZ$ are the pole values of weak-boson masses and widths.
In contrast, the calculations within the DPA are performed with real
couplings and
\begin{align}
  &\alpha =
  \frac{\sqrt{2}}{\pi}\,G_\mu\,\MW^2\,
  \left(
  1 - \frac{\MW^2}{\MZ^2}
  \right)\,.
\end{align}
The calculation is carried out with five active flavours, although bottom-induced channels do not contribute to the considered process. We do include photon-induced partonic channels that
arise at NLO EW. The $\overline{\rm MS}$ factorisation scheme is used for the treatment of initial-state collinear singularities of both QED and QCD type.
We employ \sloppy\texttt{NNPDF40\_nnlo\_as\_01180\_qed}~\cite{NNPDF:2024djq} parton-distribution functions, linked to our Monte Carlo codes via the \textsc{Lhapdf} interface \cite{Buckley:2014ana}.
The factorisation and renormalisation scales are set to the same dynamical value,
\begin{equation}
\label{eq:scale}
\mu_{\rm R} =
\mu_{\rm F} =
\sqrt{\,\pt{\Pj_1}\,\pt{\Pj_2}} \,,
\end{equation}
where $\Pj_1$ and $\Pj_2$ are the leading and sub-leading jet (sorted in transverse momentum), respectively.
Scale uncertainties are obtained from conventional 7-point scale variations by
a factor two. The scales defined in \refeq{eq:scale} are scaled by
factors
\beq
(\xi_\rF,\xi_\rR) \in \{(1/2,1/2),(1/2,1),(1,1/2),(1,1),(1,2),(2,1),(2,2)\}
\eeq
and the resulting maximal and minimal values of the cross-section are used to calculate the
scale variation.

The selections follow closely those used for a recent polarisation measurement by the CMS collaboration \cite{CMS:2020etf}.
The events are required to have one positron and one anti-muon fulfilling [$\ell_{1(2)}$ is the (sub)leading lepton],
\beq
\pt{\ell_1}>25\GeV\,,\qquad \pt{\ell_2}>20\GeV\,,\qquad |y_{\ell_{1,2}}|<2.5\,,\,\qquad M_{\Pe^+\mu^+}>20\GeV\,.
\eeq
Furthermore, we demand at least two jets, clustered with the anti-$k_{\rm t}$ algorithm \cite{Cacciari:2008gp} and recombination radius $R=0.4$, satisfying,
\beq
\pt{\Pj} > 50\GeV\,,\qquad |y_{\Pj}|<4.7\,,\qquad \Delta R_{\Pe^+\Pj}>0.4\,,\qquad \Delta R_{\mu^+\Pj}>0.4\,,
\eeq
the leading and subleading jet being in a VBS kinematic regime,
\beq
M_{\Pj_1\Pj_2}>500\GeV\,,\qquad |\Delta y_{\Pj_1\Pj_2}|>2.5\,.
\eeq
The selected events need to have a minimum missing transverse momentum
\beq
\pt{\rm miss} > 30\GeV\,
\eeq
and to satisfy the rapidity requirement,
\beq
\max_{\ell} {\left|y_\ell-\frac{y_{\Pj_1}+y_{\Pj_2}}2\right|} <0.75\,{|\Delta y_{\Pj_1\Pj_2}|}\,,\quad\ell=\Pe^+,\mu^+\,.
\eeq
Photons are recombined with jets and leptons according to the anti-$k_{\rm t}$ algorithm \cite{Cacciari:2008gp}
and recombination radius $R=0.1$.
The jet clustering and lepton dressing is applied to particles with pseudorapidity $|\eta|<5$.

The calculations in this work have been performed with
\bbmc  and checked with \mocanlo. 
Both Monte Carlo
codes have already been
used for the simulation of intermediate polarised bosons at NLO accuracy
for several di-boson production processes \cite{Denner:2020bcz,Denner:2020eck,Denner:2021csi,Denner:2022riz,Denner:2023ehn}.
The two codes use the \recola library
\cite{Actis:2012qn,Actis:2016mpe} for the calculation of tree-level and
one-loop amplitudes with fixed polarisation states for intermediate resonances.
For the calculation of one-loop integrals \recola employs the \collier
library \cite{Denner:2016kdg}. 
In the previously calculated di-boson processes many consistency checks have been performed. 
The UV finiteness of the virtual amplitudes has been tested for polarised $\PZ\PW$ production \cite{Denner:2022riz}. 
\recola amplitudes have been compared to {\sc MadLoop} \cite{Hirschi:2011pa}. 

For the process studied in the present paper, \ie same-sign $\PW^+\PW^+$ scattering,
the high-accuracy numerical results shown in \refses{sec:intres} and \ref{sec:diffres}
have been obtained with BBMC. A detailed comparison with lower-accuracy results
from MoCaNLO has been performed for the unpolarised cross-section and the purely longitudinal
cross-section in DPA. 
The results have been compared at the integrated level and at the
histogram level for the LO ($\alpha^6$) and NLO EW ($\alpha^7$)
contributions finding the same results within the integration uncertainty
of $\approx 0.4 \%$.  
For the integrated cross-section at NLO QCD ($\alpha_s \alpha^6$) the
results agree within the integration uncertainty of $\approx 3\%$. 
Moreover, the integrated cross-sections of the most important partonic
processes  and those that involve a triply-resonant contribution
are individually consistent within the numerical
uncertainty of $\lesssim 3\%$.
For the full off-shell process and the other polarisation states only
partial results have been compared. 
Moreover, the following checks have been carried out:
The implemented subtraction counterterms have been tested extensively. 
First, the correct application of the on-shell projection has been verified, in
particular for processes with additional radiation from the decay of
one of the $\PW^+$~bosons where the DPA$^{(2,3)}$ and DPA$^{(3,2)}$
enter. 
To this end, the on-shell-projected momenta of \bbmc and \mocanlo have
been compared finding no difference beyond numerical inaccuracies (at
the level of $10^{-11}$).  
Furthermore it has been checked that the IR poles of the virtual and
integrated dipole contributions cancel, in particular for the newly
implemented dipole counterterms involving resonances from \refses{sec:prod_counterterms}
and \ref{sec:decay_counterterms}.  
This has been done by varying the IR scale and showing that the sum of
virtual contributions and integrated dipoles evaluated at different
scales agree within the integration uncertainty of $\approx 1\%$.
Additionally, the local and integrated dipoles have been compared between \mocanlo and \bbmc for individual phase-space points finding perfect agreement. 
The finite parts of the massless CS local and integrated dipoles have
been tested by evaluating them with different $\alpha_{\text{dipole}}$
parameters \cite{Nagy:1998bb} and verifying the independence of the final results on
these parameters. 
For both the NLO QCD and NLO EW contribution from summed real emission and
integrated dipoles we find agreement within the numerical uncertainty
of $\approx 0.7\%$ and $\approx 14\%$, respectively. Note that the
NLO EW corrections are dominated by the virtual corrections, and the
sum of the real corrections and the integrated dipoles amounts to
only $\approx 0.5\%$ of the corresponding LO cross-section.

As stated in \refse{sec:overlap} our handling of the triply-resonant contributions results in a mismatch between the local and integrated counterterms. 
By evaluating the integrated counterterms with the decay width set to zero and with keeping the pole values of the decay widths the size of the mismatch can be explicitly calculated. 
In the setup used here with both $\PW$ bosons unpolarised we find that
the integrated cross-section of the individual integrated EW and QCD counterterm
contributions differ at most by 0.3\%. 
For the fiducial NLO cross-section this relates to an absolute difference of $2.5\cdot 10^{-4}\fb$ and a relative difference of 0.021\%.
This is much smaller then the intrinsic uncertainty of the DPA.

\subsection{Results for fiducial cross-sections}\label{sec:intres}

\begin{table*}
\begin{center}
\tabcolsep 5pt
\begin{footnotesize}
\begin{tabular}{cccccccc}
\hline\rule{0ex}{2.7ex}
\cellcolor{blue!14} state  & 
        \cellcolor{blue!14} $\sigma_{\rm LO \alpha^6}$ [fb] & 
        \cellcolor{blue!14} $\sigma_{\rm LO \alpha_{\rm s} \alpha^5}$ [fb] & 
        \cellcolor{blue!14} $\delta_{\rm \alpha_{\rm s} \alpha^5}$ & 
        \cellcolor{blue!14} $\sigma_{\rm LO \alpha_{\rm s}^2 \alpha^4}$ [fb] & 
        \cellcolor{blue!14} $\delta_{\rm \alpha_{\rm s}^2 \alpha^4}$ & 
        \cellcolor{blue!14} $\delta_{\rm \alpha_{\rm s} \alpha^5 + \alpha_{\rm s}^2 \alpha^4}$ & 
        \cellcolor{blue!14} $\sigma_{\rm LO}$ [fb] \\ 
\hline\\[-0.4cm]
full & 
        $    1.4863(1)$ &
        $  0.044877(9)$ &       $0.03$ &
        $   0.14686(2)$ &       $0.10$ & 
$0.13$ &        $    1.6780(1)$ \\ 
unp.  & 
        $   1.46455(9)$ &
        $  0.044386(8)$ &       $0.03$ &
        $   0.14664(2)$ &       $0.10$ & 
         $0.13$ &       $   1.65558(9)$ \\ 
LL   & 
        $   0.14879(1)$ &
        $  0.006120(1)$ &       $0.04$ &
        $  0.012298(2)$ &       $0.08$ & 
         $0.12$ & 
        $   0.16721(1)$ \\ 
LT   & 
        $   0.23209(2)$ &
        $  0.007284(2)$ &       $0.03$ &
        $  0.029465(6)$ &       $0.13$ & 
         $0.16$ &       $   0.26884(2)$ \\ 
TL   & 
        $   0.23208(2)$ &
        $  0.007284(2)$ &       $0.03$ &
        $  0.029471(6)$ &       $0.13$ & 
         $0.16$ & 
        $   0.26884(2)$ \\ 
TT   & 
        $   0.87702(7)$ &
        $  0.026402(6)$ &       $0.03$ &
        $   0.07938(2)$ &       $0.09$ & 
         $0.12$ & 
        $   0.98281(7)$ \\ 
int. & 
        $   -0.0254(1)$ &
        $  -0.00270(1)$ &       $0.11$ &
        $  -0.00398(3)$ &       $0.16$ & 
         $0.26$ & 
        $   -0.0321(1)$ \\ 
\hline\\[-0.3cm]
\end{tabular}
\end{footnotesize}
\caption{
  LO contributions to the integrated cross-section (in fb) for the
  process $\Pp\Pp\to\Pe^+\nu_{\Pe}\mu^+\nu_{\mu} + \Pj\Pj$ at the
  LHC.
  The numbers in columns 4, 6, and 7 give the ratio of the respective background to the LO signal $\order{\alpha^{6}}$.
}
\label{tab:integrated_cross_section_lo}
\end{center}
\end{table*}
In \refta{tab:integrated_cross_section_lo} the LO integrated
cross-sections of the VBS signal of $\order{\alpha^{6}}$, the interference
background of $\order{ \alphas\alpha^{5}}$, and the QCD background
of $\order{\alphas^{2}\alpha^{4}}$ are shown. 
In a polarisation analysis the contributions from the two backgrounds would be subtracted from the measured cross-section before the VBS signal is split into the polarised contributions.
The study of the background contributions is nonetheless important to analyse how well the chosen setup, in particular the phase-space cuts, suppresses the backgrounds. 
The interference background gives a relative contribution of only
$\approx 3\%$, while the QCD background contributes $\approx 10\%$ of the VBS signal. 
The VBS signal and the QCD background prefer different kinematic regions, 
since the dominant contribution to the different LO contributions comes from diagrams with different topologies. 
The largest contribution to the VBS signal comes from diagrams where
the two quark lines each emit a $\PW^+$ boson that then scatter off each other (see top-left diagram in \reffi{fig:diagsNbody}). For the QCD background the two quarks exchange  a $t$-channel gluon
and each emit one $\PW^+$~boson (see top-right diagram in \reffi{fig:diagsNbody}).
The effects of these different diagram topologies can be seen more clearly when analysing the differential cross-sections in the next section.

\begin{table*}
\begin{center}
\begin{tabular}{ccccc}
\hline\rule{0ex}{2.7ex}
\cellcolor{blue!14} state  & 
        \cellcolor{blue!14} $f_{\rm LO \, \alpha^6}[\%]$ & 
        \cellcolor{blue!14} $f_{\rm LO\,\alpha_{\rm s} \alpha^5}[\%]$ & 
        \cellcolor{blue!14} $f_{\rm LO\,\alpha_{\rm s}^2 \alpha^4}[\%]$ & 
        \cellcolor{blue!14} $f_{\rm LO}[\%]$ \\ 
\hline\\[-0.4cm]
full & 
        $101.5$ &
        $101.1$ &
        $100.1$ &
        $101.4$ \\ 
unp.  & 
        $100.0$ &
        $100.0$ &
        $100.0$ &
        $100.0$ \\ 
LL   & 
        $10.2$ &
        $13.8$ &
        $8.4$ &
        $10.1$ \\ 
LT   & 
        $15.8$ &
        $16.4$ &
        $20.1$ &
        $16.2$ \\ 
TL   & 
        $15.8$ &
        $16.4$ &
        $20.1$ &
        $16.2$ \\ 
TT   & 
        $59.9$ &
        $59.5$ &
        $54.1$ &
        $59.4$ \\ 
int. & 
        $-1.7$ &
        $-6.1$ &
        $-2.7$ &
        $-1.9$ \\[0.1cm]
\hline\\[-0.3cm]
\end{tabular}
\caption{
        LO polarisation fractions for the process $\Pp\Pp\to\Pe^+\nu_{\Pe}\mu^+\nu_{\mu} + \Pj\Pj$ at the LHC.}
\label{tab:polarisation_fractions_lo}
\end{center}
\end{table*}
The LO polarisation fractions, defined as the corresponding cross
sections normalised to the unpolarised DPA cross section, are listed in
\refta{tab:polarisation_fractions_lo} for the individual orders and
their sum.  The chosen reference frame to define the polarisation
vectors, the $\PW^+ \PW^+$ centre-of-mass frame, is the natural choice
for VBS.  The DPA approximates the full off-shell computation within
$1.5\%$ for the EW signal and even better for the background contributions.  This is
consistent with the intrinsic DPA accuracy of $\order{\Gamma_{\PW} /
M_{\PW}}$.  As already seen in many other di-boson processes
\cite{Denner:2020eck, Denner:2021csi, Denner:2022riz, Denner:2023ehn} the dominant
contribution comes from both \PW~bosons being transversely polarised.
The mixed polarisation states contribute at 15.8\% while the purely
longitudinal polarisation state gives the smallest contribution at
10.2\%.  As the decay particles of the both $\PW^+$ bosons are treated
equally by the phase-space cuts, we get the same polarisations
fractions for the TL and the LT states.

\begin{table*}
\begin{center}
\begin{tabular}{ccccc}
\hline\rule{0ex}{2.7ex}
\cellcolor{blue!14} state  & 
        \cellcolor{blue!14} $\sigma_{\rm LO}$ [fb] & 
        \cellcolor{blue!14} $\Delta \sigma_{\rm NLO\,EW}$ [fb] & 
        \cellcolor{blue!14} $\Delta \sigma_{\rm NLO\,QCD}$ [fb] & 
        \cellcolor{blue!14} $\sigma_{\rm NLO\,EW+QCD}$ [fb] \\ 
\hline\\[-0.3cm]
full & 
        $    1.4863(1)^{+9.2\%}_{-7.8\%}$ &
        $   -0.2084(6)$&
        $   -0.0704(7)$& 
        $     1.208(1)^{+1.6\%}_{-3.1\%}$ \\ 
unp.  & 
        $   1.46455(9)^{+9.2\%}_{-7.8\%}$ &
        $   -0.2076(2)$&
        $   -0.0733(5)$& 
        $    1.1836(5)^{+1.7\%}_{-3.3\%}$ \\ 
LL   & 
        $   0.14879(1)^{+8.3\%}_{-7.2\%}$ &
        $  -0.01505(2)$&
        $  -0.00660(7)$& 
        $   0.12715(8)^{+1.0\%}_{-2.1\%}$ \\ 
LT   & 
        $   0.23209(2)^{+9.1\%}_{-7.8\%}$ &
        $  -0.03040(4)$&
        $   -0.0098(1)$& 
        $    0.1919(1)^{+1.4\%}_{-2.8\%}$ \\ 
TL   & 
        $   0.23208(2)^{+9.1\%}_{-7.8\%}$ &
        $  -0.03051(4)$&
        $  -0.0097(1)$& 
        $    0.1918(1)^{+1.4\%}_{-2.8\%}$ \\ 
TT   & 
        $   0.87702(7)^{+9.4\%}_{-8.0\%}$ &
        $   -0.1352(1)$&
        $   -0.0474(4)$& 
        $    0.6944(4)^{+1.9\%}_{-3.7\%}$ \\ 
int. & 
        $   -0.0254(1)^{-8.9\%}_{+10.6\%}$ &
        $    0.0035(2)$&
        $    0.0002(6)$& 
        $   -0.0217(7)^{-1.6\%}_{+0.7\%}$ \\ 
[-0.3cm] \\ \hline
\cellcolor{blue!14} state & \cellcolor{blue!14} & 
        \cellcolor{blue!14} $\delta_{\rm EW}$ & 
        \cellcolor{blue!14} $\delta_{\rm QCD}$ & 
        \cellcolor{blue!14} $\delta_{\rm EW+QCD}$ \\ 
\hline\\[-0.3cm]
full & &
        $-0.140$ &
        $-0.047$ &
        $-0.188$ \\ 
unp.  & &
        $-0.142$ &
        $-0.050$ &
        $-0.192$ \\ 
LL   & &
        $-0.101$ &
        $-0.044$ &
        $-0.145$ \\ 
LT   & &
        $-0.131$ &
        $-0.042$ &
        $-0.173$ \\ 
TL   & &
        $-0.131$ &
        $-0.042$ &
        $-0.173$ \\ 
TT   & &
        $-0.154$ &
        $-0.054$ &
        $-0.208$ \\ 
int. & &
        $-0.139$ &
        $-0.007$ &
        $-0.147$ \\ 
\hline\\[-0.3cm]
\end{tabular}
\caption{NLO contributions to the integrated cross-section (in fb) for the process $\Pp\Pp\to\Pe^+\nu_{\Pe}\mu^+\nu_{\mu} + \Pj\Pj$ at the LHC.
The numbers in the lower part of the table give the
  ratio of the given $\order{\alpha^{7}}$ and
  $\order{\alphas\alpha^{6}}$ NLO corrections to the corresponding LO signal at $\order{\alpha^{6}}$.}
\label{tab:integrated_crosssection_nlo}
\end{center}
\end{table*}
The NLO corrections to the fiducial cross-section are presented  in
\refta{tab:integrated_crosssection_nlo} for the unpolarised and
polarised fiducial cross-sections, specifically  
the EW $\order{\alpha^7}$ and QCD $\order{\alpha^6\alphas}$ corrections to the VBS signal. 
Both EW and QCD corrections are negative for all polarisation
states. The cross-sections for unpolarised and transverse
$\PW^+$~bosons receive the well-known negative EW corrections of about
$-15\%$ resulting from Sudakov logarithms
\cite{Biedermann:2016yds,Biedermann:2017bss}. The corrections for the
mixed polarisations are by $2.3\%$ smaller and those for longitudinal
$\PW^+$~bosons by another $3\%$. This results from the smaller EW
Casimir operators for longitudinal vector bosons with respect to transverse
vector bosons that multiply the leading double logarithms in the EW
corrections \cite{Denner:2000jv}.
The differences in size of the QCD corrections to the polarised
cross-sections are smaller than those of the EW corrections and well
within $1\%$.

Since at LO VBS is a purely EW process, the QCD-scale uncertainty
results only from varying the factorisation scale. 
For a reasonable estimate of the size of the higher-order QCD corrections the NLO QCD corrections need to be included. 
This gives a scale uncertainty of 1--3\% for the NLO EW+QCD prediction. 
The size of the scale uncertainties depends only weakly on the polarisation. 

\begin{table*}
\begin{center}
\begin{tabular}{ccccc}
\hline\rule{0ex}{2.7ex}
\cellcolor{blue!14} state  & 
        \cellcolor{blue!14} $f_{\rm LO}[\%]$ & 
        \cellcolor{blue!14} $f_{\rm NLO\,EW}[\%]$ & 
        \cellcolor{blue!14} $f_{\rm NLO\,QCD}[\%]$ & 
        \cellcolor{blue!14} $f_{\rm NLO\,EW+QCD}[\%]$ \\ 
\hline\\[-0.4cm]
full & 
        $101.5$ &
        $101.7$ &
        $101.8$ &
        $102.0$ \\ 
unp.  & 
        $100.0$ &
        $100.0$ &
        $100.0$ &
        $100.0$ \\ 
LL   & 
        $10.2$ &
        $10.6$ &
        $10.2$ &
        $10.7$ \\ 
LT   & 
        $15.8$ &
        $16.0$ &
        $16.0$ &
        $16.2$ \\ 
TL   & 
        $15.8$ &
        $16.0$ &
        $16.0$ &
        $16.2$ \\ 
TT   & 
        $59.9$ &
        $59.0$ &
        $59.6$ &
        $58.7$ \\ 
int. & 
        $-1.7$ &
        $-1.7$ &
        $-1.8$ &
        $-1.8$ \\[0.1cm] 
\hline\\[-0.3cm]
\end{tabular}
\caption{
        NLO polarisation fractions for the process $\Pp\Pp\to\Pe^+\nu_{\Pe}\mu^+\nu_{\mu} + \Pj\Pj$ at the LHC.}
\label{tab:polarisation_fractions_nlo}
\end{center}
\end{table*}
In \refta{tab:polarisation_fractions_nlo} the polarisation fractions
including the NLO EW and QCD corrections are shown. 
The NLO EW and QCD corrections slightly increase the relative size of
the non-resonant background from 1.5\% to 2.0\%, which is 
still within the expected range of the DPA accuracy of $\order{\Gamma_{\PW} / M_{\PW}}$. 
The polarisation fractions only change at the level of $1\%$ when the NLO corrections are added preserving the general pattern seen at LO. 
The polarisation modes with at least one longitudinally polarised boson and the interferences slightly increase, while the purely transverse polarisation state is decreased. 
This is caused by the less negative NLO EW corrections for longitudinally polarised bosons already seen in \refta{tab:integrated_crosssection_nlo}. 
The interference contribution is small because of the cancellations from the integration over the decay phase space described in \refse{sec:polsel}. 
The inclusion of NLO corrections results in a slightly larger negative interference contribution.

\subsection{Results for differential distributions}\label{sec:diffres}
\begin{figure}
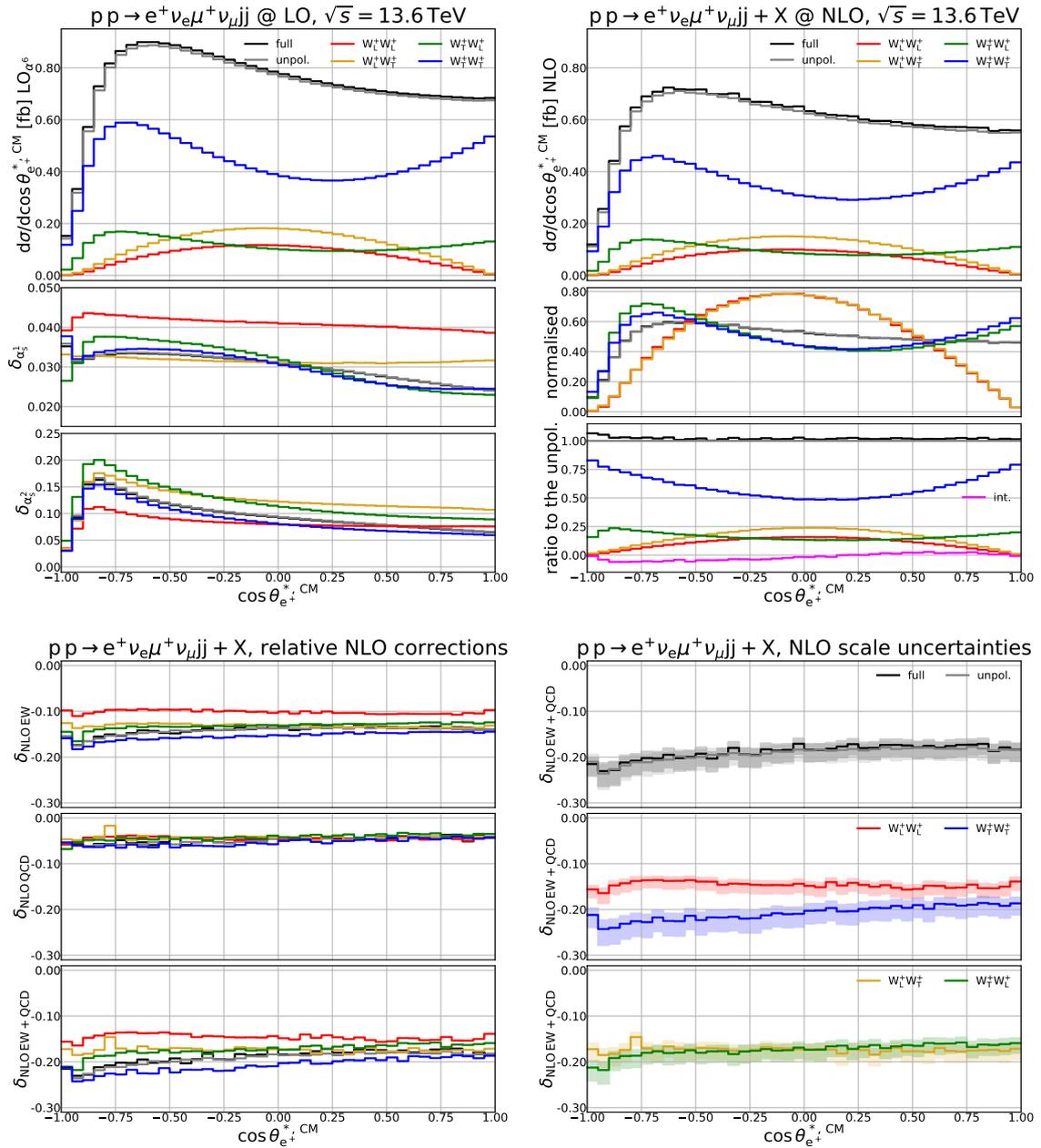

  \centering
  \subfigure{\includegraphics[scale=0.028, page=37]{Figures/BBMC_histograms_publication.pdf}}
  \subfigure{\includegraphics[scale=0.028, page=38]{Figures/BBMC_histograms_publication.pdf}}
  \subfigure{\includegraphics[scale=0.028, page=39]{Figures/BBMC_histograms_publication.pdf}}
  \subfigure{\includegraphics[scale=0.028, page=40]{Figures/BBMC_histograms_publication.pdf}}
  \caption{Distribution in the polar decay angle of the positron in
    the rest frame of the decaying $\PW^+$ boson.
    Details are described in the main text (first paragraphs of \refse{sec:diffres}).}    
  \label{fig:costhetastar} 
\end{figure}

\begin{figure}
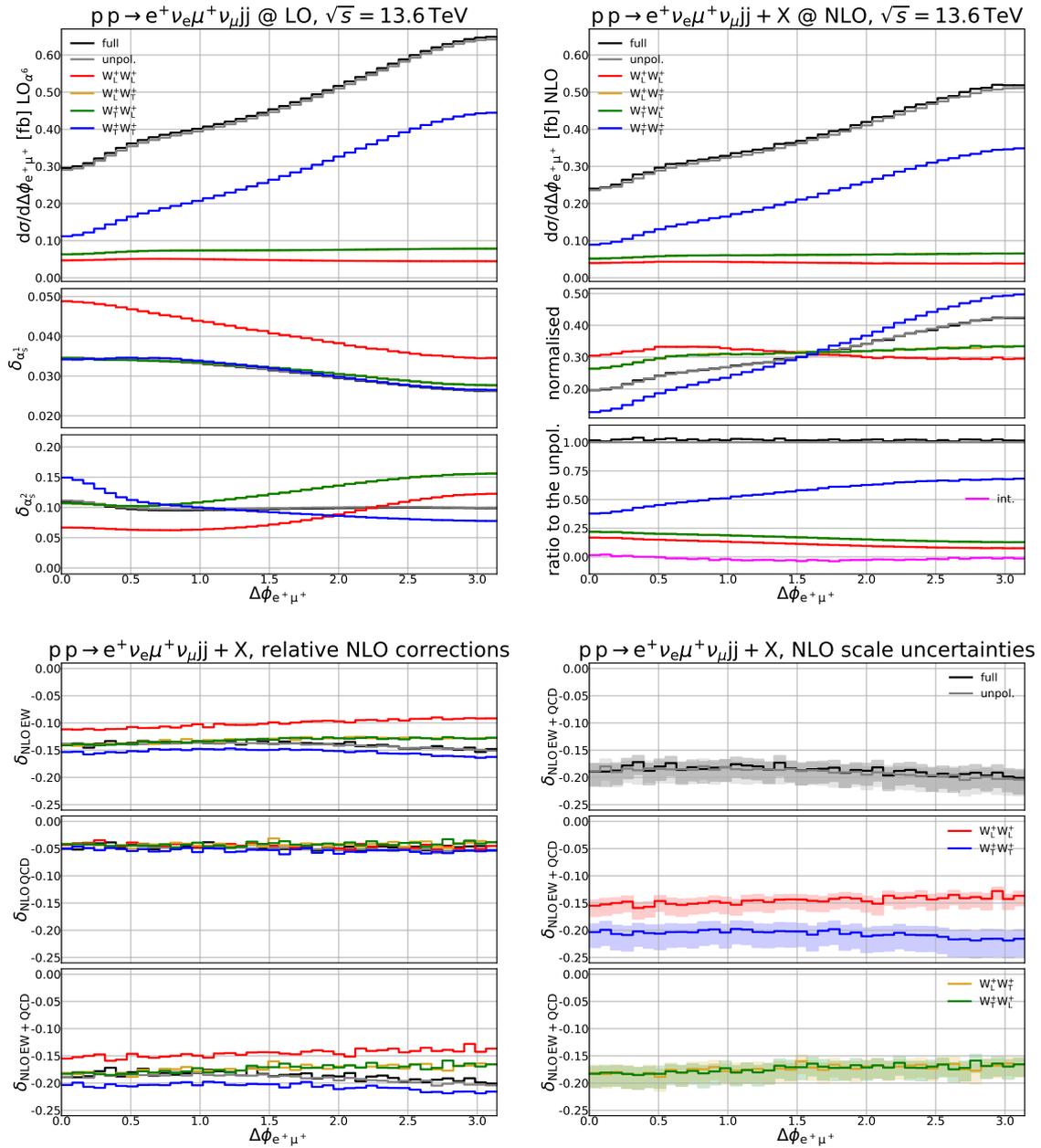

  \centering
  \subfigure{\includegraphics[scale=0.028, page=21]{Figures/BBMC_histograms_publication.pdf}}
  \subfigure{\includegraphics[scale=0.028, page=22]{Figures/BBMC_histograms_publication.pdf}}
  \subfigure{\includegraphics[scale=0.028, page=23]{Figures/BBMC_histograms_publication.pdf}}
  \subfigure{\includegraphics[scale=0.028, page=24]{Figures/BBMC_histograms_publication.pdf}}
  \caption{
    Distribution in the azimuthal-angle difference between the two charged leptons.
    Details are described in the main text (first paragraphs of \refse{sec:diffres}).
    }\label{fig:dphie+mu+}
\end{figure}

\begin{figure}
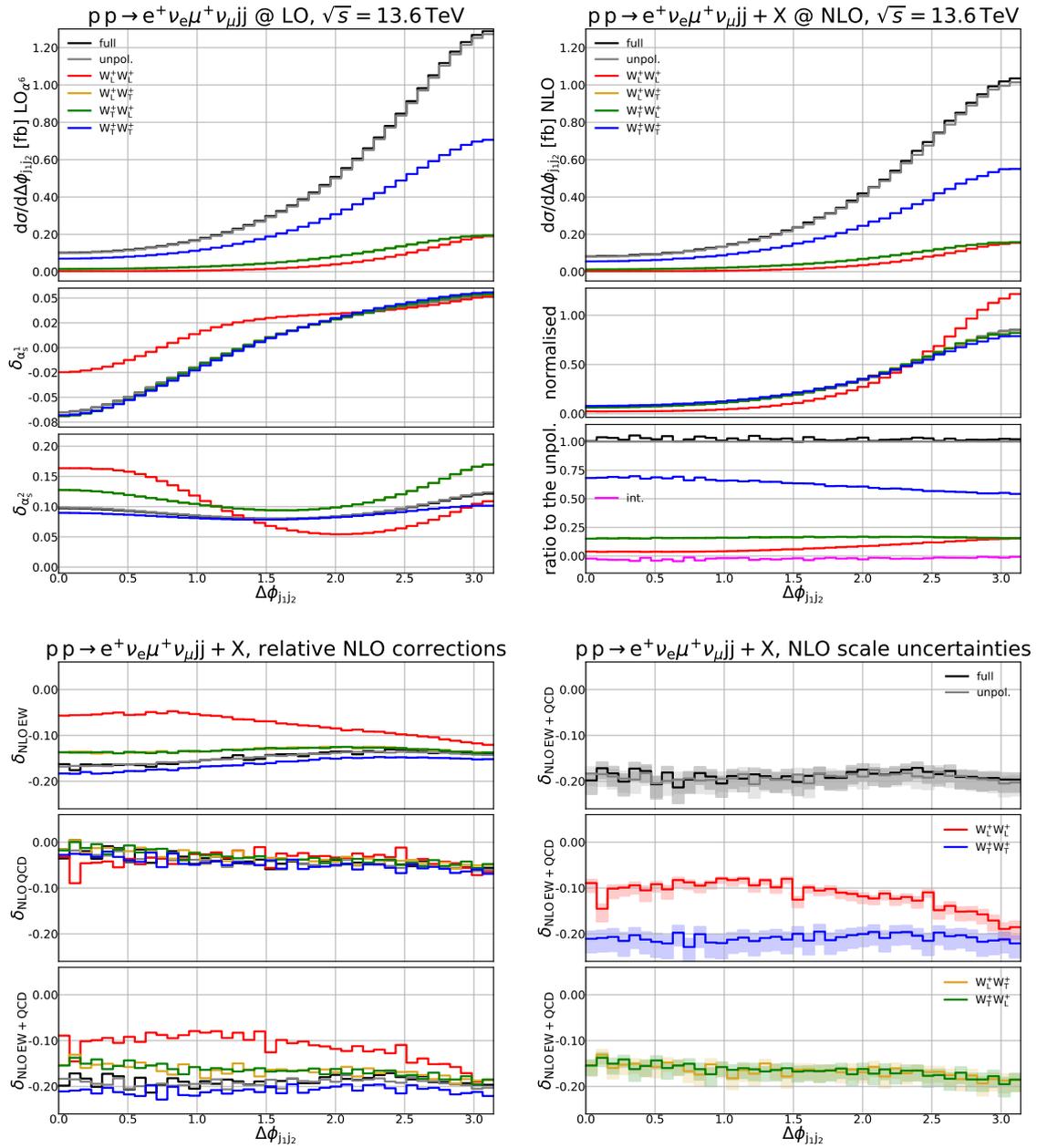

  \centering
  \subfigure{\includegraphics[scale=0.028, page=25]{Figures/BBMC_histograms_publication.pdf}}
  \subfigure{\includegraphics[scale=0.028, page=26]{Figures/BBMC_histograms_publication.pdf}}
  \subfigure{\includegraphics[scale=0.028, page=27]{Figures/BBMC_histograms_publication.pdf}}
  \subfigure{\includegraphics[scale=0.028, page=28]{Figures/BBMC_histograms_publication.pdf}}
  \caption{
    Distribution in the azimuthal-angle difference between the two tagged jets.
    Details are described in the main text (first paragraphs of \refse{sec:diffres}).    
  }\label{fig:dphij1j2} 
\end{figure}

\begin{figure}
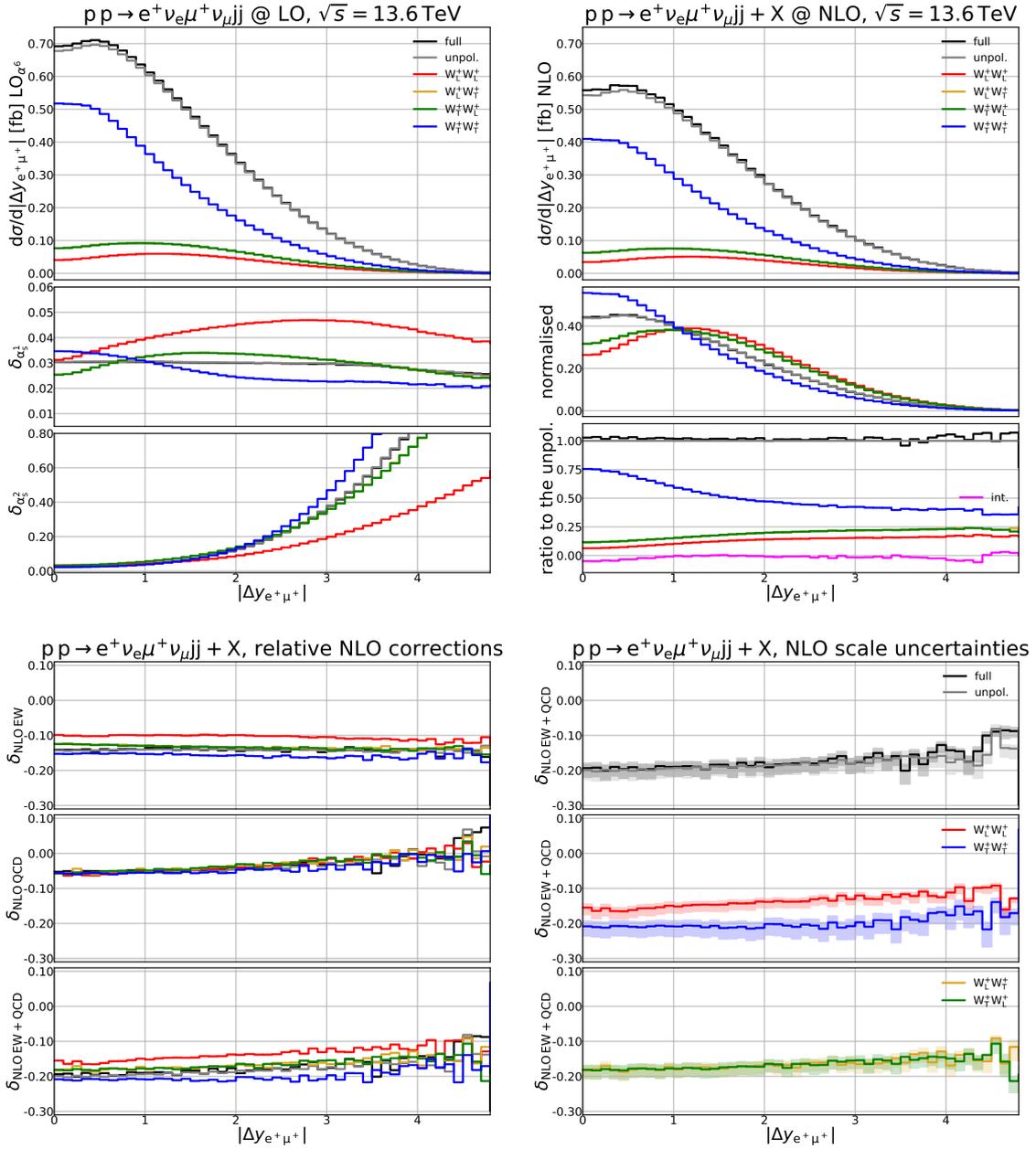

  \centering
  \subfigure{\includegraphics[scale=0.028, page=29]{Figures/BBMC_histograms_publication.pdf}}
  \subfigure{\includegraphics[scale=0.028, page=30]{Figures/BBMC_histograms_publication.pdf}}
  \subfigure{\includegraphics[scale=0.028, page=31]{Figures/BBMC_histograms_publication.pdf}}
  \subfigure{\includegraphics[scale=0.028, page=32]{Figures/BBMC_histograms_publication.pdf}}
  \caption{
    Distribution in the absolute value of the rapidity difference between the two
    charged leptons.
    Details are described in the main text (first paragraphs of \refse{sec:diffres}).    
  }\label{fig:dye+mu+} 
\end{figure}

\begin{figure}
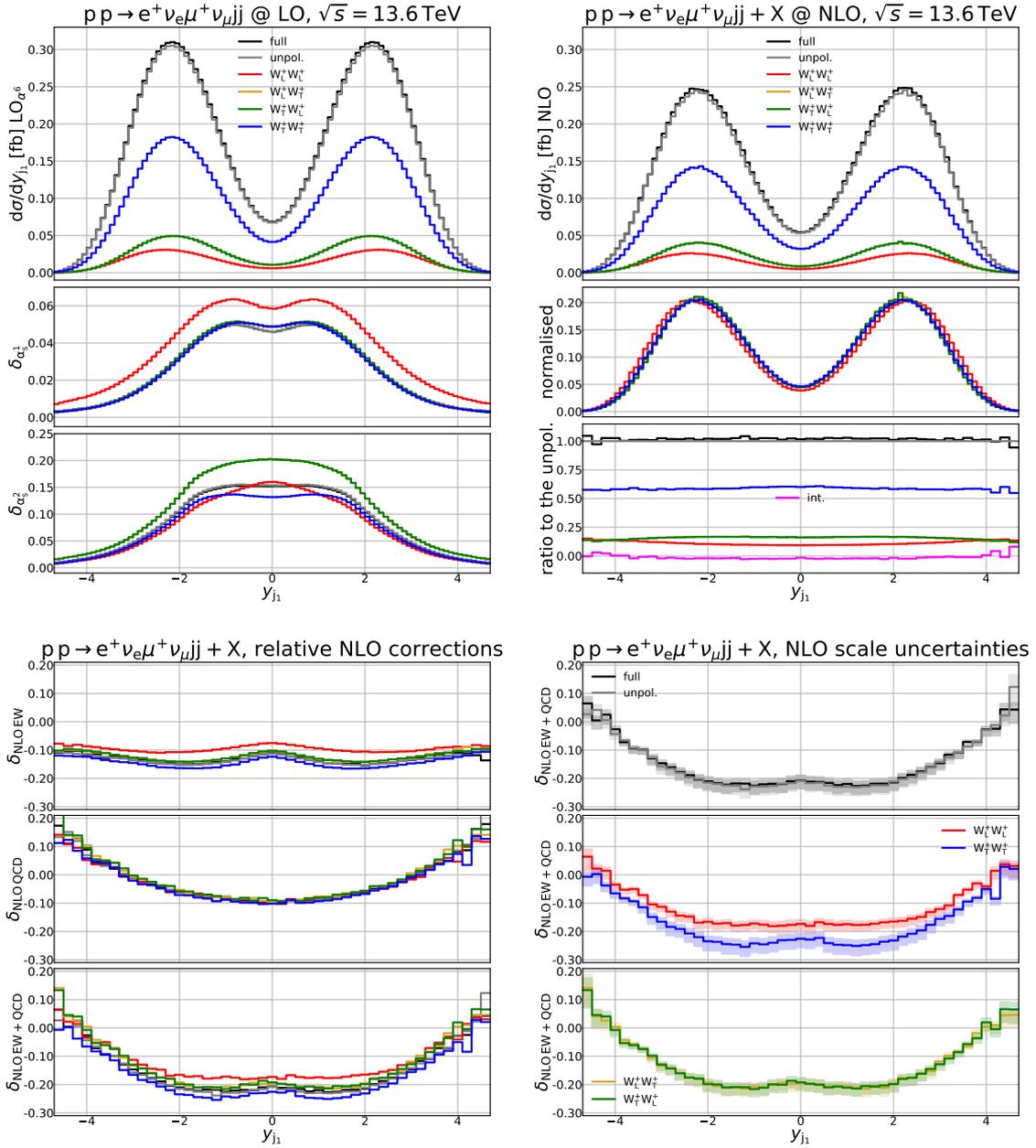

  \centering
  \subfigure{\includegraphics[scale=0.028, page=13]{Figures/BBMC_histograms_publication.pdf}}
  \subfigure{\includegraphics[scale=0.028, page=14]{Figures/BBMC_histograms_publication.pdf}}
  \subfigure{\includegraphics[scale=0.028, page=15]{Figures/BBMC_histograms_publication.pdf}}
  \subfigure{\includegraphics[scale=0.028, page=16]{Figures/BBMC_histograms_publication.pdf}}
  \caption{
    Distribution in the rapidity of the leading jet.
    Details are described in the main text (first paragraphs of \refse{sec:diffres}).    
  }\label{fig:yj1} 
\end{figure}

\begin{figure}
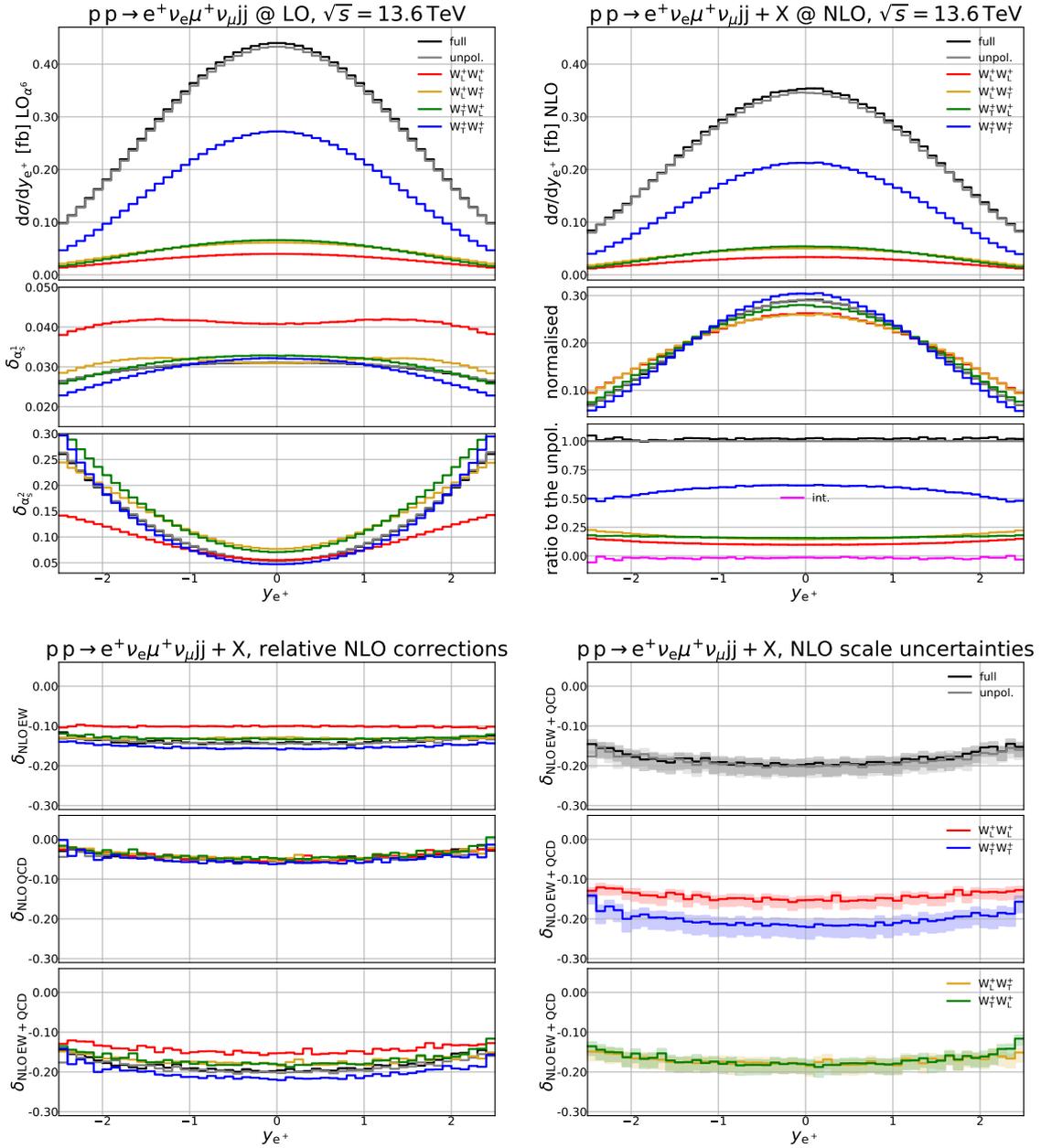

  \centering
  \subfigure{\includegraphics[scale=0.028, page=9]{Figures/BBMC_histograms_publication.pdf}}
  \subfigure{\includegraphics[scale=0.028, page=10]{Figures/BBMC_histograms_publication.pdf}}
  \subfigure{\includegraphics[scale=0.028, page=11]{Figures/BBMC_histograms_publication.pdf}}
  \subfigure{\includegraphics[scale=0.028, page=12]{Figures/BBMC_histograms_publication.pdf}}
  \caption{
    Distribution in the rapidity of the positron.
    Details are described in the main text (first paragraphs of \refse{sec:diffres}).    
  }\label{fig:ye+} 
\end{figure}

\begin{figure}
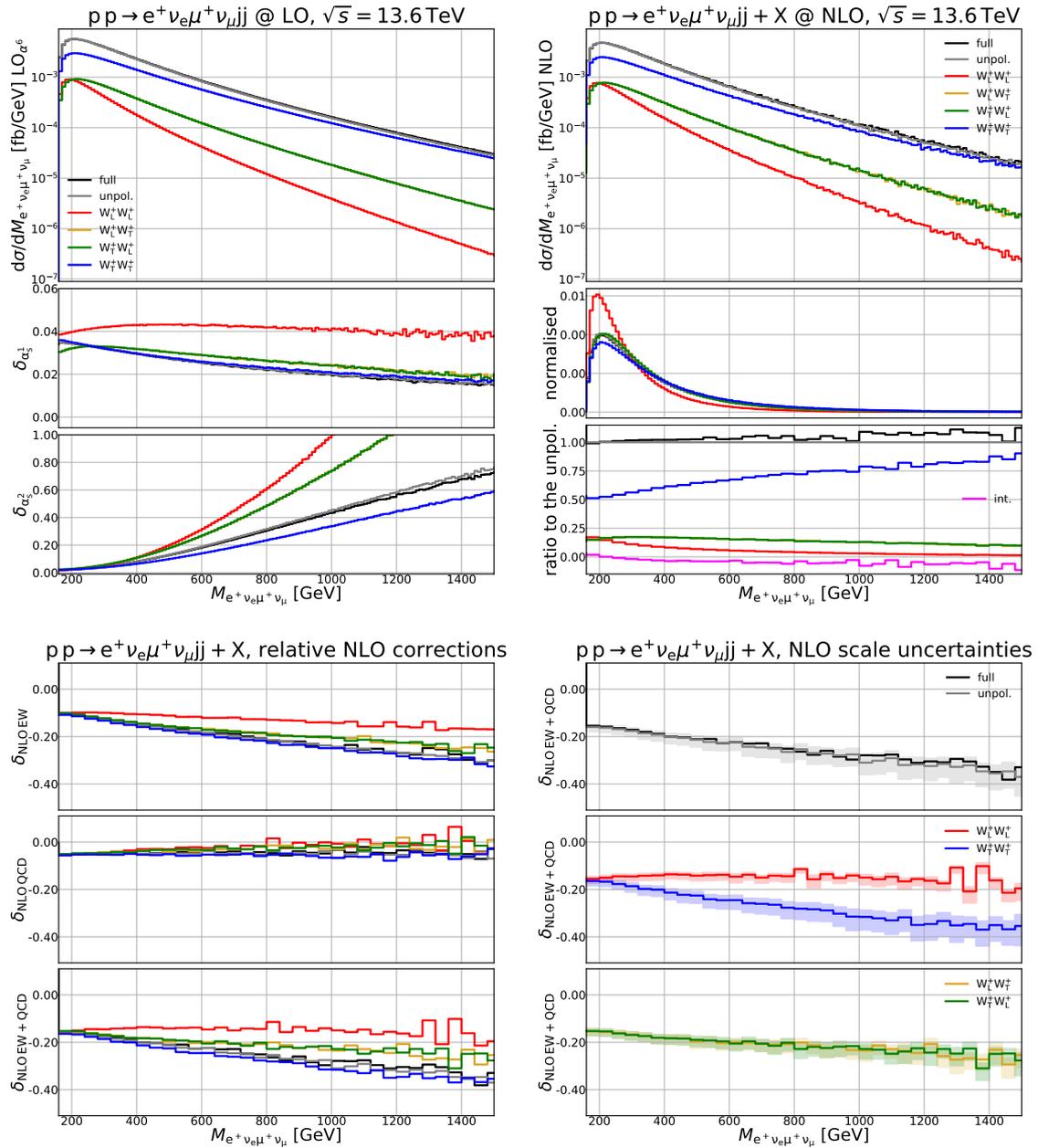

  \centering
  \subfigure{\includegraphics[scale=0.028, page=41]{Figures/BBMC_histograms_publication.pdf}}
  \subfigure{\includegraphics[scale=0.028, page=42]{Figures/BBMC_histograms_publication.pdf}}
  \subfigure{\includegraphics[scale=0.028, page=43]{Figures/BBMC_histograms_publication.pdf}}
  \subfigure{\includegraphics[scale=0.028, page=44]{Figures/BBMC_histograms_publication.pdf}}
  \caption{
    Distribution in the invariant mass of the system formed by the two charged leptons and the
    two neutrinos.
    Details are described in the main text (first paragraphs of \refse{sec:diffres}).    
  }\label{fig:M4l} 
\end{figure}

\begin{figure}
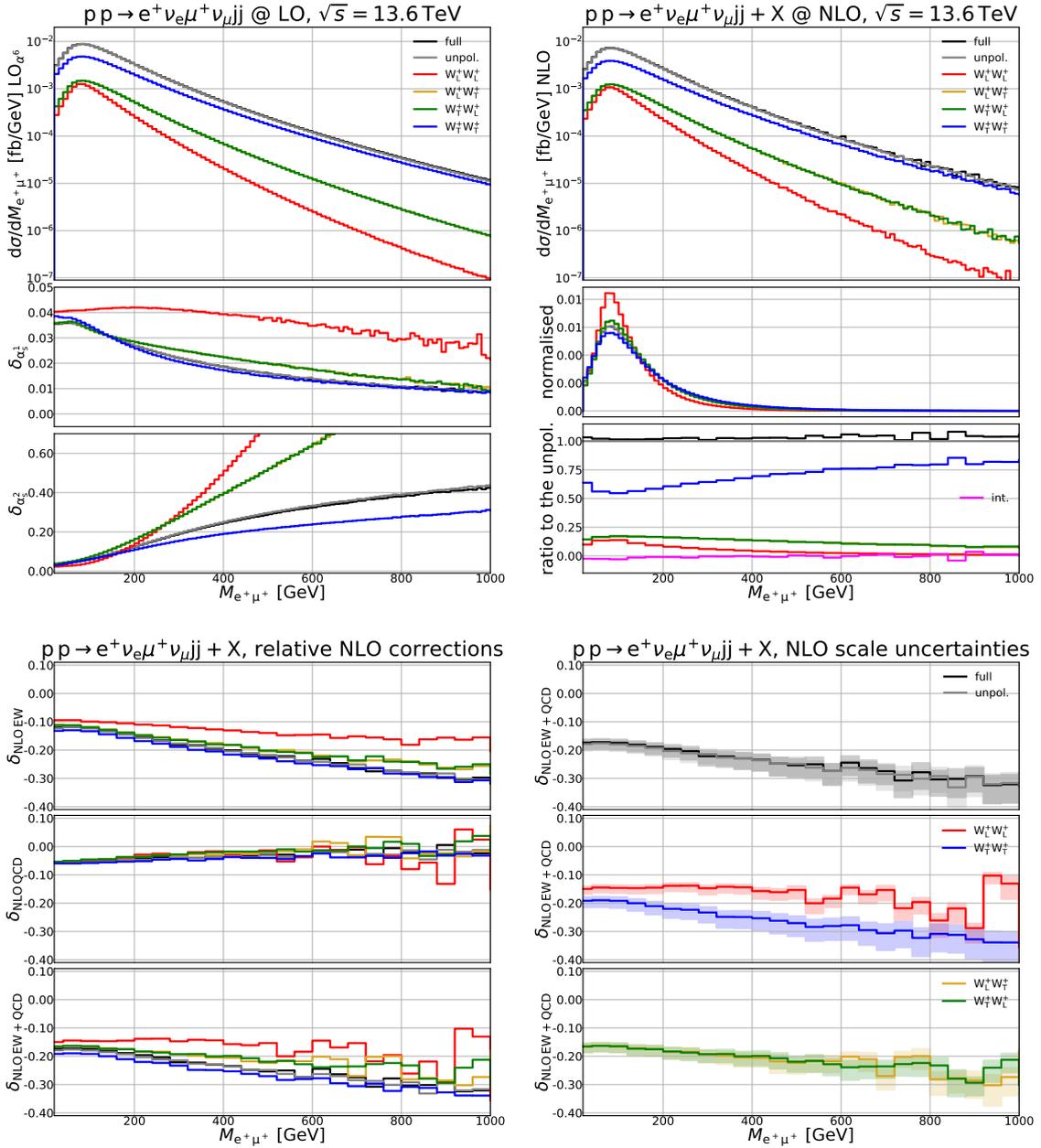

  \centering
  \subfigure{\includegraphics[scale=0.028, page=17]{Figures/BBMC_histograms_publication.pdf}}
  \subfigure{\includegraphics[scale=0.028, page=18]{Figures/BBMC_histograms_publication.pdf}}
  \subfigure{\includegraphics[scale=0.028, page=19]{Figures/BBMC_histograms_publication.pdf}}
  \subfigure{\includegraphics[scale=0.028, page=20]{Figures/BBMC_histograms_publication.pdf}}
  \caption{
    Distribution in the invariant mass of the charged-lepton pair.
    Details are described in the main text (first paragraphs of \refse{sec:diffres}).    
  }\label{fig:Me+mu+} 
\end{figure}

\begin{figure}
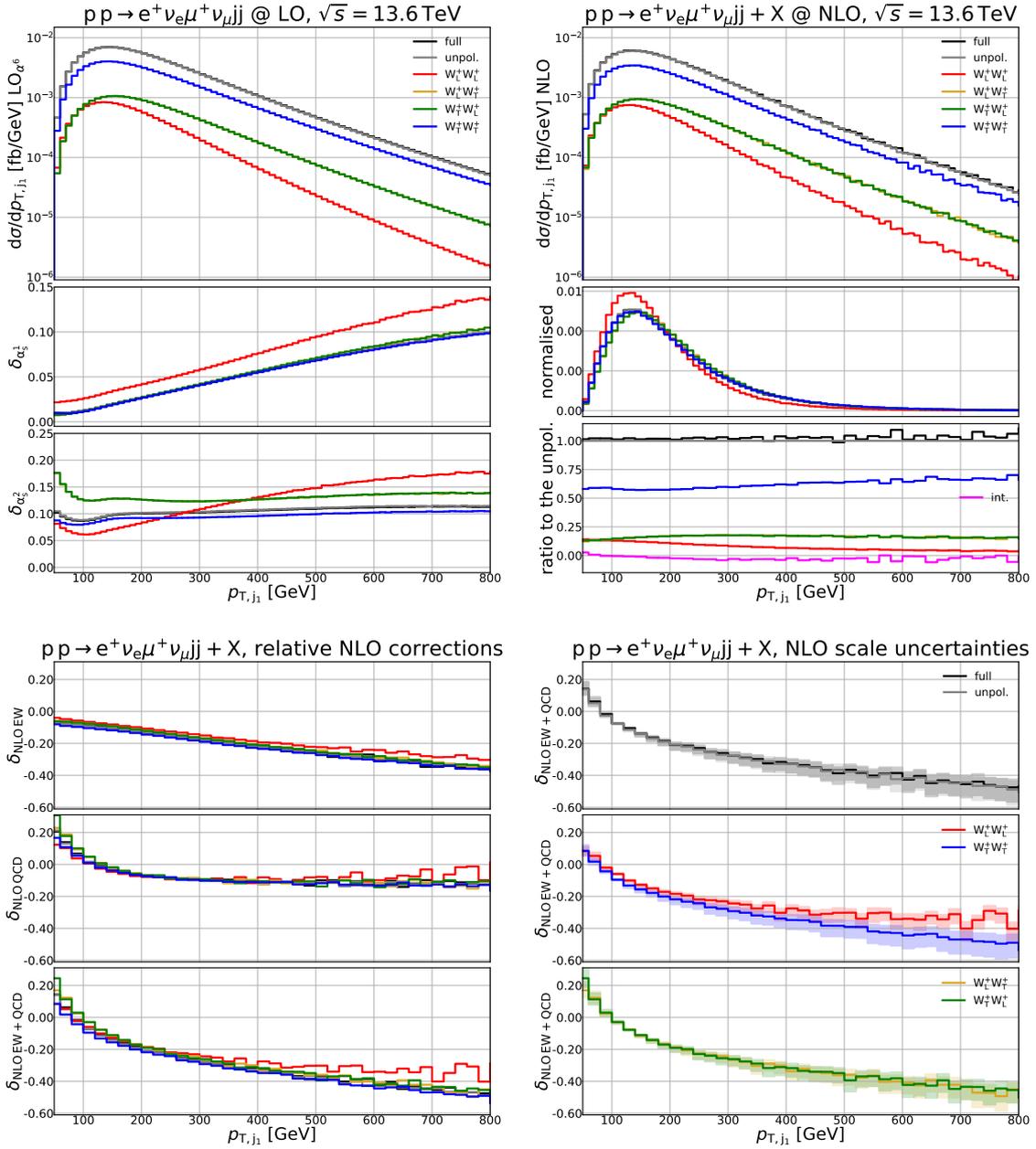

  \centering
  \subfigure{\includegraphics[scale=0.028, page=5]{Figures/BBMC_histograms_publication.pdf}}
  \subfigure{\includegraphics[scale=0.028, page=6]{Figures/BBMC_histograms_publication.pdf}}
  \subfigure{\includegraphics[scale=0.028, page=7]{Figures/BBMC_histograms_publication.pdf}}
  \subfigure{\includegraphics[scale=0.028, page=8]{Figures/BBMC_histograms_publication.pdf}}
  \caption{
    Distribution in the transverse momentum of the leading jet.
    Details are described in the main text (first paragraphs of \refse{sec:diffres}).    
  }\label{fig:ptj1} 
\end{figure}

\begin{figure}
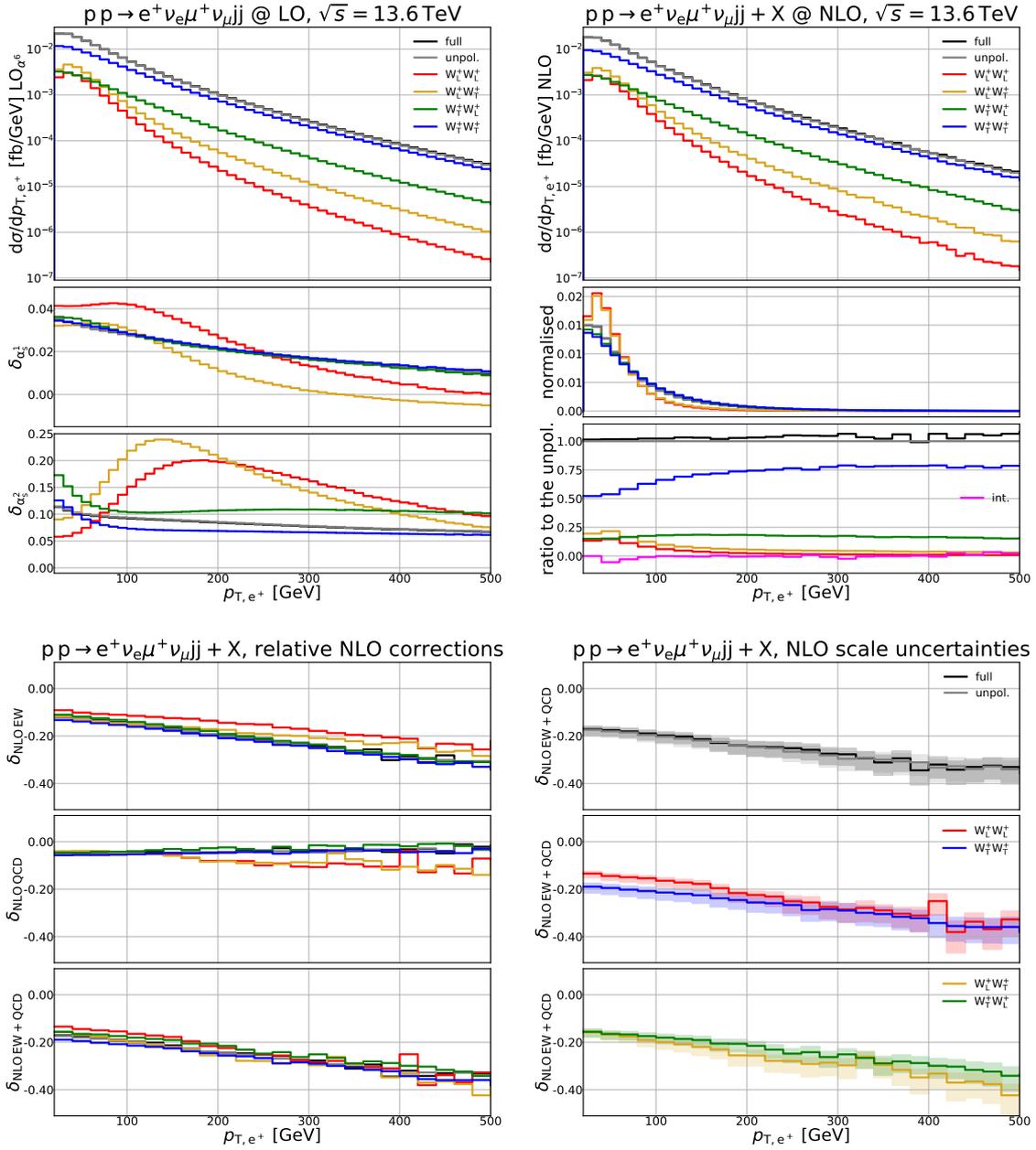

  \centering
  \subfigure{\includegraphics[scale=0.028, page=1]{Figures/BBMC_histograms_publication.pdf}}
  \subfigure{\includegraphics[scale=0.028, page=2]{Figures/BBMC_histograms_publication.pdf}}
  \subfigure{\includegraphics[scale=0.028, page=3]{Figures/BBMC_histograms_publication.pdf}}
  \subfigure{\includegraphics[scale=0.028, page=4]{Figures/BBMC_histograms_publication.pdf}}
  \caption{
    Distribution in the transverse momentum of the positron.
    Details are described in the main text (first paragraphs of \refse{sec:diffres}).    
  }\label{fig:pte+} 
\end{figure}

\begin{figure}
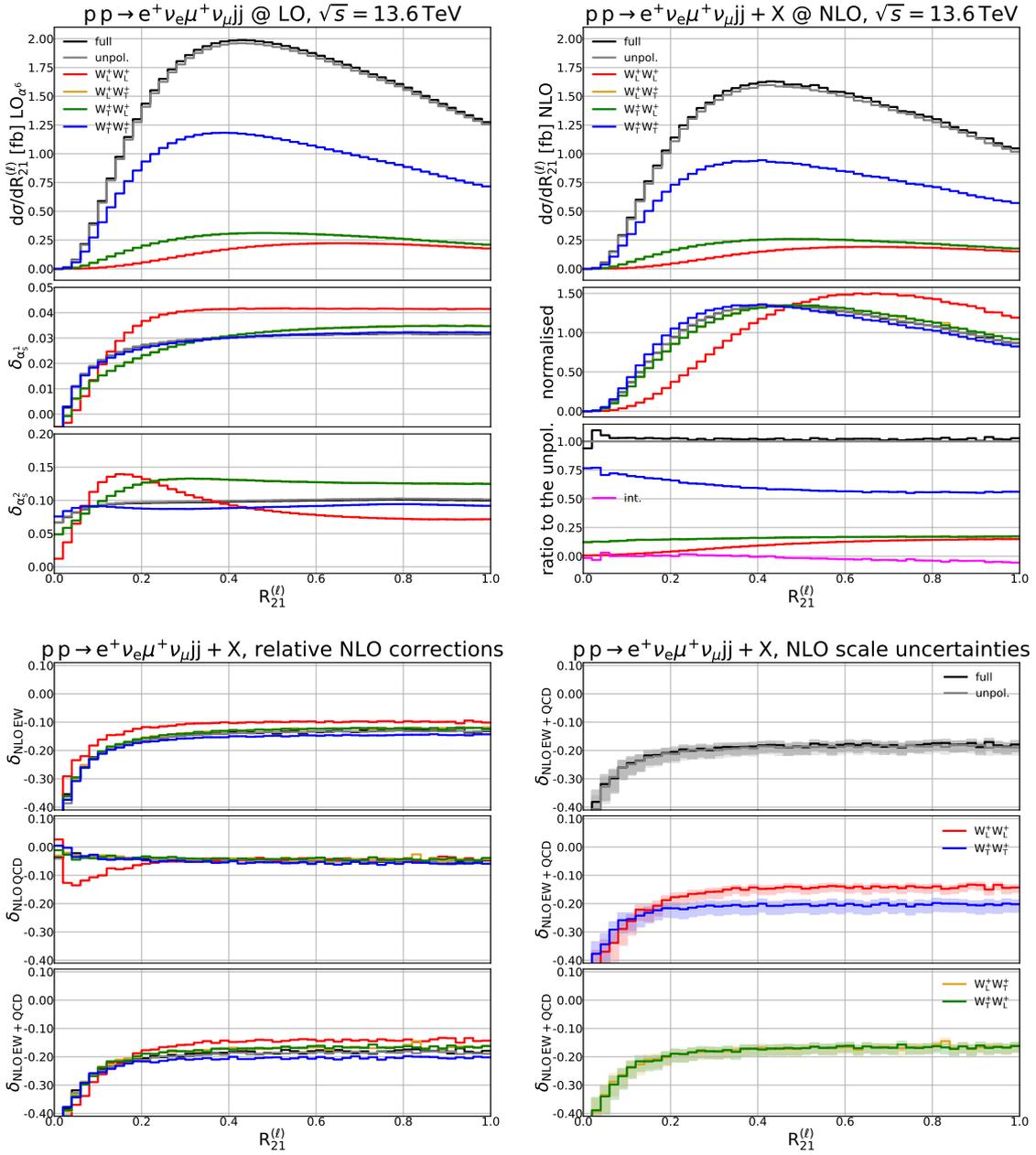

  \centering
  \subfigure{\includegraphics[scale=0.028, page=33]{Figures/BBMC_histograms_publication.pdf}}
  \subfigure{\includegraphics[scale=0.028, page=34]{Figures/BBMC_histograms_publication.pdf}}
  \subfigure{\includegraphics[scale=0.028, page=35]{Figures/BBMC_histograms_publication.pdf}}
  \subfigure{\includegraphics[scale=0.028, page=36]{Figures/BBMC_histograms_publication.pdf}}
  \caption{
    Distribution in the ratio of the transverse momenta of the
    subleading and leading lepton.
    Details are described in the main text (first paragraphs of \refse{sec:diffres}).    
  }\label{fig:R21l} 
\end{figure}

To fully understand the differences between the various polarisation states
it is necessary to study the cross-section at the differential level.
In \reffis{fig:costhetastar}--\ref{fig:R21l} we present differential results
for several observables, obtained in
the fiducial setup defined in \refse{sec:setup}.

For each distribution we show four plots with three panels each.
The upper-left plot contains LO results. Its top panel displays the
absolute distributions at $\order{\alpha^6}$ resulting from the full off-shell
calculation (black), the DPA calculation (grey, dubbed ``unpol.''), and the contribution
of the two longitudinal \PW~bosons (red), a longitudinal $\PW^+$~boson
decaying into $\Pe^+\nu_\Pe$ and a transverse
$\PW^+$~boson decaying into $\mu^+\nu_\mu$ (yellow), a transverse and
a longitudinal boson (green), and two 
transverse \PW~bosons (blue). The middle and bottom panels present the
contributions of orders $\order{\alphas\alpha^5}$ and $\order{\alphas^2\alpha^4}$, respectively,
normalised to the corresponding $\order{\alpha^6}$ results. 
The upper-right plot shows in its top panel the absolute distributions at NLO, \ie
including besides the $\order{\alpha^6}$ the corrections of orders $\order{\alpha^7}$
and $\order{\alphas\alpha^6}$ combined additively. The middle panel contains the same
curves but normalised to the corresponding fiducial cross-sections. The bottom
panel shows the same distributions normalised to the unpolarised DPA cross-section.
This allows to read off the NLO polarisation
fractions and the contributions beyond the DPA. In addition, the
interference contribution is displayed in magenta.
The lower-left plot displays the relative NLO corrections, \ie the EW
corrections of $\order{\alpha^7}$ in the top panel, the QCD
corrections of $\order{\alphas\alpha^6}$ in the middle panel, and the
sum of both in the bottom panel, all normalised to the
corresponding LO ($\order{\alpha^6}$).
The panels in the lower-right plot
accomodate the same NLO corrections already shown in the lower-left one
but supplemented with the corresponding scale variations. The results are obtained by
normalising the NLO cross-sections with 7-point QCD-scale variation by the
LO cross-section at the central scale, defined in \refeq{eq:scale}.

We begin by investigating angular distributions.
In \reffi{fig:costhetastar} the differential cross-section with
respect to the cosine of the decay angle of the positron in the
centre-of-mass frame of the decaying boson is presented. 
This distribution is directly sensitive to the
polarisation of the decaying vector boson. 
For $\PW^+ \PW^+$ scattering, it is not measurable as it would require the reconstruction of individual momenta of the two neutrinos. 
However, one can still learn much from this plot, as there are measurable observables with related features. 
The decay products of a transversely polarised $\PW^+$ boson are preferably emitted in or opposite to the boson direction. 
This causes the dip in the central region for the TT and TL polarisation states. 
For longitudinally polarised $\PW^+$ bosons the decay particles are preferably emitted orthogonal to the boson direction. 
Therefore, the LL and LT polarisation states feature a peak in the central region. 
The region $-1.0 < \cos \theta^{*, \rm CM}_{\Pe^+} \lesssim -0.75$ is strongly
affected by the transverse-momentum cut on the positron distorting the
shape expected in a fully inclusive setup. 
As a consequence of the cuts, the distributions for a longitudinal
$\PW$~boson are not symmetric with respect to $ \cos \theta^{*, \rm CM}_{\Pe^+}= 0$. 
The normalised shapes of the TT and TL
polarisation states at NLO exhibit small differences. 
The TL polarisation state has a more prominent peak near $\cos
\theta^{*, \rm CM}_{\Pe^+} = -0.7$, while the TT polarisation state
has a steeper increase towards $\cos \theta^{*, \rm CM}_{\Pe^+} = 1$. 
For a  longitudinally polarised  $\PW^+$ boson (LL, LT) the shapes are
almost identical. 
As already seen at the integrated level the interference contribution is small. 
While it is negative for $\cos \theta^{*, \rm CM}_{\Pe^+} \leq
0.25$, it becomes positive for larger $\cos \theta^{*, \rm CM}_{\Pe^+}$.
The relative contributions of the LO backgrounds are not flat, \ie
they have a shape different from the signal and tend to distort
the distributions towards smaller $\cos \theta^{*, \rm CM}_{\Pe^+}$. 
The relative NLO corrections both from EW and QCD are very close
to constant mirroring the effects seen at the integrated level. 
Only in the region $\cos \theta^{*, \rm CM}_{\Pe^+} < -0.75$, where
the phase-space cuts have the largest effect, small
deviations from the constant EW corrections for transverse
\PW~bosons become visible. The QCD-scale uncertainties relative to the
central values are roughly constant over the whole decay-angle
spectrum.
This holds also for other angular
distributions shown below.

Figure \ref{fig:dphie+mu+} depicts the differential cross-section with
respect to the angular separation of the positron and the anti-muon.  
As already observed at LO \cite{Ballestrero:2020qgv},
the normalised shapes of the distributions differ considerably between
the different polarisation states making this observable well suited
to distinguish the polarised signals.  
There is a clear preference for the positron and anti-muon to be emitted
in opposite directions when both $\PW^+$~bosons are transversely
polarised.  This is again related to the fact that the leptons tend to
be aligned with the \PW~bosons they originate from if the latter are
transverse.
While the distribution is completely flat for the LL mode, 
the mixed polarisation states show only a very small preference to be emitted in opposite hemispheres. 
As a consequence, the polarisation fraction of the TT mode increases to
$68\%$ for large $\Delta\phi_{\Pe^+\mu^+}$. The contribution of the QCD
background varies between $6$ and $16\%$ for the various polarised
modes and the contribution of the interference between $2.5$ and
$5\%$. While the QCD corrections are pretty flat, the EW corrections
for the individual polarised modes each vary within $2\%$.

Figure \ref{fig:dphij1j2} shows the differential cross-section with respect
to the azimuthal-angle separation between the leading and subleading jet. 
The two tagging jets are preferably emitted with large angular separations. 
This preference is even more pronounced when both \PW~bosons are
longitudinally polarised as seen in its distinct normalised shape. 
Accordingly, the longitudinal polarisation fraction increases from $4\%$ to $15\%$
with increasing $\Delta\phi_{\Pj_1\Pj_2}$.
The contribution of the QCD background varies between $5$ and $17\%$
for the various polarised modes and the contribution of the
EW-QCD interference between $-7$ and $+6\%$.
While the relative NLO EW corrections to the LL polarisation state are
almost constant at $-5\%$ for $\Delta\phi_{\Pj_1\Pj_2} < 1.0$, they  
increase in size to $-12\%$ with growing
$\Delta\phi_{\Pj_1\Pj_2}$. For the other polarisation modes the NLO EW
corrections vary within $3\%$
with a tendency to increase with the angular
difference. The relative QCD corrections become slightly 
more negative with growing $\Delta\phi_{\Pj_1\Pj_2}$.

In \reffi{fig:dye+mu+} we consider the differential cross-section with respect to
the absolute value of the rapidity difference between the positron and the anti-muon. 
From the normalised shapes it is obvious that this observable is well suited to discriminate between polarisation states. 
The fraction of the TT mode is largest for small $|\Delta y_{\rm \Pe^+
  \mu^+}|$, while those of the other modes increase with the variable.
The EW corrections depend only slightly on the rapidity difference,
while the QCD corrections decrease in size from $-6\%$ to $\approx 0\%$. 

In \reffi{fig:yj1} the differential cross-section with respect to the
rapidity of the leading jet is displayed. 
The curves for all polarisation states feature two peaks at $y_{\Pj_1} \approx \pm 2.2$. 
The typical VBS kinematics favours jets that are emitted in directions close to the beam axis.
The LL polarisation state leads to a slightly different shape compared
to the other polarisation states with the two peaks being shifted to
more forward/backward directions.  
There are noticeable shape differences between the LO signal and the
irreducible backgrounds, since
the latter prefer smaller leading-jet rapidities.
The polarisation fractions are almost independent of
$y_{\Pj_1}$. While the NLO EW corrections show only a mild variation
within about 4\%,
the NLO QCD corrections range between $-10\%$ for small $y_{\Pj_1}$ to
more than $+10\%$ at large $y_{\Pj_1}$. As for other angular
distributions, the relative QCD-scale uncertainty is
rather independent of the variable.

Figure \ref{fig:ye+} shows the differential cross-section with respect to the positron rapidity. 
The distribution is peaked in the central region as a
consequence of the fact that in VBS the two bosons are predominantly emitted in the central region. 
The QCD background has a very different shape from the signal one and
actually has two peaks at $y_{\Pe^+} \approx \pm 1.6$ (not directly
visible in the plots). 
This is caused by the structure of the QCD diagrams like the one shown in \reffi{fig:diagsNbody} 
(top right). They involve a $t$-channel gluon
exchange, and the $\PW^+$~bosons are emitted from the different quark
lines, preferably in forward and backward direction.
There are slight shape differences between the various polarised signals. 
A positron resulting from a transversely polarised $\PW^+$ boson
tends to be more central than one resulting from a longitudinally polarised one.
The EW corrections are independent of the positron rapidity if the positron comes from the decay of a
longitudinal boson, while a variation of $2\%$ is found in the case of a transverse boson. The QCD
corrections are pretty independent on the $\PW$~polarisation.

We turn to distributions in dimensionful variables starting with
invariant-mass distributions.
Figure \ref{fig:M4l} shows the differential cross-section with respect to the invariant mass of the four-lepton system, which is not measurable at the LHC owing to the presence of the two neutrinos. 
The results are anyway helpful in understanding many features found for other
correlated observables. 
The LO shapes of the signal and the two background contributions show clear differences. 
While the signal curves feature a sharp peak at $M_{4l} = 200\GeV$,
the QCD background is strongly enhanced at higher invariant masses. 
This is caused by the different topologies of the contributing Feynman diagrams. 
The dominant contributions to the signal result from diagrams that
feature the VBS topology, like the one depicted in the top left of \reffi{fig:diagsNbody}.
Since the tagging jets carry most of the
energy, this results in a four-lepton invariant mass of the order of a
few hundred GeV \cite{Biedermann:2016yds}. 
As mentioned already in the discussion of the $y_{\Pe^+}$
  distribution, the diagrams that contribute to the QCD background instead feature a
$t$-channel gluon exchange with \PW~bosons emitted from different
quark lines preferably in the direction of the beams (see top right of \reffi{fig:diagsNbody}).  
This favours the production of two \PW~bosons with a larger invariant mass. 
For high invariant masses, the rate at which the distributions drop depends
on the polarisation of the $\PW^+$ bosons. 
As can be best seen from the normalised shapes, the TT mode falls off the slowest, while the LL polarisation state
drops the fastest and
the mixed polarisation states have a drop in between. 
This feature has already been observed at LO for VBS processes before
\cite{Ballestrero:2017bxn,Ballestrero:2019qoy}. 
The different fall-off rates cause the polarisation fractions of the TT polarisation state to increase and of the LL polarisation state to decrease with the invariant mass. 
The production of pairs of longitudinally polarised $\PW$ bosons
happens primarily at small invariant masses.
Shifting the attention to the relative NLO corrections, one observes
that the NLO EW corrections become more negative with increasing invariant mass 
owing to the growing significance of the Sudakov logarithms.
As already emphasised at the integrated level, the size of the NLO EW
corrections depends on the polarisation.  For the production of a pair
of longitudinal $\PW^+$ bosons they are smaller than for the other
polarisation states and grow at a smaller rate. 
The reason for this is again the smaller EW Casimir operator for
longitudinal vector bosons.
The NLO QCD corrections remain fairly constant across the considered
range of invariant masses and polarisation states with a variation
within $\approx 5 \%$.  
The relative QCD-scale
uncertainties grow with increasing invariant mass. The growth is
larger for transverse vector bosons than for longitudinal vector
bosons. This feature is also seen in the distributions for other dimensionful
quantities shown below.

Figure \ref{fig:Me+mu+} shows the differential cross-section with respect to the invariant mass of the two charged leptons. 
This observable is correlated with the four-lepton invariant mass shown in \reffi{fig:M4l} and has very similar features. 
Thus, the distributions drop with different speeds for the different polarisation modes.
Also the steep increase of the relative contribution of the QCD
background is analogous to the one seen for the four-lepton invariant mass. 
Moreover, the polarisation fractions of the longitudinal modes become smaller
with increasing  invariant mass. Also the behaviour of the relative EW
and QCD corrections and the corresponding scale dependence is quite
similar to those in \reffi{fig:M4l}.

In \reffi{fig:ptj1} the differential cross-section with respect to the
transverse momentum of the hardest jet is presented.
The irreducible LO background becomes more important with increasing
$p_{\rT,\Pj_1}$, in particular for the LL mode.
The purely longitudinal polarisation state falls off faster in
the high-transverse momentum region than the other polarisation
states.  
The NLO EW corrections become more negative with increasing transverse
momentum. The difference between the EW corrections for the different
polarisation modes is smaller than for the four-lepton invariant mass
or the lepton-pair invariant mass. 
The NLO QCD corrections are pretty constant above
$p_{\rT,\Pj_1}\approx250\GeV$, but rise for low transverse momenta, where the LO
cross-section is suppressed.
This feature has already been observed in other VBS studies \cite{Biedermann:2017bss,Denner:2019tmn,Denner:2022pwc}.

Figure \ref{fig:pte+} is devoted to the differential cross-section with respect to the
transverse momentum of the positron. 
The QCD background and to a lesser extent the interference background have different shapes then the LO signal. 
This is particularly significant for the LL and LT polarisation
states, while for the TT and TL polarisation states the shape differences between the signal and background are smaller. 
As can be seen from the normalised shapes, the distributions for a
positron resulting from the decay of a transverse vector boson falls
less steeply than those from a longitudinal vector boson, which
actually has a peak near  $p_{\rm T, \Pe^+} = 40 \GeV$. This can be
explained by the fact that the positron is preferably aligned with the
direction of a transverse mother $\PW$ boson, while it tends to be
orthogonal to the direction of a longitudinal mother. Accordingly, the
LL polarisation fraction drops fast with increasing $p_{\rm T, \Pe^+}$.
The relative NLO EW corrections to the states with longitudinally
polarised vector bosons decaying into the positron are smaller than
those with transversely polarised bosons.  
The reason is again the smaller prefactor of the Sudakov logarithms,
pointed out for the integrated level and the four-lepton
invariant-mass distribution.
Also the NLO QCD corrections are somewhat different for positrons
resulting from transverse or longitudinal vector bosons. 
As a result the differences between transverse and
longitudinal vector bosons tend to cancel in the sum of NLO EW and QCD corrections.

We finally consider in \reffi{fig:R21l} the differential cross-section
with respect to the ratio of
the transverse momenta of the subleading charged lepton and the leading charged lepton,
\beq
R_{21}^{(\ell)} = \frac{\pt{\ell_2}}{\pt{\ell_1}}\,.
\eeq
This observable is well suited to distinguish the purely longitudinal polarisation state from the other polarised modes. 
The transverse momentum of the two charged leptons is more similar when the $\PW^+$~bosons are longitudinally polarised. 
The normalised distribution for the LL polarisation state has a maximum
at $R^{(\ell)}_{21} \approx 0.65$, while the other polarised signals
have maxima near $R^{(\ell)}_{21} \approx 0.40$. 
The distribution for the TT mode is only slightly shifted towards
lower $R_{21}^{(\ell)}$ compared to the mixed modes.
These results can be understood from \reffi{fig:pte+}. Since the
$p_{\rT,\Pe^+}$ spectrum is much softer for leptons resulting from
longitudinal \PW~bosons than for those from transverse \PW~bosons, the
momenta of leading and subleading lepton tend to be more equal for the
LL mode as compared to modes with transverse vector bosons, where the
leading lepton (typically resulting from a transverse vector boson)
can have a higher transverse momentum more easily. 
Accordingly the polarisation fraction of the LL mode increases from very small
values to more than $15\%$ near  $R_{21}^{(\ell)}=1$. Note that also the
interference contribution reaches $-5\%$ there.
Both the NLO QCD and EW corrections are almost constant for $R^{(\ell)}_{21} \gtrsim 0.4$.
For $R^{(\ell)}_{21} \lesssim 0.2$ the EW
corrections grow negative and reach $-40\%$. This is a consequence of
the EW Sudakov logarithms that dominate at large lepton transverse
momenta (see \reffi{fig:pte+}).  Since the subleading charged lepton
must have a transverse momentum of at least $20\GeV$ owing to the
phase-space cuts, the leading lepton must have a very large transverse
momentum in the low $R^{ (\ell)}_{21}$ region. Also the negative QCD
corrections for the LL mode below $0.2$ result from the larger
negative QCD corrections for high transverse momenta of leptons
originating from longitudinal \PW~bosons.

\section{Conclusion}\label{sec:conclu}
The importance of vector-boson scattering (VBS) in unravelling the
electroweak (EW) symmetry-breaking mechanism is
witnessed by a wide experimental programme targeting this fundamental
process in hadronic collisions at the LHC,
as well as by many theoretical studies addressing both
higher-order corrections in the SM and new-physics effects that give
relevant phenomenological consequences at the LHC energies.
While the full off-shell VBS processes are known up to NLO EW+QCD accuracy
in all production mechanisms, the modelling of intermediate bosons with
definite polarisation modes is still limited to LO accuracy.
These predictions are currently used by experimental analyses aiming
at the measurement of longitudinal VBS.
Therefore an effort is needed to achieve a better theoretical control in view of
the upcoming Run-3 and High-Luminosity data sets.
In this work we have partially filled this gap, achieving for the first time NLO EW+QCD
accuracy for doubly polarised $\PW^+\PW^+$ scattering in the fully leptonic decay channel.
This results from a non-trivial extension of methods applied in inclusive di-boson production,
relying on the pole expansion and the definition of polarised signals at the amplitude level.

We have computed polarised and unpolarised cross-sections at LO and including additively
NLO EW and QCD corrections, both at integrated level and differentially in several LHC
observables and Monte Carlo-truth variables.

Both radiative corrections are negative with the usual dynamical choice for the central
factorisation and renormalisation scale. While relative QCD corrections, which amount at
integrated level to about $-5\%$, are quite independent of the polarisation state, the
NLO EW corrections give different results for longitudinal and transverse modes, both at
the normalisation level and at the shape level. The LL state receives $-10\%$ EW corrections,
while other states receive between $-13\%$ and $-15\%$ corrections,
in line with literature results for the off-shell process.

The polarisation fractions mildly change from LO to NLO,
with the LL contribution being around $10\%$,
the LT and TL ones $16\%$ each,
and the TT state giving the dominant contribution ($59\%$).
The relative impact of the non-resonant background ($+2\%$)
and of the polarisation interferences ($-2\%$)
remains almost unchanged when adding NLO corrections.

NLO EW corrections distort the shapes of differential distributions 
both in angular observables and in energy-dependent ones, in
particular with the LL state being
affected differently than the other states.
On the contrary, the QCD corrections change the shapes of
distributions in a very similar way for all doubly polarised states.
A marked discrimination power between polarisation modes
is found in typical leptonic angular distributions, but also in
jet observables, like the azimuthal-angle distance between the VBS tagging jets.
Also some invariant-mass and transverse-momentum distributions can be included in experimental
analyses, either as polarisation discriminants by themselves or as inputs to modern neural-network
or boosted-decision-tree scores. Interestingly, the ratio between the subleading- and leading-lepton
transverse momenta gives a LL shape which noticeably differs from all other modes. 

With this work, we have improved the perturbative description in the EW and QCD couplings for
polarised cross-sections and distributions in VBS, achieving the same state-of-the-art accuracy as the
one of full off-shell processes. This paves the way for subsequent NLO calculations targeting other
VBS production mechanisms and including parton-shower effects.
The presented calculation provides SM predictions that could be used in ATLAS and CMS analyses of
Run-3 VBS data.

\section*{Acknowledgements}
We would like to thank Sandro Uccirati and Jean-Nicolas Lang for their
maintenance of \recola and Daniele Lombardi and Christopher Schwan for useful discussions.
This work is supported by the German Federal Ministry for Education and Research (BMBF)
under contract no.~05H21WWCAA
and the German Research Foundation (DFG) under reference number~DE 623/8-1.
The authors acknowledge support from the COMETA COST Action CA22130.


\appendix
\section{Mismatch in the local and integrated counterterms}
\label{sec:appendix_dpa_cs}

In \refse{sec:prod_counterterms} we have documented our approach for the local subtraction of
IR singularities associated to the production sub-process, highlighting the importance
of applying first the CS mappings and second the DPA on-shell projection.
In this appendix we show that reversing the order of the two operations (DPA first, CS second)
leads to a mismatch between local and integrated dipoles.

Choosing the same dipole as in \refse{sec:prod_counterterms}, the
partially subtracted real contribution for the production sub-process
with two \emph{on-shell} $\PW$~bosons with momenta $\tilde{k}_{12}$,
$\tilde{k}_{34}$ reads
\begin{align}\label{eq:dpa_2_prod}
        (\mc R-\mc D)_{\rm prod}         \propto{}&
\sum_{\lambda_{12}\lambda_{34}}\biggr[
\left|\cM^{(5)}_{\rP,\mu\nu}\left(Q; \tilde{k}_{12}, \tilde{k}_{34},
    k_5,k_6,k_7\right)
  {\tilde{\varepsilon}}^{\mu,*}_{12}\tilde{\varepsilon}^{\nu,*}_{34}\right|^2 \notag\\
&\qquad\times\rd\Phi_5\Bigl(Q;\tilde{k}_{12},\tilde{k}_{34}, k_5,k_6,k_7\Bigr)\notag\\
                & {}-\mc D_{[12]7,5}(\bar{\tilde{k}}_{12},\bar{k}_5;\tilde{y}, \tilde{z}, \tilde{\phi})\,{\left|\cM^{(4)}_{\rP,\mu\nu}\left(Q; \bar{\tilde{k}}_{12}, \tilde{k}_{34}, \bar{k}_5,k_6\right) {\tilde{\varepsilon}}^{\mu,*}_{12}\tilde{\varepsilon}^{\nu,*}_{34}\right|^2}\notag\\
                &\qquad \times
\rd\Phi_{\rm rad}\left(\bar{\tilde{k}}_{12}+\bar{k}_5;\tilde{z},\tilde{y},\tilde{\phi}\right)
\rd\Phi_4\left(Q;\bar{\tilde{k}}_{12},\tilde{k}_{34}, \bar{k}_5,k_6\right)\biggl]\,,
\end{align}
where the dipole $\mc D_{[12]7,5}$ regulates the IR singularities
associated to a massive final-state emitter (the $\PW$~boson with
momentum $\tilde{k}_{12}$) and a massless final-state spectator with
momentum $k_5$.

The analogue of \refeq{eq:dpa_2_prod}, but including $\PW$-boson
decays and therefore carrying out the local subtractions with
DPA-projected momenta in the off-shell phase space, reads
\begin{align}\label{eq:rsubtr_2}
                (\mc R-\mc D)_{\rm DPA} {}&{} \propto 
\sum_{\lambda_{12}\lambda_{34}}\biggr[
{\left|\cM^{(5)}_{\rP,\mu\nu}\left(Q; \tilde{k}_{12}, \tilde{k}_{34}, k_5,k_6,k_7\right) {\tilde{\varepsilon}}^{\mu,*}_{12}\tilde{\varepsilon}^{\nu,*}_{34}\right|^2}\notag\\
                & \qquad \times\,\rd\Phi_5\left(Q;{k}_{12},{k}_{34}, k_5,k_6,k_7\right)\notag\\
                &\,-{{\mc D}_{[12]7,5}(\bar{\tilde{k}}_{12},\bar{k}_5;\tilde{y}, \tilde{z}, \tilde{\phi})\,\left|\cM^{(4)}_{\rP,\mu\nu}\left(Q; \bar{\tilde{k}}_{12}, \tilde{k}_{34}, \bar{k}_5,k_6\right) {\tilde{\varepsilon}}^{\mu,*}_{12}\tilde{\varepsilon}^{\nu,*}_{34}\right|^2}\notag\\
                &\qquad  \times\,\rd\Phi_{\rm rad}\left(\bar{k}_{12}+\bar{k}_5;z,y,\phi\right)\rd\Phi_4\left(Q;\bar{k}_{12},{k}_{34}, \bar{k}_5,k_6\right)\biggl] \notag\\
                &\, \times 
\frac{1}{\bw{k_{12}}\,\bw{k_{34}}}
\left|{\tilde{\varepsilon}}^\mu_{12}\cM_{\rD,\mu}^{(2)}\left( \tilde{k}_{12}; \tilde{k}_1,\tilde{k}_2\right) \right|^2\,\left|{\tilde{\varepsilon}}^\mu_{34}\cM_{\rD,\mu}^{(2)}\left( \tilde{k}_{34}; \tilde{k}_3,\tilde{k}_4\right) \right|^2\notag\\
                &\,\times \frac{\rd k_{12}^2}{2\pi}\frac{\rd k_{34}^2}{2\pi} \rd\Phi_2\left(k_{12};{k}_1,k_2\right) \rd\Phi_2\left(k_{34};{k}_3,k_4\right)\,.      
\end{align}
Comparing \refeqs{eq:dpa_2_prod}--\eqref{eq:rsubtr_2}, it is easy
to spot a difference between the phase-space measures in the real and in the dipole contribution.
In particular,
\beq
\rd\Phi_5\left(Q;\tilde{k}_{12},\tilde{k}_{34}, k_5,k_6,k_7\right)
\quad\rightarrow
\quad\rd\Phi_5\left(Q;{k}_{12},{k}_{34}, k_5,k_6,k_7\right)
\,,\label{eq:mismatchR}
\eeq
and
\begin{align}\label{eq:mismatchD}
&\rd\Phi_{\rm rad}\left(\bar{\tilde{k}}_{12}+\bar{k}_5;\tilde{z},\tilde{y},\tilde{\phi}\right)
\rd\Phi_4\left(Q;\bar{\tilde{k}}_{12},\tilde{k}_{34}, \bar{k}_5,k_6\right)\nnb
\quad\rightarrow \\
&\rd\Phi_{\rm rad}\left(\bar{k}_{12}+\bar{k}_5;z,y,\phi\right)\,\,\rd\Phi_4\left(Q;\bar{k}_{12},{k}_{34}, \bar{k}_5,k_6\right)          \, . 
\end{align}
Since in the DPA, the full phase space is used, the transformation in \refeq{eq:mismatchR} leads to a Jacobian associated to the on-shell projection, which is of order $\mc O(\MW^2/k^2_{12})$. Although this can have a sizeable impact on the fully differential phase-space measure, it represents an effect beyond the DPA upon integration over the $\PW$-boson off-shellness $k^2_{12}$, since the far off-shell kinematic configurations are suppressed by the Breit-Wigner modulation of the real matrix element.
For \refeq{eq:mismatchD} the situation is more complicated because
for dipoles with an intermediate resonance as emitter and/or spectator
the DPA on-shell projection and the CS mapping do not commute.
In the evaluation of the local counterterm, the IR kernel and the
underlying Born matrix element are evaluated starting from momenta for
the real-radiation process that are first projected on~shell and then undergo the CS mapping.
In order to stick to literature results \cite{Catani:2002hc} for both
local and integrated counterterms, the integrated ones are evaluated
with a LO phase-space kinematics that is projected on~shell,
intrinsically corresponding to first applying the subtraction mapping
to the real kinematics and then projecting the momenta on-shell. Since
CS mappings and DPA projections do not commute, this procedure leads to a mismatch between local and integrated counterterms which can be numerically sizeable, spoiling the subtraction procedure by effects that are not formally beyond the DPA accuracy.
This mismatch could be prevented upon including the Jacobian of the
DPA on-shell mapping in the analytic $d$-dimensional integration of the local counterterms in the DPA, which is, however, hampered by more complicated integrands than those of \citere{Catani:2002hc}.
On the other hand, reversing the order of the CS-mapping and DPA
on-shell projection is
more convenient, because the subtraction procedure is not spoiled and the integrated-dipole structures known in the literature can be safely recycled with mismatches that are beyond the intrinsic DPA accuracy.

\section{Production dipoles}
\label{sec:production_counterterms}
As detailed in \refse{sec:prod_counterterms}, after partial fractioning the real-photon radiation
off $\PW$-boson propagators, the subtraction of QED
IR divergences associated to the production process
($qq \longrightarrow \PW^+\PW^+qq + \gamma$) requires a class of dipoles where at least one $\PW$~boson
plays the role of the emitter or the spectator. Owing to the finite
mass of the W~boson the corresponding singularities result only from
soft but not from collinear configurations and are therefore spin independent.
This allows us to use the massive-fermion dipoles from
\citeres{Catani:2002hc,Schonherr:2017qcj} for these counterterms. In
the following, we detail their structure in the general case of
charged resonances (not necessarily $\PW$ bosons). 

\subsection{Final-state massive emitter and final-state massive spectator}

We start with the dipole where both the emitter (label $j$) and the
spectator (label $k$) are a massive charged resonance.
The results can be obtained as a special case from Section 5.1 of
\citere{Catani:2002hc} up to the translation from QCD to QED.
Let us call $Q = p_i + p_j + p_k$ the sum of the emitter momentum ($p_j$), the spectator momentum ($p_k$) and the radiated-photon momentum ($p_i$).
The emitted photon (label $i$) is always massless, \ie $p_i^2 = m_i^2 = 0$.
Since, according to the discussion in \refse{sec:prod_counterterms},
the CS mappings are applied to the off-shell kinematics, in the
following $p_j^2$ and $p_k^2$ are in general not equal to the squared
pole masses of the resonant particles. Moreover, we choose to set
$m_{ij}^2$ equal to $p_j^2$.
Then, the CS mapping takes the following form,
\begin{equation}
        \begin{split}
                \bar{p}_k &= \sqrt{\dfrac{\lambda\left(Q^2, p_j^2, p_k^2\right)}{\lambda\left(Q^2, \left(p_i + p_j\right)^2, p_k^2\right)}} \left( p_k - \dfrac{Q\!\cdot\! p_k}{Q^2}Q \right) + \dfrac{Q^2 + p_k^2 - p_j^2}{2 Q^2} Q, \\
                \bar{p}_j &= Q - \bar{p}_k\,,
        \end{split}
\end{equation}
where $\lambda(x,y,z) = x^2+y^2+z^2-2xy-2xz-2yz$.
The radiation variables are defined as
\beq
z\equiv z_i =
\frac{p_i\!\cdot\!p_k}{p_i\!\cdot\!p_k+p_j\!\cdot\!p_k}\,,\quad z_j =
1-z_i\,,\quad y\equiv y_{ij,k} = \frac{p_i\!\cdot\!p_j}{p_i\!\cdot\!p_j+p_i\!\cdot\!p_k+p_j\!\cdot\!p_k}.
\label{eq:CS_mapping_massive_massiv_final_final_invariants}
\eeq
We introduce rescaled squared momenta for the massive emitter and spectator,
\begin{equation}\label{eq:CS_mapping_massive_final_final_mu}
        \begin{split}
                \mu_{j}^2 &= \dfrac{p_j^2}{Q^2}\,,\qquad \mu_{k}^2 = \dfrac{p_k^2}{Q^2}\,
        \end{split}
\end{equation}
and the quantities
\begin{equation}\label{eq:CS_mapping_massive_final_final_vijk}
        \begin{split}
                \bar{v}_{ij,k} &= \dfrac{\sqrt{\lambda\left(1, \mu_{j}^2, \mu_{k}^2\right)}}{1 - \mu_{j}^2 - \mu_{k}^2}
          = \dfrac{\sqrt{1 + \left(\mu_{j}^2\right)^2 + \left(\mu_{k}^2\right)^2 - 2 \left(\mu_{j}^2 + \mu_{k}^2 + \mu_{j}^2 \mu_{k}^2\right)}}{1 - \mu_{j}^2 - \mu_{k}^2}\,,\\
                v_{ij,k} &= \dfrac{\sqrt{\left[2 \mu_{k}^2 + \left(1 - \mu_{j}^2 - \mu_{k}^2\right)\left(1 - y\right)\right]^2 - 4 \mu_{k}^2}}{\left(1 - \mu_{j}^2 - \mu_{k}^2\right) \left(1 -y\right)}\,. \\
        \end{split}
\end{equation}
The local dipole reads
\begin{equation}
        \begin{split}
                {\mc D}_{ij,k} =& (8\pi\alpha)\mu^{2\epsilon}
                \dfrac{\theta_{[ij]}\theta_k Q_{[ij]}Q_k}{-y Q^2 \left(1 - \mu_{j}^2 - \mu_{k}^2\right)}\\
                &\, \times\left[
                \dfrac{2}{1 - \left(1 - z\right)\left(1-y\right)} - \dfrac{\bar{v}_{ij,k}}{v_{ij,k}}
                \left(2 - z \left(1-\epsilon \right) + \dfrac{2 \mu_{j}^2}{y\left(1 - \mu_{j}^2 - \mu_{k}^2\right)}\right)
                \right]\,,
        \end{split}
\end{equation}
where $\alpha$ is the EW coupling, $\mu$ is the
IR-regularisation scale, $\theta_a$ equals $+1$ or $-1$ if
particle $a$ is in the final or initial state, respectively, and
$Q_a$ is the relative electric charge of particle $a$. Note that the
label $[ij]$ is associated to the mapped emitter, which in the case of photon emission
has always the same flavour, charge, and mass as the original emitter $j$. 
The corresponding integrated dipole, obtained through integration over the radiation measure in dimensional regularisation, reads
\begin{equation}\label{eq:integFFMM}
        \begin{split}
          \int\!\rd\Phi^{(4-2\epsilon)}_{\rm rad}&{\mc D}_{ij,k}
          \,= \left(-\theta_{[ij]}\theta_k Q_{[ij]}Q_k \right)
          \frac{\alpha}{2\pi}\frac{(4\pi)^\epsilon}{\Gamma(1-\epsilon)}\left(\frac{\mu^2}{Q^2}\right)^\epsilon\\
                & \times\Biggl[ \dfrac{1}{\epsilon} \left(\dfrac{\ln(\rho)}{\bar{v}_{ij,k}} + 1 \right) \, +\dfrac{1}{\bar{v}_{ij,k}} \biggl( - 2\ln(\rho) \ln\left( 1 - (\mu_j + \mu_k)^2 \right) - \ln(\rho_j)^2 \\
                &\, - \ln( \rho_k)^2 + \dfrac{\pi^2}{3} + 4\text{Li}_2(-\rho) - 4\text{Li}_2(1-\rho) - \text{Li}_2(1 - \rho_j^2)- \text{Li}_2(1 -\rho_k^2)\biggr)\\
        &\, +\dfrac{1}{2} \ln(\mu_j^2) - 2 - 2 \ln \left(\left(1 -\mu_k\right)^2 - \mu_j^2\right) + \ln\left(1 - \mu_k\right)\\
        &\, - \dfrac{2 \mu_j^2}{1 - \mu_j^2 - \mu_k^2} \ln\left(\dfrac{\mu_j}{1-\mu_k}\right) + 5 - \dfrac{\mu_k}{1- \mu_k} - \dfrac{2 \mu_k\left(1 - 2 \mu_k\right)}{1 - \mu_j^2 - \mu_k^2} + O(\epsilon)\Bigg]\,,
        \end{split}
\end{equation}
where,
\begin{equation}
    \rho_n =
    \sqrt{\dfrac{1- \bar{v}_{ij,k} + \dfrac{2\mu_n^2}{1 - \mu_j^2 - \mu_k^2}}{1+ \bar{v}_{ij,k} + \dfrac{2\mu_n^2}{1 - \mu_j^2 - \mu_k^2}}}\,,
    \quad n=j,k\,,
    \qquad
    \rho = \sqrt{\dfrac{1- \bar{v}_{ij,k} }{1+ \bar{v}_{ij,k} }}\,.
\end{equation}
We observe that in processes with $\PW$-boson pairs the reduced-Born momenta entering the local and the integrated dipole are treated with the DPA$^{(2,2)}$ (if the kinematics is above threshold),
and therefore both rescaled squared masses can be set to
\begin{equation}
        \mu_{j}^2 = \mu_{k}^2 = \dfrac{M_{\PW}^2}{Q^2}\,.
\end{equation}
It is worth noting that in \refeq{eq:integFFMM} only a single $\epsilon$ pole is present, as the two masses lead to the absence of collinear-photon singularities.

\subsection{Final-state massless emitter and final-state massive spectator}

We now consider dipoles with a massive charged resonance as spectator (with momentum $p_k$ and mass $m_{k}$, equal to $\MW$ in our VBS process) and a massless fermion as emitter (with momentum $p_j$ and $m_{j}=0$).
The results can again be obtained as a special case from Section 5.1 of
\citere{Catani:2002hc}. Using the same notation as in the previous subsection, the mapped momenta read,
\begin{align}
\bar{p}_k ={}&\frac{Q}2\left(1+\mu_{k}^2\right) + \frac12 \frac{Q^2-p_{k}^2}{\sqrt{(Q\!\cdot\! p_k)^2-Q^2 p_{k}^2}}\left(p_k-Q\frac{Q\!\cdot\! p_k}{Q^2}\right) \,,\nnb\\
\bar{p}_j ={}& Q - \bar{p}_k\,.
\label{eq:mapping_massless_massive_final_final}
\end{align}
Using the quantities defined in
\refeqs{eq:CS_mapping_massive_massiv_final_final_invariants}--\eqref{eq:CS_mapping_massive_final_final_vijk}
and setting $\mu_j = 0$, the local and integrated dipoles are given by
\beq\label{eq:FF0M_subtr}
    {\mc D}_{ij,k}=(8\pi\alpha)\mu^{2\epsilon}\,
    \frac{\theta_{[ij]}\theta_k Q_{[ij]}Q_k}{-y\,Q^2(1-\mu_{k}^2)}\,\left[\frac{2}{1-(1-z)(1-y)}-\frac1{v_{ij,k}}
      \left(2-z\,(1-\epsilon)
      \right)\right]
\eeq
and
\begin{align}\label{eq:FF0M_integ}
\int\!\rd\Phi^{(4-2\epsilon)}_{\rm rad}{\mc D}_{ij,k}
={}&
(-\theta_{[ij]}\theta_k Q_{[ij]}Q_k)
\frac{\alpha}{2\pi}\frac{(4\pi)^\epsilon}{\Gamma(1-\epsilon)}\left(\frac{\mu^2}{Q^2}\right)^\epsilon\nnb\\
&\times
\biggl[\frac{1}{\epsilon ^2}-\frac2{\epsilon } \log \left(1-\mu_k^2\right)
-\frac{5 \pi ^2}{6}+2 \log ^2\left(1-\mu_k^2\right)+2
\text{Li}_2\left(1-\mu_k^2\right)\nnb\\
&\hspace*{0.3cm}{}+
\frac{3}{2\epsilon}+2 +\frac3{1+\mu_k}-3\log(1-\mu_k) +\mc{O}(\epsilon)
\biggr]\,.
\end{align}
Since the emitter is a final-state massless fermion, this dipole regularises soft-collinear configurations of the radiated photon, leading to a squared $\epsilon$ pole in dimensional regularisation.
Note that in $\PW$-boson processes in the PA we have $\mu^2_k = \MW^2/Q^2$.

\subsection{Final-state massive emitter and final-state massless spectator}\label{subsec:FM_Fm}

We also need to consider dipoles with one massive emitter ($m_j=\MW$,
in the $\PW$-boson case) and a massless fermion as spectator
($m_{k}=0$), both in the final state. These results follow from
those of Section 5.1 of \citere{Catani:2002hc}.
The mapped momenta are
\begin{equation}
\bar{p}_k = p_k \frac{Q^2-p_{j}^2}{2\,Q\!\cdot\! p_k},\qquad
\bar{p}_j = Q-p_k \frac{Q^2-p_{j}^2}{2\,Q\!\cdot\! p_k}
\,.
\end{equation}
Reversing the role of labels $j$ and $k$ in the previous subsection, this local dipole reads
\beq\label{eq:FFM0_subtr}
    {\mc D}_{ij,k}=(8\pi\alpha)\mu^{2\epsilon}\,
    \frac{\theta_{[ij]}\theta_k Q_{[ij]}Q_k}{-y\,Q^2(1-\mu_{j}^2)}\,\left[\frac{2}{1-(1-z)(1-y)}
      %
      -2
      +z\,(1-\epsilon)
      - \frac{2\mu_{j}^2}{y\,(1-\mu_{j}^2)}
      \right],
\eeq
while its integrated counterpart is given by
\begin{align}\label{eq:FFM0_integ}
\int\!\rd\Phi^{(4-2\epsilon)}_{\rm rad}{\mc D}_{ij,k}
={}&
(-\theta_{[ij]}\theta_k Q_{[ij]}Q_k)
\frac{\alpha}{2\pi}\frac{(4\pi)^\epsilon}{\Gamma(1-\epsilon)}\left(\frac{\mu^2}{Q^2}\right)^\epsilon\nnb\\
&\times
\biggl[
\frac{2 }{\epsilon }\log\mu_j
-2\log^2\mu_j
-4 \log\mu_j \log \left(1-\mu_j^2\right)
-4 \text{Li}_2\left(1-\mu_j^2\right)
\nnb\\
&\hspace*{0.3cm}{} +
\frac{1}{\epsilon }+3+\frac{3\mu_j^2-1 }{\mu_j^2-1}\log\mu_j-2 \log \left(1-\mu_j^2\right)+\mc{O}(\epsilon)
\biggr]\,,
\end{align}
where the absence of the squared pole is again due to the absence of configurations where the photon is collinear to the massive emitter.

\subsection{Final-state massive emitter and initial-state massless spectator}
\noindent
So far we have focused on dipoles with both the emitter and the spectator in the final state. 
In our calculation of $\PW\PW$ scattering and in general in any production process with charged resonances,
we also need to consider the case where the charged resonance (always in the final state) is the emitter (momentum $p_j$)
and an initial-state parton is the spectator (momentum $p_a$). 
These results are derived from those of Section 5.2 of \citere{Catani:2002hc}.
Using $Q=p_i+p_j-p_a = \bar{p}_j-\bar{p}_a$, the mapped momenta read
\begin{equation}
\bar{p}_j = p_i+p_j-(1-x_{ij,a})\,p_a, \qquad
\bar{p}_a = x_{ij,a} \,p_a\,.
\label{eq:CS_mapping_final_initial}
\end{equation} 
Analogously to the final--final dipoles, the CS mapping is applied to the off-shell phase-space points. 
The radiation variables $x_{ij,a},\,z_i,\,z_j$ are defined in the same way as in the massless case \cite{Catani:1996vz,Catani:2002hc},
\beq\label{eq:IFm_mapp}
       x \equiv  x_{ij,a} = \frac{p_i\!\cdot\!p_a+p_j\!\cdot\!p_a-p_i\!\cdot\!p_j}{p_i\!\cdot\!p_a+p_j\!\cdot\!p_a}\,,\quad 
       z \equiv  z_i = \frac{p_i\!\cdot\!p_a}{p_i\!\cdot\!p_a+p_j\!\cdot\!p_a}\,,\quad 
        z_j = 1-z_i\,.
\eeq
It is convenient to define the quantity
\beq
        \mu^2_{j} = \dfrac{p_{j}^2}{{2 \bar{p}_{j} \!\cdot\!p_a}}= \dfrac{p_{j}^2}{{2 \left(p_i \!\cdot\! p_a + p_j \!\cdot\! p_a \right)}}
        = {x}\dfrac{p_{j}^2}{{p_{j}^2 - Q^2}}\,.
        \label{eq:CS_mapping_invariants_final_initial}
\eeq      
To evaluate the local-counterterm kernel in the DPA, the on-shell-projected radiation variables are used when the real-kinematics phase-space point can be projected on-shell in the DPA. 
Otherwise the analogous variables are constructed with the original off-shell kinematics.
The local dipole reads
\begin{equation}\label{eq:FI0M_subtr}
    {\mc D}_{ij,a}\,=\,(8\pi\alpha)\mu^{2\epsilon}\,
    \frac{\theta_{[ij]}\theta_a Q_{[ij]}Q_a}{\left(1 - x\right)\left(Q^2 - p_{j}^2\right) }\, \left(
      \dfrac{2}{1 - x + z} + z \left(1-\epsilon\right) - 2 -\dfrac{2 p_{j}^2}{p_{j}^2-Q^2}\dfrac{x}{1-x}
      \right)
      \,,
\end{equation}
and its integration leads to
\begin{align}
\label{eq:FI0M_integ}
  \int\!\rd\Phi^{(4-2\epsilon)}_{\rm rad}{\mc D}_{ia,j}
         ={}& \theta_{[ij]}\theta_a Q_{[ij]}Q_a \dfrac{\alpha}{2 \pi} \dfrac{(4\pi)^2}{\Gamma \left( 1 - \epsilon\right)} \left(\dfrac{\mu^2}{2 \bar{p}_{j} \!\cdot\! p_a}\right)^\epsilon \notag\\
  &  \times\Biggl\{ \left[ \dfrac{1-x}{2\left(1-x + \mu_j^2\right)^2} - \dfrac{2}{1-x} \Bigl( 1 + \log \left(1-x +  \mu_j^2\right)\Bigr)\right]_+\notag\\
  &\, + \left(\dfrac{2}{1-x}\right)_+ \log(2+  \mu_j^2 - x )\notag\\
         &  {}+ \delta(1-x) \Biggl[ \dfrac{1}{\epsilon} \left( \log \left( \dfrac{\mu_j^2}{1 + \mu_j^2} \right) + 1 \right) - \dfrac{1}{2}\log^2 \left( \mu_j^2 \right) + \dfrac{1}{2} \log \left( \mu_j^2 \right)\notag\\
         &{} + \dfrac{3}{2} - \dfrac{2 \pi^2}{3}  + \dfrac{1}{2}
         \log^2(1+ \mu_j^2) - 2 \log( \mu_j^2) \log(1+ \mu_j^2) \notag\\
         &\, - 4 \text{Li}_2(- \mu_j^2) + \dfrac{1}{2} \log(1+ \mu_j^2) + \dfrac{ \mu_j^2}{2 \left( 1+ \mu_j^2\right)}\Biggr]+ \mc{O}(\epsilon) \Biggr\}\,,
\end{align}
where the plus distribution is defined as
\begin{equation}
        \int_{-1}^1 \!a(x)_+ \,b(x)\, \rd x = \int_{-1}^1\! a(x) \bigl[ b(x) - b (1) \bigr] \rd x\,.
\end{equation}
In the DPA calculation of $\PW^+\PW^+$ scattering, the
integrated-dipole evaluation is carried out with on-shell-projected
Born momenta, which implies
\begin{equation}\label{eq:mujfsmeis0s}
          \mu^2_j \,=\, \mu_{\PW}^2 =
          x\dfrac{\MW^2}{\MW^2-Q^2}
\end{equation}
in \refeq{eq:FI0M_integ}.

\subsection{Initial-state massless emitter and final-state massive spectator}

The last case that we treat is the one where the charged massive resonance is the spectator (momentum $p_j$) and an initial-state parton is the emitter (momentum $p_a$),
which follows from the results of Section 5.3 of  \citere{Catani:2002hc}.
Upon defining $Q=p_i+p_j-p_a = \bar{p}_j-\bar{p}_a$, the off-shell mapped momenta read
\begin{equation}
        \bar{p}_j = p_i + p_j - \left( 1 - x_{ij,a} \right) p_a \,, \qquad
        \bar{p}_{a} = x_{ij,a} p_a\,, \\
\end{equation}
where the radiation variables $x_{ij,a},\,z_i,\,z_j$  are defined as in \refeq{eq:IFm_mapp},
and the rescaled squared momentum of the spectating resonance $j$ is defined as in \refeq{eq:CS_mapping_invariants_final_initial}. 
The local dipole takes the form
\begin{equation}
        {\mc D}_{ia,j}=(8\pi\alpha)\mu^{2\epsilon} \, \dfrac{\theta_{[ia]}\theta_j Q_{[ia]}Q_j}{z \left(Q^2-p_{j}^2 \right) } 
        \left( \dfrac{2}{1 - x + z} - (1 + x) - \epsilon \left(1- x \right) \right)\,,
\end{equation}
while the integrated counterpart can be written as
\begin{align}
\label{eq:FI0M_integ_3}
  \int\!\rd\Phi^{(4-2\epsilon)}_{\rm rad}&{\mc D}_{ia,j}
        = \theta_{[ia]}\theta_j Q_{[ia]}Q_j \dfrac{\alpha}{2 \pi} \dfrac{(4 \pi)^\epsilon}{\Gamma \left( 1 - \epsilon\right)} \left(\dfrac{ \mu^2}{2 \bar{p}_j \!\cdot\!p_a}\right)^\epsilon \notag\\
        & \times\biggl\{ -\dfrac{2}{\epsilon} \left(\dfrac{1}{1-x}\right)_+ + 4\left(\dfrac{\log(1-x)}{1-x}\right)_+ 
         + 2 \left(\dfrac{1}{1-x}\right)_+ \log\left(\dfrac{2-x}{2-x+\mu_{j}^2}\right)\notag\\
  &\quad +\dfrac{1}{\epsilon} \left(1+x\right) -\left(1 + x\right) \log(1-x) + \left(1- x\right)
   -\left(1 + x\right) \log\left(\dfrac{1-x}{1-x+ \mu_{j}^2} \right)\notag\\
  &\, - \dfrac2{1-x}\log(2-x)
  \, + \delta(1-x)
  \biggl[ \dfrac{1}{\epsilon^2} + \dfrac{1}{\epsilon} \log(1+ \mu_{j}^2) + \dfrac{1}{2} \log^2 (1+ \mu_{j}^2)\notag\\
    &\, + 2 \mathrm{Li}_2 \left( \dfrac{1}{1+\mu_{j}^2} \right) - \dfrac{\pi^2}{6}
    \biggr]\, + \mc{O}(\epsilon)\biggr\}\,.
\end{align}
The application of this dipole class to the DPA of  $\PW^+\PW^+$
scattering works in the same manner as described at the end of the
previous subsection, \ie using \refeq{eq:mujfsmeis0s}.

\subsection{Additional dipoles with a massive resonant spectator}
In the case of an initial-state photon splitting into a fermion--anti-fermion pair, a massive charged resonance in the final state
could in principle be used as the spectator absorbing the recoil of
the mapping. It is always possible to avoid such a situation,
by choosing the other initial-state parton or a final-state massless particle as spectator. In the considered VBS process, we simply
pick as a spectator one of the external quarks.
It is worth mentioning that this argument does not apply to QCD. Owing to colour correlation,
an initial-state splitting of a gluon into a massless-quark pair requires one dipole for each other external coloured parton
(acting as spectator), including possible top quarks in the final
state. In this case, which is known in the literature \cite{Catani:2002hc}, it is
unavoidable to pick the massive coloured resonance in the final state as a spectator.

\bibliographystyle{JHEPmod} 
\bibliography{vbs_w+w+}

\end{document}

%% file: macros.tex
\def\refeq#1{\mbox{Eq.~\eqref{#1}}}
\def\refeqs#1{\mbox{Eqs.~\eqref{#1}}}
\def\reffi#1{\mbox{Figure~\ref{#1}}}
\def\reffis#1{\mbox{Figures~\ref{#1}}}
\def\refta#1{\mbox{Table~\ref{#1}}}

\def\refse#1{\mbox{Section~\ref{#1}}}
\def\refses#1{\mbox{Sections~\ref{#1}}}

\def\citere#1{\mbox{Ref.~\cite{#1}}}
\def\citeres#1{\mbox{Refs.~\cite{#1}}}

\newcommand{\newc}{\newcommand}
\newc{\beq}{\begin{equation}}
\newc{\eeq}{\end{equation}}
\newc{\beqn}{\begin{eqnarray}}
\newc{\eeqn}{\end{eqnarray}}
\newc{\bit}{\begin{itemize}}
\newc{\eit}{\end{itemize}}
\newc{\ben}{\begin{enumerate}}
\newc{\een}{\end{enumerate}}
\newc{\bce}{\begin{center}}
\newc{\ece}{\end{center}}
\newc{\bfi}{\begin{figure}}
\newc{\efi}{\end{figure}}

\newcommand{\cM}{\ensuremath{\mathcal{M}}\xspace}

\newcommand{\ri}{\mathrm i}

\newcommand{\rd}{\mathrm d}

\newcommand{\rD}{{\mathrm{D}}}
\newcommand{\rP}{{\mathrm{P}}}
\newcommand{\rT}{{\mathrm{T}}}
\newcommand{\rR}{{\mathrm{R}}}
\newcommand{\rL}{{\mathrm{L}}}
\newcommand{\rF}{{\mathrm{F}}}

\newcommand{\ie}{{i.e.}\ }

\newcommand{\GeV}{\ensuremath{\,\text{GeV}}\xspace}
\newcommand{\TeV}{\ensuremath{\,\text{TeV}}\xspace}
\newcommand{\fb}{{\ensuremath\unskip\,\text{fb}}\xspace}

\newcommand{\PH}{\ensuremath{\text{H}}\xspace}
\newcommand{\Pj}{\ensuremath{\text{j}}\xspace}
\newcommand{\Pp}{\ensuremath{\text{p}}}
\newcommand{\Pe}{\ensuremath{\text{e}}\xspace}

\newcommand{\Pt}{\ensuremath{\text{t}}\xspace}

\newcommand{\PW}{\ensuremath{\text{W}}\xspace}
\newcommand{\PZ}{\ensuremath{\text{Z}}\xspace}

\newcommand{\pt}[1]{p_{\rT,{#1}}}

\newcommand{\Mt}{\ensuremath{m_\Pt}\xspace}
\newcommand{\MH}{\ensuremath{M_\PH}\xspace}
\newcommand{\MWOS}{\ensuremath{M_\PW^\text{OS}}\xspace}
\newcommand{\MW}{\ensuremath{M_\PW}\xspace}
\newcommand{\MZOS}{\ensuremath{M_\PZ^\text{OS}}\xspace}
\newcommand{\MZ}{\ensuremath{M_\PZ}\xspace}

\newcommand{\Gt}{\ensuremath{\Gamma_\Pt}\xspace}
\newcommand{\GH}{\ensuremath{\Gamma_\PH}\xspace}
\newcommand{\GZ}{\ensuremath{\Gamma_\PZ}\xspace}
\newcommand{\GZOS}{\ensuremath{\Gamma_\PZ^\text{OS}}\xspace}
\newcommand{\GW}{\ensuremath{\Gamma_\PW}\xspace}
\newcommand{\GWOS}{\ensuremath{\Gamma_\PW^\text{OS}}\xspace}

\newcommand{\GF}{\ensuremath{G_\mu}}
\newcommand{\alphas}{\ensuremath{\alpha_\text{s}}\xspace}

\newcommand{\MVOS}{\ensuremath{M_{V}^\text{OS}}\xspace}%
\newcommand{\GVOS}{\ensuremath{\Gamma_{V}^\text{OS}}\xspace}%

\newcommand{\order}[1]{\ensuremath{\mathcal{O}{\left(#1\right)}}\xspace}

\newcommand{\recola}{{\sc Recola}\xspace}

\newcommand{\Sherpa}{{\sc Sherpa}\xspace}

\newcommand{\mocanlo}{{\sc MoCaNLO}\xspace}
\newcommand{\bbmc}{{\sc BBMC}\xspace}
\newcommand{\collier}{{\sc Collier}\xspace}

\newcommand{\madgraphnlos}{{\sc\small MG5\_aMC@NLO}\xspace}

\newcommand{\phantommc}{{\sc\small PHANTOM}\xspace}

\newcolumntype{.}{D{.}{.}{-1}}
\newcolumntype{d}[1]{D{.}{.}{#1}}
\colorlet{tableoverheadcolor}{gray!37.5}
\colorlet{tableheadcolor}{gray!25}
\colorlet{tablerowcolor}{gray!12.5}

\def\draftdate{\relax}
\def\mda{\relax}
\def\mua{\relax}
\def\mla{\relax}
\def\draft{
\def\thtystars{******************************}
\def\sixtystars{\thtystars\thtystars}
\typeout{}
\typeout{\sixtystars**}
\typeout{* Draft mode!
         For final version remove \protect\draft\space in source file *}
\typeout{\sixtystars**}
\typeout{}
\def\draftdate{\today}
\def\mua{\marginpar[\boldmath\hfil$\uparrow$]%
                   {\boldmath$\uparrow$\hfil}\color{black}%
                    \typeout{marginpar: $\uparrow$}\ignorespaces}
\def\mda{\color{red}\marginpar[\boldmath\hfil$\downarrow$]%
                   {\boldmath$\downarrow$\hfil}%
                    \typeout{marginpar: $\downarrow$}\ignorespaces}
\def\mla{\marginpar[\boldmath\hfil$\rightarrow$]%
                   {\boldmath$\leftarrow $\hfil}%
                    \typeout{marginpar: $\leftrightarrow$}\ignorespaces}
\def\Mua{\marginpar[\boldmath\hfil$\Uparrow$]%
                   {\boldmath$\Uparrow$\hfil}\color{black}%
                    \typeout{marginpar: $\uparrow$}\ignorespaces}
\def\Mda{\color{red}\marginpar[\boldmath\hfil$\Downarrow$]%
                   {\boldmath$\Downarrow$\hfil}%
                    \typeout{marginpar: $\downarrow$}\ignorespaces}
\def\Mla{\marginpar[\boldmath\hfil\textcolor{red}{$\Rightarrow$}]%
                   {\boldmath\textcolor{red}{$\Leftarrow $}\hfil}%
                    \typeout{marginpar: $\leftrightarrow$}\ignorespaces}
\overfullrule 5pt
\oddsidemargin 15mm
\marginparwidth 29mm
}


%% file: vbs_ww.bbl
\providecommand{\href}[2]{#2}\begingroup\raggedright\begin{thebibliography}{100}

\bibitem{Anders:2018oin}
C.~F. Anders et~al., {\it {Vector boson scattering: Recent experimental and
  theory developments}},  {\em Rev. Phys.} {\bf 3} (2018) 44--63,
  [\href{http://arxiv.org/abs/1801.04203}{{\tt arXiv:1801.04203}}].

\bibitem{Covarelli:2021gyz}
R.~Covarelli, M.~Pellen, and M.~Zaro, {\it {Vector-Boson scattering at the LHC:
  Unraveling the electroweak sector}},  {\em Int. J. Mod. Phys. A} {\bf 36}
  (2021) 2130009, [\href{http://arxiv.org/abs/2102.10991}{{\tt
  arXiv:2102.10991}}].

\bibitem{BuarqueFranzosi:2021wrv}
D.~Buarque~Franzosi et~al., {\it {Vector boson scattering processes: Status and
  prospects}},  {\em Rev. Phys.} {\bf 8} (2022) 100071,
  [\href{http://arxiv.org/abs/2106.01393}{{\tt arXiv:2106.01393}}].

\bibitem{Cornwall:1974km}
J.~M. Cornwall, D.~N. Levin, and G.~Tiktopoulos, {\it {Derivation of Gauge
  Invariance from High-Energy Unitarity Bounds on the S Matrix}},  {\em Phys.
  Rev. D} {\bf 10} (1974) 1145. [Erratum: Phys.Rev.D 11, 972 (1975)].

\bibitem{Vayonakis:1976vz}
C.~E. Vayonakis, {\it {Born Helicity Amplitudes and Cross-Sections in
  Nonabelian Gauge Theories}},  {\em Lett. Nuovo Cim.} {\bf 17} (1976) 383.

\bibitem{Lee:1977yc}
B.~W. Lee, C.~Quigg, and H.~B. Thacker, {\it {The Strength of Weak Interactions
  at Very High-Energies and the Higgs Boson Mass}},  {\em Phys. Rev. Lett.}
  {\bf 38} (1977) 883--885.

\bibitem{Chanowitz:1985hj}
M.~S. Chanowitz and M.~K. Gaillard, {\it {The TeV Physics of Strongly
  Interacting W's and Z's}},  {\em Nucl. Phys. B} {\bf 261} (1985) 379--431.

\bibitem{ATLAS:2016snd}
{\bf ATLAS} Collaboration, M.~Aaboud et~al., {\it {Measurement of
  $W^{\pm}W^{\pm}$ vector-boson scattering and limits on anomalous quartic
  gauge couplings with the ATLAS detector}},  {\em Phys. Rev. D} {\bf 96}
  (2017) 012007, [\href{http://arxiv.org/abs/1611.02428}{{\tt
  arXiv:1611.02428}}].

\bibitem{CMS:2017fhs}
{\bf CMS} Collaboration, A.~M. Sirunyan et~al., {\it {Observation of
  electroweak production of same-sign W boson pairs in the two jet and two
  same-sign lepton final state in proton-proton collisions at $\sqrt{s}
  ={}$13\,TeV}},  {\em Phys. Rev. Lett.} {\bf 120} (2018) 081801,
  [\href{http://arxiv.org/abs/1709.05822}{{\tt arXiv:1709.05822}}].

\bibitem{ATLAS:2019cbr}
{\bf ATLAS} Collaboration, M.~Aaboud et~al., {\it {Observation of electroweak
  production of a same-sign $W$ boson pair in association with two jets in $pp$
  collisions at $\sqrt{s}={}$13\,TeV with the ATLAS detector}},  {\em Phys.
  Rev. Lett.} {\bf 123} (2019) 161801,
  [\href{http://arxiv.org/abs/1906.03203}{{\tt arXiv:1906.03203}}].

\bibitem{CMS:2020gfh}
{\bf CMS} Collaboration, A.~M. Sirunyan et~al., {\it {Measurements of
  production cross sections of WZ and same-sign WW boson pairs in association
  with two jets in proton-proton collisions at $\sqrt{s}={}$13\,TeV}},  {\em
  Phys. Lett. B} {\bf 809} (2020) 135710,
  [\href{http://arxiv.org/abs/2005.01173}{{\tt arXiv:2005.01173}}].

\bibitem{ATLAS:2023sua}
{\bf ATLAS} Collaboration, G.~Aad et~al., {\it {Measurement and interpretation
  of same-sign W boson pair production in association with two jets in pp
  collisions at $ \sqrt{s}={}$13\,TeV with the ATLAS detector}},  {\em JHEP}
  {\bf 04} (2024) 026, [\href{http://arxiv.org/abs/2312.00420}{{\tt
  arXiv:2312.00420}}].

\bibitem{CMS:2017zmo}
{\bf CMS} Collaboration, A.~M. Sirunyan et~al., {\it {Measurement of vector
  boson scattering and constraints on anomalous quartic couplings from events
  with four leptons and two jets in proton\textendash{}proton collisions at
  $\sqrt{s}={}$13\,TeV}},  {\em Phys. Lett. B} {\bf 774} (2017) 682--705,
  [\href{http://arxiv.org/abs/1708.02812}{{\tt arXiv:1708.02812}}].

\bibitem{ATLAS:2020nlt}
{\bf ATLAS} Collaboration, G.~Aad et~al., {\it {Observation of electroweak
  production of two jets and a Z-boson pair}},  {\em Nature Phys.} {\bf 19}
  (2023) 237--253, [\href{http://arxiv.org/abs/2004.10612}{{\tt
  arXiv:2004.10612}}].

\bibitem{ATLAS:2023dkz}
{\bf ATLAS} Collaboration, G.~Aad et~al., {\it {Differential cross-section
  measurements of the production of four charged leptons in association with
  two jets using the ATLAS detector}},  {\em JHEP} {\bf 01} (2024) 004,
  [\href{http://arxiv.org/abs/2308.12324}{{\tt arXiv:2308.12324}}].

\bibitem{ATLAS:2018mxa}
{\bf ATLAS} Collaboration, M.~Aaboud et~al., {\it {Observation of electroweak
  $W^{\pm}Z$ boson pair production in association with two jets in $pp$
  collisions at $\sqrt{s} ={}$13\,TeV with the ATLAS detector}},  {\em Phys.
  Lett. B} {\bf 793} (2019) 469--492,
  [\href{http://arxiv.org/abs/1812.09740}{{\tt arXiv:1812.09740}}].

\bibitem{CMS:2019uys}
{\bf CMS} Collaboration, A.~M. Sirunyan et~al., {\it {Measurement of
  electroweak WZ boson production and search for new physics in WZ + two jets
  events in pp collisions at $\sqrt{s}={}$13\,TeV}},  {\em Phys. Lett. B} {\bf
  795} (2019) 281--307, [\href{http://arxiv.org/abs/1901.04060}{{\tt
  arXiv:1901.04060}}].

\bibitem{ATLAS:2024ini}
{\bf ATLAS} Collaboration, G.~Aad et~al., {\it {Measurements of electroweak
  $W^{\pm}Z$ boson pair production in association with two jets in $pp$
  collisions at $\sqrt{s} ={}$ 13\,TeV with the ATLAS detector}},  {\em JHEP}
  {\bf 06} (2024) 192, [\href{http://arxiv.org/abs/2403.15296}{{\tt
  arXiv:2403.15296}}].

\bibitem{CMS:2022woe}
{\bf CMS} Collaboration, A.~Tumasyan et~al., {\it {Observation of electroweak
  $W^+W^-$ pair production in association with two jets in proton-proton
  collisions at $\sqrt{s}={}$13\,TeV}},  {\em Phys. Lett. B} {\bf 841} (2023)
  137495, [\href{http://arxiv.org/abs/2205.05711}{{\tt arXiv:2205.05711}}].

\bibitem{ATLAS:2024ett}
{\bf ATLAS} Collaboration, G.~Aad et~al., {\it {Observation of electroweak
  production of $W^+W^-$ in association with jets in proton-proton collisions
  at $\sqrt{s}={}$13\,TeV with the ATLAS Detector}},  {\em JHEP} {\bf 07}
  (2024) 254, [\href{http://arxiv.org/abs/2403.04869}{{\tt arXiv:2403.04869}}].

\bibitem{CMS:2014mra}
{\bf CMS} Collaboration, V.~Khachatryan et~al., {\it {Study of vector boson
  scattering and search for new physics in events with two same-sign leptons
  and two jets}},  {\em Phys. Rev. Lett.} {\bf 114} (2015) 051801,
  [\href{http://arxiv.org/abs/1410.6315}{{\tt arXiv:1410.6315}}].

\bibitem{ATLAS:2016nmw}
{\bf ATLAS} Collaboration, M.~Aaboud et~al., {\it {Search for anomalous
  electroweak production of $WW/WZ$ in association with a high-mass dijet
  system in $pp$ collisions at $\sqrt{s}=8$ TeV with the ATLAS detector}},
  {\em Phys. Rev. D} {\bf 95} (2017) 032001,
  [\href{http://arxiv.org/abs/1609.05122}{{\tt arXiv:1609.05122}}].

\bibitem{CMS:2019qfk}
{\bf CMS} Collaboration, A.~M. Sirunyan et~al., {\it {Search for anomalous
  electroweak production of vector boson pairs in association with two jets in
  proton-proton collisions at 13\,TeV}},  {\em Phys. Lett. B} {\bf 798} (2019)
  134985, [\href{http://arxiv.org/abs/1905.07445}{{\tt arXiv:1905.07445}}].

\bibitem{ATLAS:2019thr}
{\bf ATLAS} Collaboration, G.~Aad et~al., {\it {Search for the electroweak
  diboson production in association with a high-mass dijet system in
  semileptonic final states in $pp$ collisions at $\sqrt{s}={}$13\,TeV with the
  ATLAS detector}},  {\em Phys. Rev. D} {\bf 100} (2019) 032007,
  [\href{http://arxiv.org/abs/1905.07714}{{\tt arXiv:1905.07714}}].

\bibitem{CMS:2021qzz}
{\bf CMS} Collaboration, A.~Tumasyan et~al., {\it {Evidence for WW/WZ vector
  boson scattering in the decay channel \ensuremath{\ell}\ensuremath{\nu}qq
  produced in association with two jets in proton-proton collisions at
  $\sqrt{s}={}$13\,TeV}},  {\em Phys. Lett. B} {\bf 834} (2022) 137438,
  [\href{http://arxiv.org/abs/2112.05259}{{\tt arXiv:2112.05259}}].

\bibitem{CMS:2020etf}
{\bf CMS} Collaboration, A.~M. Sirunyan et~al., {\it {Measurements of
  production cross sections of polarized same-sign W boson pairs in association
  with two jets in proton-proton collisions at $\sqrt{s} =$ 13 TeV}},  {\em
  Phys. Lett. B} {\bf 812} (2021) 136018,
  [\href{http://arxiv.org/abs/2009.09429}{{\tt arXiv:2009.09429}}].

\bibitem{Jager:2006zc}
B.~J{\"a}ger, C.~Oleari, and D.~Zeppenfeld, {\it {Next-to-leading order QCD
  corrections to W$^+$W$^-$ production via vector-boson fusion}},  {\em JHEP}
  {\bf 07} (2006) 015, [\href{http://arxiv.org/abs/hep-ph/0603177}{{\tt
  hep-ph/0603177}}].

\bibitem{Jager:2006cp}
B.~J{\"a}ger, C.~Oleari, and D.~Zeppenfeld, {\it {Next-to-leading order QCD
  corrections to Z boson pair production via vector-boson fusion}},  {\em Phys.
  Rev. D} {\bf 73} (2006) 113006,
  [\href{http://arxiv.org/abs/hep-ph/0604200}{{\tt hep-ph/0604200}}].

\bibitem{Bozzi:2007ur}
G.~Bozzi, B.~J{\"a}ger, C.~Oleari, and D.~Zeppenfeld, {\it {Next-to-leading
  order QCD corrections to $W^+ Z$ and $W^- Z$ production via vector-boson
  fusion}},  {\em Phys. Rev. D} {\bf 75} (2007) 073004,
  [\href{http://arxiv.org/abs/hep-ph/0701105}{{\tt hep-ph/0701105}}].

\bibitem{Jager:2009xx}
B.~J{\"a}ger, C.~Oleari, and D.~Zeppenfeld, {\it {Next-to-leading order QCD
  corrections to $W^+ W^+ jj$ and $W^- W^- jj$ production via weak-boson
  fusion}},  {\em Phys. Rev. D} {\bf 80} (2009) 034022,
  [\href{http://arxiv.org/abs/0907.0580}{{\tt arXiv:0907.0580}}].

\bibitem{Denner:2012dz}
A.~Denner, L.~Ho\v{s}ekov\'a, and S.~Kallweit, {\it {NLO QCD corrections to
  $W^+ W^+ jj$ production in vector-boson fusion at the LHC}},  {\em Phys. Rev.
  D} {\bf 86} (2012) 114014, [\href{http://arxiv.org/abs/1209.2389}{{\tt
  arXiv:1209.2389}}].

\bibitem{Biedermann:2017bss}
B.~Biedermann, A.~Denner, and M.~Pellen, {\it {Complete NLO corrections to
  $W^{+}$$W^{+}$ scattering and its irreducible background at the LHC}},  {\em
  JHEP} {\bf 10} (2017) 124, [\href{http://arxiv.org/abs/1708.00268}{{\tt
  arXiv:1708.00268}}].

\bibitem{Ballestrero:2018anz}
A.~Ballestrero et~al., {\it {Precise predictions for same-sign W-boson
  scattering at the LHC}},  {\em Eur. Phys. J. C} {\bf 78} (2018) 671,
  [\href{http://arxiv.org/abs/1803.07943}{{\tt arXiv:1803.07943}}].

\bibitem{Denner:2019tmn}
A.~Denner, S.~Dittmaier, P.~Maierh{\"o}fer, M.~Pellen, and C.~Schwan, {\it {QCD
  and electroweak corrections to WZ scattering at the LHC}},  {\em JHEP} {\bf
  06} (2019) 067, [\href{http://arxiv.org/abs/1904.00882}{{\tt
  arXiv:1904.00882}}].

\bibitem{Denner:2020zit}
A.~Denner, R.~Franken, M.~Pellen, and T.~Schmidt, {\it {NLO QCD and EW
  corrections to vector-boson scattering into ZZ at the LHC}},  {\em JHEP} {\bf
  11} (2020) 110, [\href{http://arxiv.org/abs/2009.00411}{{\tt
  arXiv:2009.00411}}].

\bibitem{Denner:2021hsa}
A.~Denner, R.~Franken, M.~Pellen, and T.~Schmidt, {\it {Full NLO predictions
  for vector-boson scattering into Z bosons and its irreducible background at
  the LHC}},  {\em JHEP} {\bf 10} (2021) 228,
  [\href{http://arxiv.org/abs/2107.10688}{{\tt arXiv:2107.10688}}].

\bibitem{Denner:2022pwc}
A.~Denner, R.~Franken, T.~Schmidt, and C.~Schwan, {\it {NLO QCD and EW
  corrections to vector-boson scattering into $W^{+}W^{-}$ at the LHC}},  {\em
  JHEP} {\bf 06} (2022) 098, [\href{http://arxiv.org/abs/2202.10844}{{\tt
  arXiv:2202.10844}}].

\bibitem{Dittmaier:2023nac}
S.~Dittmaier, P.~Maierh\"ofer, C.~Schwan, and R.~Winterhalder, {\it {Like-sign
  W-boson scattering at the LHC \textemdash{} approximations and full
  next-to-leading-order predictions}},  {\em JHEP} {\bf 11} (2023) 022,
  [\href{http://arxiv.org/abs/2308.16716}{{\tt arXiv:2308.16716}}].

\bibitem{Biedermann:2016yds}
B.~Biedermann, A.~Denner, and M.~Pellen, {\it {Large electroweak corrections to
  vector-boson scattering at the Large Hadron Collider}},  {\em Phys. Rev.
  Lett.} {\bf 118} (2017) 261801, [\href{http://arxiv.org/abs/1611.02951}{{\tt
  arXiv:1611.02951}}].

\bibitem{Jager:2011ms}
B.~J{\"a}ger and G.~Zanderighi, {\it {NLO corrections to electroweak and QCD
  production of $W^+W^+$ plus two jets in the POWHEGBOX}},  {\em JHEP} {\bf 11}
  (2011) 055, [\href{http://arxiv.org/abs/1108.0864}{{\tt arXiv:1108.0864}}].

\bibitem{Jager:2013mu}
B.~J{\"a}ger and G.~Zanderighi, {\it {Electroweak $W^+W^-jj$ prodution at NLO
  in QCD matched with parton shower in the POWHEG-BOX}},  {\em JHEP} {\bf 04}
  (2013) 024, [\href{http://arxiv.org/abs/1301.1695}{{\tt arXiv:1301.1695}}].

\bibitem{Jager:2013iza}
B.~J\"ager, A.~Karlberg, and G.~Zanderighi, {\it {Electroweak $ZZjj$ production
  in the Standard Model and beyond in the POWHEG-BOX V2}},  {\em JHEP} {\bf 03}
  (2014) 141, [\href{http://arxiv.org/abs/1312.3252}{{\tt arXiv:1312.3252}}].

\bibitem{Rauch:2016upa}
M.~Rauch and S.~Pl\"atzer, {\it {Parton Shower Matching Systematics in
  Vector-Boson-Fusion WW Production}},  {\em Eur. Phys. J. C} {\bf 77} (2017)
  293, [\href{http://arxiv.org/abs/1605.07851}{{\tt arXiv:1605.07851}}].

\bibitem{Jager:2018cyo}
B.~J{\"a}ger, A.~Karlberg, and J.~Scheller, {\it {Parton-shower effects in
  electroweak $WZjj$ production at the next-to-leading order of QCD}},  {\em
  Eur. Phys. J. C} {\bf 79} (2019) 226,
  [\href{http://arxiv.org/abs/1812.05118}{{\tt arXiv:1812.05118}}].

\bibitem{Jager:2024sij}
B.~J\"ager, A.~Karlberg, and S.~Reinhardt, {\it {QCD effects in electroweak
  $WZjj$ production at current and future hadron colliders}},  {\em Eur. Phys.
  J. C} {\bf 84} (2024) 587, [\href{http://arxiv.org/abs/2403.12192}{{\tt
  arXiv:2403.12192}}].

\bibitem{Baglio:2024gyp}
J.~Baglio et~al., {\it {Release Note -- VBFNLO 3.0}},
  \href{http://arxiv.org/abs/2405.06990}{{\tt arXiv:2405.06990}}.

\bibitem{Jager:2024eet}
B.~J\"ager and S.~L.~P. Chavez, {\it {Electroweak $W^+ W^+$ production in
  association with three jets at NLO QCD matched with parton shower}},
  \href{http://arxiv.org/abs/2408.12314}{{\tt arXiv:2408.12314}}.

\bibitem{Chiesa:2019ulk}
M.~Chiesa, A.~Denner, J.-N. Lang, and M.~Pellen, {\it {An event generator for
  same-sign W-boson scattering at the LHC including electroweak corrections}},
  {\em Eur. Phys. J.} {\bf C79} (2019) 788,
  [\href{http://arxiv.org/abs/1906.01863}{{\tt arXiv:1906.01863}}].

\bibitem{Brass:2018hfw}
S.~Brass, C.~Fleper, W.~Kilian, J.~Reuter, and M.~Sekulla, {\it {Transversal
  Modes and Higgs Bosons in Electroweak Vector-Boson Scattering at the LHC}},
  {\em Eur. Phys. J. C} {\bf 78} (2018) 931,
  [\href{http://arxiv.org/abs/1807.02512}{{\tt arXiv:1807.02512}}].

\bibitem{Zhang:2018shp}
C.~Zhang and S.-Y. Zhou, {\it {Positivity bounds on vector boson scattering at
  the LHC}},  {\em Phys. Rev. D} {\bf 100} (2019) 095003,
  [\href{http://arxiv.org/abs/1808.00010}{{\tt arXiv:1808.00010}}].

\bibitem{Gomez-Ambrosio:2018pnl}
R.~Gomez-Ambrosio, {\it {Studies of Dimension-Six EFT effects in Vector Boson
  Scattering}},  {\em Eur. Phys. J. C} {\bf 79} (2019) 389,
  [\href{http://arxiv.org/abs/1809.04189}{{\tt arXiv:1809.04189}}].

\bibitem{Chaudhary:2019aim}
G.~Chaudhary, et~al., {\it {EFT triangles in the same-sign $WW$ scattering
  process at the HL-LHC and HE-LHC}},  {\em Eur. Phys. J. C} {\bf 80} (2020)
  181, [\href{http://arxiv.org/abs/1906.10769}{{\tt arXiv:1906.10769}}].

\bibitem{Dedes:2020xmo}
A.~Dedes, P.~Koz\'ow, and M.~Szleper, {\it {Standard model EFT effects in
  vector-boson scattering at the LHC}},  {\em Phys. Rev. D} {\bf 104} (2021)
  013003, [\href{http://arxiv.org/abs/2011.07367}{{\tt arXiv:2011.07367}}].

\bibitem{Ethier:2021ydt}
J.~J. Ethier, R.~Gomez-Ambrosio, G.~Magni, and J.~Rojo, {\it {SMEFT analysis of
  vector boson scattering and diboson data from the LHC Run II}},  {\em Eur.
  Phys. J. C} {\bf 81} (2021) 560, [\href{http://arxiv.org/abs/2101.03180}{{\tt
  arXiv:2101.03180}}].

\bibitem{Bellan:2021dcy}
R.~Bellan et~al., {\it {A sensitivity study of VBS and diboson WW to
  dimension-6 EFT operators at the LHC}},  {\em JHEP} {\bf 05} (2022) 039,
  [\href{http://arxiv.org/abs/2108.03199}{{\tt arXiv:2108.03199}}].

\bibitem{Cappati:2022skp}
A.~Cappati, R.~Covarelli, P.~Torrielli, and M.~Zaro, {\it {Sensitivity to new
  physics in final states with multiple gauge and Higgs bosons}},  {\em JHEP}
  {\bf 09} (2022) 038, [\href{http://arxiv.org/abs/2205.15959}{{\tt
  arXiv:2205.15959}}].

\bibitem{Delgado:2013hxa}
R.~L. Delgado, A.~Dobado, and F.~J. Llanes-Estrada, {\it {One-loop $W_LW_L$ and
  $Z_LZ_L$ scattering from the electroweak Chiral Lagrangian with a light
  Higgs-like scalar}},  {\em JHEP} {\bf 02} (2014) 121,
  [\href{http://arxiv.org/abs/1311.5993}{{\tt arXiv:1311.5993}}].

\bibitem{Delgado:2015kxa}
R.~L. Delgado, A.~Dobado, and F.~J. Llanes-Estrada, {\it {Unitarity,
  analyticity, dispersion relations, and resonances in strongly interacting
  $W_LW_L$, $Z_LZ_L$, and hh scattering}},  {\em Phys. Rev. D} {\bf 91} (2015)
  075017, [\href{http://arxiv.org/abs/1502.04841}{{\tt arXiv:1502.04841}}].

\bibitem{Fabbrichesi:2015hsa}
M.~Fabbrichesi, M.~Pinamonti, A.~Tonero, and A.~Urbano, {\it {Vector boson
  scattering at the LHC: A study of the WW $\to$ WW channels with the Warsaw
  cut}},  {\em Phys. Rev. D} {\bf 93} (2016) 015004,
  [\href{http://arxiv.org/abs/1509.06378}{{\tt arXiv:1509.06378}}].

\bibitem{Delgado:2017cls}
R.~L. Delgado, et~al., {\it {Production of vector resonances at the LHC via
  WZ-scattering: a unitarized EChL analysis}},  {\em JHEP} {\bf 11} (2017) 098,
  [\href{http://arxiv.org/abs/1707.04580}{{\tt arXiv:1707.04580}}].

\bibitem{Kozow:2019txg}
P.~Koz\'ow, L.~Merlo, S.~Pokorski, and M.~Szleper, {\it {Same-sign WW
  Scattering in the HEFT: Discoverability vs. EFT Validity}},  {\em JHEP} {\bf
  07} (2019) 021, [\href{http://arxiv.org/abs/1905.03354}{{\tt
  arXiv:1905.03354}}].

\bibitem{Delgado:2019ucx}
R.~L. Delgado, C.~Garcia-Garcia, and M.~J. Herrero, {\it {Dynamical vector
  resonances from the EChL in VBS at the LHC: the WW case}},  {\em JHEP} {\bf
  11} (2019) 065, [\href{http://arxiv.org/abs/1907.11957}{{\tt
  arXiv:1907.11957}}].

\bibitem{Quezada-Calonge:2022lop}
C.~Quezada-Calonge, A.~Dobado, and J.~J. Sanz-Cillero, {\it {Relevance of
  fermion loops for W+W- scattering at the LHC}},  {\em Phys. Rev. D} {\bf 107}
  (2023) 093006, [\href{http://arxiv.org/abs/2207.01458}{{\tt
  arXiv:2207.01458}}].

\bibitem{Ballestrero:2015jca}
A.~Ballestrero and E.~Maina, {\it {Interference Effects in Higgs production
  through Vector Boson Fusion in the Standard Model and its Singlet
  Extension}},  {\em JHEP} {\bf 01} (2016) 045,
  [\href{http://arxiv.org/abs/1506.02257}{{\tt arXiv:1506.02257}}].

\bibitem{BuarqueFranzosi:2017prc}
D.~Buarque~Franzosi and P.~Ferrarese, {\it {Implications of Vector Boson
  Scattering Unitarity in Composite Higgs Models}},  {\em Phys. Rev. D} {\bf
  96} (2017) 055037, [\href{http://arxiv.org/abs/1705.02787}{{\tt
  arXiv:1705.02787}}].

\bibitem{Searcy:2015apa}
J.~Searcy, L.~Huang, M.-A. Pleier, and J.~Zhu, {\it {Determination of the $WW$
  polarization fractions in $pp \to W^\pm W^\pm jj$ using a deep machine
  learning technique}},  {\em Phys. Rev. D} {\bf 93} (2016) 094033,
  [\href{http://arxiv.org/abs/1510.01691}{{\tt arXiv:1510.01691}}].

\bibitem{Lee:2018xtt}
J.~Lee, et~al., {\it {Polarization fraction measurement in same-sign WW
  scattering using deep learning}},  {\em Phys. Rev. D} {\bf 99} (2019) 033004,
  [\href{http://arxiv.org/abs/1812.07591}{{\tt arXiv:1812.07591}}].

\bibitem{Lee:2019nhm}
J.~Lee, et~al., {\it {Polarization fraction measurement in ZZ scattering using
  deep learning}},  {\em Phys. Rev. D} {\bf 100} (2019) 116010,
  [\href{http://arxiv.org/abs/1908.05196}{{\tt arXiv:1908.05196}}].

\bibitem{Grossi:2020orx}
M.~Grossi, J.~Novak, B.~Kersevan, and D.~Rebuzzi, {\it {Comparing traditional
  and deep-learning techniques of kinematic reconstruction for polarization
  discrimination in vector boson scattering}},  {\em Eur. Phys. J. C} {\bf 80}
  (2020) 1144, [\href{http://arxiv.org/abs/2008.05316}{{\tt
  arXiv:2008.05316}}].

\bibitem{Kim:2021gtv}
T.~Kim and A.~Martin, {\it {A $W^\pm$ polarization analyzer from Deep Neural
  Networks}},  \href{http://arxiv.org/abs/2102.05124}{{\tt arXiv:2102.05124}}.

\bibitem{Li:2021cbp}
J.~Li, C.~Zhang, and R.~Zhang, {\it {Polarization measurement for the
  dileptonic channel of W+W- scattering using generative adversarial network}},
   {\em Phys. Rev. D} {\bf 105} (2022) 016005,
  [\href{http://arxiv.org/abs/2109.09924}{{\tt arXiv:2109.09924}}].

\bibitem{Grossi:2023fqq}
M.~Grossi, M.~Incudini, M.~Pellen, and G.~Pelliccioli, {\it {Amplitude-assisted
  tagging of longitudinally polarised bosons using wide neural networks}},
  {\em Eur. Phys. J. C} {\bf 83} (2023) 759,
  [\href{http://arxiv.org/abs/2306.07726}{{\tt arXiv:2306.07726}}].

\bibitem{Aaboud:2019gxl}
{\bf ATLAS} Collaboration, M.~Aaboud et~al., {\it {Measurement of $W^{\pm}Z$
  production cross sections and gauge boson polarisation in $pp$ collisions at
  $\sqrt{s} ={}$13\,TeV with the ATLAS detector}},  {\em Eur. Phys. J. C} {\bf
  79} (2019) 535, [\href{http://arxiv.org/abs/1902.05759}{{\tt
  arXiv:1902.05759}}].

\bibitem{CMS:2021icx}
{\bf CMS} Collaboration, A.~Tumasyan et~al., {\it {Measurement of the inclusive
  and differential WZ production cross sections, polarization angles, and
  triple gauge couplings in pp collisions at $ \sqrt{s}=$13\,TeV}},  {\em JHEP}
  {\bf 07} (2022) 032, [\href{http://arxiv.org/abs/2110.11231}{{\tt
  arXiv:2110.11231}}].

\bibitem{ATLAS:2022oge}
{\bf ATLAS} Collaboration, G.~Aad et~al., {\it {Observation of gauge boson
  joint-polarisation states in $W^{\pm}Z$ production from $pp$ collisions at
  $\sqrt{s} = 13$ TeV with the ATLAS detector}},  {\em Phys. Lett. B} {\bf 843}
  (2023) 137895, [\href{http://arxiv.org/abs/2211.09435}{{\tt
  arXiv:2211.09435}}].

\bibitem{ATLAS:2023zrv}
{\bf ATLAS} Collaboration, G.~Aad et~al., {\it {Evidence of pair production of
  longitudinally polarised vector bosons and study of CP properties in ZZ
  \textrightarrow{} 4\ensuremath{\ell} events with the ATLAS detector at $
  \sqrt{s} $ = 13 TeV}},  {\em JHEP} {\bf 12} (2023) 107,
  [\href{http://arxiv.org/abs/2310.04350}{{\tt arXiv:2310.04350}}].

\bibitem{ATLAS:2024qbd}
{\bf ATLAS} Collaboration, G.~Aad et~al., {\it {Studies of the energy
  dependence of diboson polarization fractions and the Radiation Amplitude Zero
  effect in WZ production with the ATLAS detector}},
  \href{http://arxiv.org/abs/2402.16365}{{\tt arXiv:2402.16365}}.

\bibitem{Azzi:2019yne}
P.~Azzi et~al., {\it {Report from Working Group 1}: {Standard Model Physics at
  the HL-LHC and HE-LHC}},  in {\em {Report on the Physics at the HL-LHC,and
  Perspectives for the HE-LHC}} (A.~Dainese, et~al., eds.), vol.~7, pp.~1--220.
\newblock CERN, 12, 2019.
\newblock \href{http://arxiv.org/abs/1902.04070}{{\tt arXiv:1902.04070}}.

\bibitem{Roloff:2021kdu}
J.~Roloff, V.~Cavaliere, M.-A. Pleier, and L.~Xu, {\it {Sensitivity to
  longitudinal vector boson scattering in semileptonic final states at the
  HL-LHC}},  {\em Phys. Rev. D} {\bf 104} (2021) 093002,
  [\href{http://arxiv.org/abs/2108.00324}{{\tt arXiv:2108.00324}}].

\bibitem{Denner:2020bcz}
A.~Denner and G.~Pelliccioli, {\it {Polarized electroweak bosons in $W^+W^-$
  production at the LHC including NLO QCD effects}},  {\em JHEP} {\bf 09}
  (2020) 164, [\href{http://arxiv.org/abs/2006.14867}{{\tt arXiv:2006.14867}}].

\bibitem{Denner:2020eck}
A.~Denner and G.~Pelliccioli, {\it {NLO QCD predictions for doubly-polarized WZ
  production at the LHC}},  {\em Phys. Lett. B} {\bf 814} (2021) 136107,
  [\href{http://arxiv.org/abs/2010.07149}{{\tt arXiv:2010.07149}}].

\bibitem{Poncelet:2021jmj}
R.~Poncelet and A.~Popescu, {\it {NNLO QCD study of polarised $W^+ W^-$
  production at the LHC}},  {\em JHEP} {\bf 07} (2021) 023,
  [\href{http://arxiv.org/abs/2102.13583}{{\tt arXiv:2102.13583}}].

\bibitem{Denner:2022riz}
A.~Denner, C.~Haitz, and G.~Pelliccioli, {\it {NLO QCD corrections to polarized
  diboson production in semileptonic final states}},  {\em Phys. Rev. D} {\bf
  107} (2023) 053004, [\href{http://arxiv.org/abs/2211.09040}{{\tt
  arXiv:2211.09040}}].

\bibitem{Denner:2021csi}
A.~Denner and G.~Pelliccioli, {\it {NLO EW and QCD corrections to polarized ZZ
  production in the four-charged-lepton channel at the LHC}},  {\em JHEP} {\bf
  10} (2021) 097, [\href{http://arxiv.org/abs/2107.06579}{{\tt
  arXiv:2107.06579}}].

\bibitem{Le:2022lrp}
D.~N. Le and J.~Baglio, {\it {Doubly-polarized WZ hadronic cross sections at
  NLO QCD + EW accuracy}},  {\em Eur. Phys. J. C} {\bf 82} (2022) 917,
  [\href{http://arxiv.org/abs/2203.01470}{{\tt arXiv:2203.01470}}].

\bibitem{Le:2022ppa}
D.~N. Le, J.~Baglio, and T.~N. Dao, {\it {Doubly-polarized WZ hadronic
  production at NLO QCD+EW: calculation method and further results}},  {\em
  Eur. Phys. J. C} {\bf 82} (2022) 1103,
  [\href{http://arxiv.org/abs/2208.09232}{{\tt arXiv:2208.09232}}].

\bibitem{Dao:2023pkl}
T.~N. Dao and D.~N. Le, {\it {Enhancing the doubly-longitudinal polarization in
  WZ production at the LHC}},  {\em Commun. in Phys.} {\bf 33} (2023) 223,
  [\href{http://arxiv.org/abs/2302.03324}{{\tt arXiv:2302.03324}}].

\bibitem{Denner:2023ehn}
A.~Denner, C.~Haitz, and G.~Pelliccioli, {\it {NLO EW corrections to
  polarised~$W^+W^-$ production and decay at the LHC}},  {\em Phys. Lett. B}
  {\bf 850} (2024) 138539, [\href{http://arxiv.org/abs/2311.16031}{{\tt
  arXiv:2311.16031}}].

\bibitem{Dao:2023kwc}
T.~N. Dao and D.~N. Le, {\it {NLO electroweak corrections to doubly-polarized
  $W^+W^-$ production at the LHC}},  {\em Eur. Phys. J. C} {\bf 84} (2024) 244,
  [\href{http://arxiv.org/abs/2311.17027}{{\tt arXiv:2311.17027}}].

\bibitem{Hoppe:2023uux}
M.~Hoppe, M.~Sch\"onherr, and F.~Siegert, {\it {Polarised cross sections for
  vector boson production with Sherpa}},  {\em JHEP} {\bf 04} (2024) 001,
  [\href{http://arxiv.org/abs/2310.14803}{{\tt arXiv:2310.14803}}].

\bibitem{Pelliccioli:2023zpd}
G.~Pelliccioli and G.~Zanderighi, {\it {Polarised-boson pairs at the LHC with
  NLOPS accuracy}},  {\em Eur. Phys. J. C} {\bf 84} (2024) 16,
  [\href{http://arxiv.org/abs/2311.05220}{{\tt arXiv:2311.05220}}].

\bibitem{Javurkova:2024bwa}
M.~Javurkova, R.~Ruiz, R.~C.~L. de~S\'a, and J.~Sandesara, {\it {Polarized ZZ
  pairs in gluon fusion and vector boson fusion at the LHC}},  {\em Phys. Lett.
  B} {\bf 855} (2024) 138787, [\href{http://arxiv.org/abs/2401.17365}{{\tt
  arXiv:2401.17365}}].

\bibitem{Ballestrero:2017bxn}
A.~Ballestrero, E.~Maina, and G.~Pelliccioli, {\it {$W$ boson polarization in
  vector boson scattering at the LHC}},  {\em JHEP} {\bf 03} (2018) 170,
  [\href{http://arxiv.org/abs/1710.09339}{{\tt arXiv:1710.09339}}].

\bibitem{BuarqueFranzosi:2019boy}
D.~Buarque~Franzosi, O.~Mattelaer, R.~Ruiz, and S.~Shil, {\it {Automated
  predictions from polarized matrix elements}},  {\em JHEP} {\bf 04} (2020)
  082, [\href{http://arxiv.org/abs/1912.01725}{{\tt arXiv:1912.01725}}].

\bibitem{Ballestrero:2019qoy}
A.~Ballestrero, E.~Maina, and G.~Pelliccioli, {\it {Polarized vector boson
  scattering in the fully leptonic WZ and ZZ channels at the LHC}},  {\em JHEP}
  {\bf 09} (2019) 087, [\href{http://arxiv.org/abs/1907.04722}{{\tt
  arXiv:1907.04722}}].

\bibitem{Ballestrero:2020qgv}
A.~Ballestrero, E.~Maina, and G.~Pelliccioli, {\it {Different polarization
  definitions in same-sign $WW$ scattering at the LHC}},  {\em Phys. Lett. B}
  {\bf 811} (2020) 135856, [\href{http://arxiv.org/abs/2007.07133}{{\tt
  arXiv:2007.07133}}].

\bibitem{Ballestrero:2007xq}
A.~Ballestrero, A.~Belhouari, G.~Bevilacqua, V.~Kashkan, and E.~Maina, {\it
  {PHANTOM: A Monte Carlo event generator for six parton final states at high
  energy colliders}},  {\em Comput. Phys. Commun.} {\bf 180} (2009) 401--417,
  [\href{http://arxiv.org/abs/0801.3359}{{\tt arXiv:0801.3359}}].

\bibitem{Alwall:2014hca}
J.~Alwall, et~al., {\it {The automated computation of tree-level and
  next-to-leading order differential cross sections, and their matching to
  parton shower simulations}},  {\em JHEP} {\bf 07} (2014) 079,
  [\href{http://arxiv.org/abs/1405.0301}{{\tt arXiv:1405.0301}}].

\bibitem{Bothmann:2019yzt}
{\bf Sherpa} Collaboration, E.~Bothmann et~al., {\it {Event Generation with
  Sherpa 2.2}},  {\em SciPost Phys.} {\bf 7} (2019) 034,
  [\href{http://arxiv.org/abs/1905.09127}{{\tt arXiv:1905.09127}}].

\bibitem{Denner:2024ufg}
A.~Denner, M.~Pellen, M.~Sch\"onherr, and S.~Schumann, {\it {Tri-boson and WH
  production in the $\mathrm{W}^+\mathrm{W}^+\mathrm{j}\mathrm{j}$ channel:
  predictions at full NLO accuracy and beyond}},  {\em JHEP} {\bf 08} (6, 2024)
  043, [\href{http://arxiv.org/abs/2406.11516}{{\tt arXiv:2406.11516}}].

\bibitem{Denner:2024xul}
A.~Denner, D.~Lombardi, and C.~Schwan, {\it {Double-pole approximation for
  leading-order semi-leptonic vector-boson scattering at the LHC}},  {\em JHEP}
  {\bf 08} (2024) 146, [\href{http://arxiv.org/abs/2406.12301}{{\tt
  arXiv:2406.12301}}].

\bibitem{Stuart:1991cc}
R.~G. Stuart, {\it {General renormalization of the gauge invariant perturbation
  expansion near the $Z^0$ resonance}},  {\em Phys. Lett. B} {\bf 272} (1991)
  353--358.

\bibitem{Stuart:1991xk}
R.~G. Stuart, {\it {Gauge invariance, analyticity and physical observables at
  the $Z^0$ resonance}},  {\em Phys. Lett. B} {\bf 262} (1991) 113--119.

\bibitem{Aeppli:1993rs}
A.~Aeppli, G.~J. van Oldenborgh, and D.~Wyler, {\it {Unstable particles in one
  loop calculations}},  {\em Nucl. Phys.} {\bf B428} (1994) 126--146,
  [\href{http://arxiv.org/abs/hep-ph/9312212}{{\tt hep-ph/9312212}}].

\bibitem{Denner:2000bj}
A.~Denner, S.~Dittmaier, M.~Roth, and D.~Wackeroth, {\it {Electroweak radiative
  corrections to ${e}^+ {e}^- \to {W W} \to$ 4 fermions in double pole
  approximation: The RACOONWW approach}},  {\em Nucl. Phys.} {\bf B587} (2000)
  67--117, [\href{http://arxiv.org/abs/hep-ph/0006307}{{\tt hep-ph/0006307}}].

\bibitem{Beenakker:1997bp}
W.~Beenakker, A.~P. Chapovsky, and F.~A. Berends, {\it {Non-factorizable
  corrections to W~pair production}},  {\em Phys. Lett.} {\bf B411} (1997)
  203--210, [\href{http://arxiv.org/abs/hep-ph/9706339}{{\tt hep-ph/9706339}}].

\bibitem{Denner:1997ia}
A.~Denner, S.~Dittmaier, and M.~Roth, {\it {Nonfactorizable photonic
  corrections to ${e}^+ {e}^- \to {W W} \to $ 4 fermions}},  {\em Nucl. Phys.}
  {\bf B519} (1998) 39--84, [\href{http://arxiv.org/abs/hep-ph/9710521}{{\tt
  hep-ph/9710521}}].

\bibitem{Beenakker:1997ir}
W.~Beenakker, A.~P. Chapovsky, and F.~A. Berends, {\it {Non-factorizable
  corrections to W-pair production: Methods and analytic results}},  {\em Nucl.
  Phys.} {\bf B508} (1997) 17--63,
  [\href{http://arxiv.org/abs/hep-ph/9707326}{{\tt hep-ph/9707326}}].

\bibitem{Dittmaier:2015bfe}
S.~Dittmaier and C.~Schwan, {\it {Non-factorizable photonic corrections to
  resonant production and decay of many unstable particles}},  {\em Eur. Phys.
  J.} {\bf C76} (2016) 144, [\href{http://arxiv.org/abs/1511.01698}{{\tt
  arXiv:1511.01698}}].

\bibitem{Fadin:1993dz}
V.~S. Fadin, V.~A. Khoze, and A.~D. Martin, {\it {Interference radiative
  phenomena in the production of heavy unstable particles}},  {\em Phys. Rev.}
  {\bf D49} (1994) 2247--2256.

\bibitem{Melnikov:1993np}
K.~Melnikov and O.~I. Yakovlev, {\it {Top near threshold: all $\alpha_s$
  corrections are trivial}},  {\em Phys. Lett.} {\bf B324} (1994) 217--223,
  [\href{http://arxiv.org/abs/hep-ph/9302311}{{\tt hep-ph/9302311}}].

\bibitem{Fadin:1993kt}
V.~S. Fadin, V.~A. Khoze, and A.~D. Martin, {\it {How suppressed are the
  radiative interference effects in heavy instable particle production?}},
  {\em Phys. Lett.} {\bf B320} (1994) 141--144,
  [\href{http://arxiv.org/abs/hep-ph/9309234}{{\tt hep-ph/9309234}}].

\bibitem{Catani:2002hc}
S.~Catani, S.~Dittmaier, M.~H. Seymour, and Z.~Tr\'ocs\'anyi, {\it {The dipole
  formalism for next-to-leading order QCD calculations with massive partons}},
  {\em Nucl. Phys.} {\bf B627} (2002) 189--265,
  [\href{http://arxiv.org/abs/hep-ph/0201036}{{\tt hep-ph/0201036}}].

\bibitem{Basso:2015gca}
L.~Basso, S.~Dittmaier, A.~Huss, and L.~Oggero, {\it {Techniques for the
  treatment of IR divergences in decay processes at NLO and application to the
  top-quark decay}},  {\em Eur. Phys. J. C} {\bf 76} (2016) 56,
  [\href{http://arxiv.org/abs/1507.04676}{{\tt arXiv:1507.04676}}].

\bibitem{Schonherr:2017qcj}
M.~Sch{\"o}nherr, {\it {An automated subtraction of NLO EW infrared
  divergences}},  {\em Eur. Phys. J.} {\bf C78} (2018) 119,
  [\href{http://arxiv.org/abs/1712.07975}{{\tt arXiv:1712.07975}}].

\bibitem{Catani:1996vz}
S.~Catani and M.~H. Seymour, {\it {A general algorithm for calculating jet
  cross-sections in NLO QCD}},  {\em Nucl. Phys.} {\bf B485} (1997) 291--419,
  [\href{http://arxiv.org/abs/hep-ph/9605323}{{\tt hep-ph/9605323}}]. [Erratum:
  Nucl. Phys. {\bf B510} (1998) 503].

\bibitem{Veltman:1989vw}
M.~J.~G. Veltman and F.~J. Yndurain, {\it {Radiative Corrections to WW
  Scattering}},  {\em Nucl. Phys. B} {\bf 325} (1989) 1--17.

\bibitem{Dawson:1989up}
S.~Dawson and S.~Willenbrock, {\it {Radiative Corrections to Longitudinal
  Vector Boson Scattering}},  {\em Phys. Rev. D} {\bf 40} (1989) 2880.

\bibitem{Passarino:1990hk}
G.~Passarino, {\it {WW scattering and perturbative unitarity}},  {\em Nucl.
  Phys. B} {\bf 343} (1990) 31--59.

\bibitem{Denner:1997kq}
A.~Denner and T.~Hahn, {\it {Radiative Corrections to $W^+ W^-\to W^+ W^-$ in
  the Electroweak Standard Model}},  {\em Nucl. Phys. B} {\bf 525} (1998)
  27--50, [\href{http://arxiv.org/abs/hep-ph/9711302}{{\tt hep-ph/9711302}}].

\bibitem{Actis:2012qn}
S.~Actis, A.~Denner, L.~Hofer, A.~Scharf, and S.~Uccirati, {\it {Recursive
  generation of one-loop amplitudes in the Standard Model}},  {\em JHEP} {\bf
  04} (2013) 037, [\href{http://arxiv.org/abs/1211.6316}{{\tt
  arXiv:1211.6316}}].

\bibitem{Actis:2016mpe}
S.~Actis, et~al., {\it {RECOLA: REcursive Computation of One-Loop Amplitudes}},
   {\em Comput. Phys. Commun.} {\bf 214} (2017) 140--173,
  [\href{http://arxiv.org/abs/1605.01090}{{\tt arXiv:1605.01090}}].

\bibitem{Workman:2022ynf}
{\bf Particle Data Group} Collaboration, R.~L. Workman and Others, {\it {Review
  of Particle Physics}},  {\em PTEP} {\bf 2022} (2022) 083C01.

\bibitem{Bardin:1988xt}
D.~Bardin, A.~Leike, T.~Riemann, and M.~Sachwitz, {\it {Energy-dependent width
  effects in $e^+e^-$-annihilation near the Z-boson pole}},  {\em Physics
  Letters B} {\bf 206} (1988) 539--542.

\bibitem{Dittmaier:2001ay}
S.~Dittmaier and M.~Kr{\"a}mer, {\it {Electroweak radiative corrections to
  W-boson production at hadron colliders}},  {\em Phys. Rev.} {\bf D65} (2002)
  073007, [\href{http://arxiv.org/abs/hep-ph/0109062}{{\tt hep-ph/0109062}}].

\bibitem{Denner:2005fg}
A.~Denner, S.~Dittmaier, M.~Roth, and L.~H. Wieders, {\it {Electroweak
  corrections to charged-current ${e}^+ {e}^- \to$ 4 fermion processes:
  Technical details and further results}},  {\em Nucl. Phys.} {\bf B724} (2005)
  247--294, [\href{http://arxiv.org/abs/hep-ph/0505042}{{\tt hep-ph/0505042}}].
  [Erratum: Nucl. Phys. {\bf B854} (2012) 504].

\bibitem{Denner:2006ic}
A.~Denner and S.~Dittmaier, {\it {The complex-mass scheme for perturbative
  calculations with unstable particles}},  {\em Nucl. Phys. Proc. Suppl.} {\bf
  160} (2006) 22--26, [\href{http://arxiv.org/abs/hep-ph/0605312}{{\tt
  hep-ph/0605312}}].

\bibitem{Denner:2019vbn}
A.~Denner and S.~Dittmaier, {\it {Electroweak Radiative Corrections for
  Collider Physics}},  {\em Phys. Rept.} {\bf 864} (2020) 1--163,
  [\href{http://arxiv.org/abs/1912.06823}{{\tt arXiv:1912.06823}}].

\bibitem{NNPDF:2024djq}
{\bf NNPDF} Collaboration, R.~D. Ball et~al., {\it {Photons in the proton:
  implications for the LHC}},  {\em Eur. Phys. J. C} {\bf 84} (2024) 540,
  [\href{http://arxiv.org/abs/2401.08749}{{\tt arXiv:2401.08749}}].

\bibitem{Buckley:2014ana}
A.~Buckley, et~al., {\it {LHAPDF6: parton density access in the LHC precision
  era}},  {\em Eur. Phys. J.} {\bf C75} (2015) 132,
  [\href{http://arxiv.org/abs/1412.7420}{{\tt arXiv:1412.7420}}].

\bibitem{Cacciari:2008gp}
M.~Cacciari, G.~P. Salam, and G.~Soyez, {\it {The anti-$k_t$ jet clustering
  algorithm}},  {\em JHEP} {\bf 04} (2008) 063,
  [\href{http://arxiv.org/abs/0802.1189}{{\tt arXiv:0802.1189}}].

\bibitem{Denner:2016kdg}
A.~Denner, S.~Dittmaier, and L.~Hofer, {\it {COLLIER: a fortran-based Complex
  One-Loop LIbrary in Extended Regularizations}},  {\em Comput. Phys. Commun.}
  {\bf 212} (2017) 220--238, [\href{http://arxiv.org/abs/1604.06792}{{\tt
  arXiv:1604.06792}}].

\bibitem{Hirschi:2011pa}
V.~Hirschi, et~al., {\it {Automation of one-loop QCD corrections}},  {\em JHEP}
  {\bf 05} (2011) 044, [\href{http://arxiv.org/abs/1103.0621}{{\tt
  arXiv:1103.0621}}].

\bibitem{Nagy:1998bb}
Z.~Nagy and Z.~Tr\'ocs\'anyi, {\it {Next-to-leading order calculation of
  four-jet observables in electron-positron annihilation}},  {\em Phys. Rev.}
  {\bf D59} (1999) 014020, [\href{http://arxiv.org/abs/hep-ph/9806317}{{\tt
  hep-ph/9806317}}]. [Erratum: Phys. Rev. {\bf D62} (2000) 099902].

\bibitem{Denner:2000jv}
A.~Denner and S.~Pozzorini, {\it {One-loop leading logarithms in electroweak
  radiative corrections. 1. Results}},  {\em Eur. Phys.~J.} {\bf C18} (2001)
  461--480, [\href{http://arxiv.org/abs/hep-ph/0010201}{{\tt hep-ph/0010201}}].

\end{thebibliography}\endgroup
